\documentclass[preprint2,tighten]{aastex62_reep}

\usepackage{epstopdf}
\usepackage{subfigure}  
\usepackage{amsmath}
\usepackage{gensymb}

\usepackage{longtable}

\shorttitle{GOES Flare Duration}
\shortauthors{Reep \& Knizhnik}

\begin{document}

\title{What determines the X-ray intensity and duration of a solar flare?}

\author[0000-0003-4739-1152]{Jeffrey W. Reep}
\affiliation{Space Science Division, Naval Research Laboratory, Washington, DC 20375, USA; \href{mailto:jeffrey.reep@nrl.navy.mil}{jeffrey.reep@nrl.navy.mil}}

\author[0000-0002-2544-2927]{Kalman J. Knizhnik}
\affiliation{National Research Council Post-doctoral Associate, Space Science Division, Naval Research Laboratory, Washington, DC 20375, USA}

\begin{abstract}
Solar flares are observed and classified according to their intensity measured with the GOES X-ray Sensors.  We show that the duration of a flare, as measured by the full width at half maximum (FWHM) in GOES is not related to the size of the flare as measured by GOES intensity.  The durations of X-class flares range from a few minutes to a few hours, and the same is true of M- and C-class flares.  In this work, we therefore examine the statistical relationships between the basic properties of flares -- temperature, emission measure, energy, \textit{etc.} -- in comparison to both their size and duration.  We find that the size of the flare is directly related to all of these basic properties, as previously found by many authors.  The duration is not so clear.  When examining the whole data set, the duration appears to be independent of all of these properties.  In larger flares, however, there are direct correlations between the GOES FWHM and magnetic reconnection flux and ribbon area.  We discuss the possible explanations, finding that this discrepancy may be due to large uncertainties in small flares, though we cannot rule out the possibility that the driving physical processes are different in smaller flares than larger ones.  We discuss the implications of this result and how it relates to the magnetic reconnection process that releases energy in flares.
\end{abstract}
\keywords{Sun: flares, Sun: activity, Sun: X-rays, stars: flares}

\section{Introduction}
\label{sec:introduction}

Solar flares are driven by magnetic reconnection, releasing energy that drives heating, particle acceleration, and magnetohydrodynamic wave excitation.  The chromosphere is strongly heated, raising the temperature and driving an increase in pressure that causes the ablation of material into the corona, brightening up flaring loops in the extreme ultraviolet and soft X-rays (SXRs).  The SXR brightening is routinely measured with NOAA's Geostationary Operational Environmental Satellite (GOES, \citealt{donnelly1977}) network, with the X-ray Sensor (XRS) on board.  XRS measures spatially unresolved light curves in two wavelength bands: 1--8\,\AA\ (1.5--12 keV) and 0.5--4\,\AA\ (3--24 keV).  The flux levels in GOES XRS are used to report flares and classify their size according to their brightness in the 1--8\,\AA\ band.  

GOES XRS measurements therefore contain fundamental parameters of solar flares: their size as measured by peak X-ray flux, their duration, their rate of occurrence, and the solar background levels.  Furthermore, the ratio of fluxes in the two XRS channels can be used to derive time-varying temperatures and emission measures (EMs), giving estimates of the plasma parameters that produce the flares \citep{thomas1985,garcia1994,white2005}.  

Because the observations with GOES are spatially unresolved, however, many other properties of flares cannot be determined.  For example, spatial resolution is required to determine the location on the Sun, size of the chromospheric footpoint brightenings (ribbons), volume, magnetic flux and geometry, \textit{etc}.  In a recent paper, \citet{kazachenko2017} has derived many of these properties using a combination of data taken with the Atmospheric Imaging Assembly (AIA, \citealt{lemen2012}) and Helioseismic and Magnetic Imager (HMI, \citealt{schou2012}) onboard the Solar Dynamics Observatory (SDO, \citealt{pesnell2012}).  \citet{kazachenko2017} have created a database, the RibbonDB, of 3137 flares observed by AIA and HMI, that includes the locations, active region area, ribbon area, and magnetic flux swept out by the flares.  

In this work, we combine the RibbonDB with measurements by GOES of fundamental flare properties.  We examine relationships between the size, duration, and plasma properties of the flares in order to better understand what drives these flares.  We derive the flux, full width at half maximum (FWHM), temperature, and EM from the GOES light curves.  We also combine the volume measurements from the RibbonDB with the GOES data to derive thermal energy content in the flares.  In Section \ref{sec:methodology}, we explain the basic methodology.  In Section \ref{sec:flux}, we discuss the relationships between the size (GOES class) of flares and basic properties.  In Section \ref{sec:duration}, we discuss the relationships between flare duration and basic properties.   Finally, in Section \ref{sec:discussion}, we discuss the implications of our findings.

\section{Flare Selection and Methodology}
\label{sec:methodology}

We have examined 2956 flares, using a subset of the RibbonDB \citep{kazachenko2017}.  We choose events that are sufficiently isolated in time from other events (that is, the FWHM of the GOES X-ray light curves is well-defined in both channels) and do not have any data gaps in the GOES data.  The RibbonDB additionally only includes flares greater than C1.0 (pre-background subtraction) that occurred within 45\degree\ of disk center, over the time period April 2010 -- April 2016.  After background subtraction, there were 15 X-class flares (0.5\%), 242 M-class flares (8.2\%), and 2699 C-class or smaller flares (91.3\%).

\begin{figure}
    \centering
    \includegraphics[width=\linewidth]{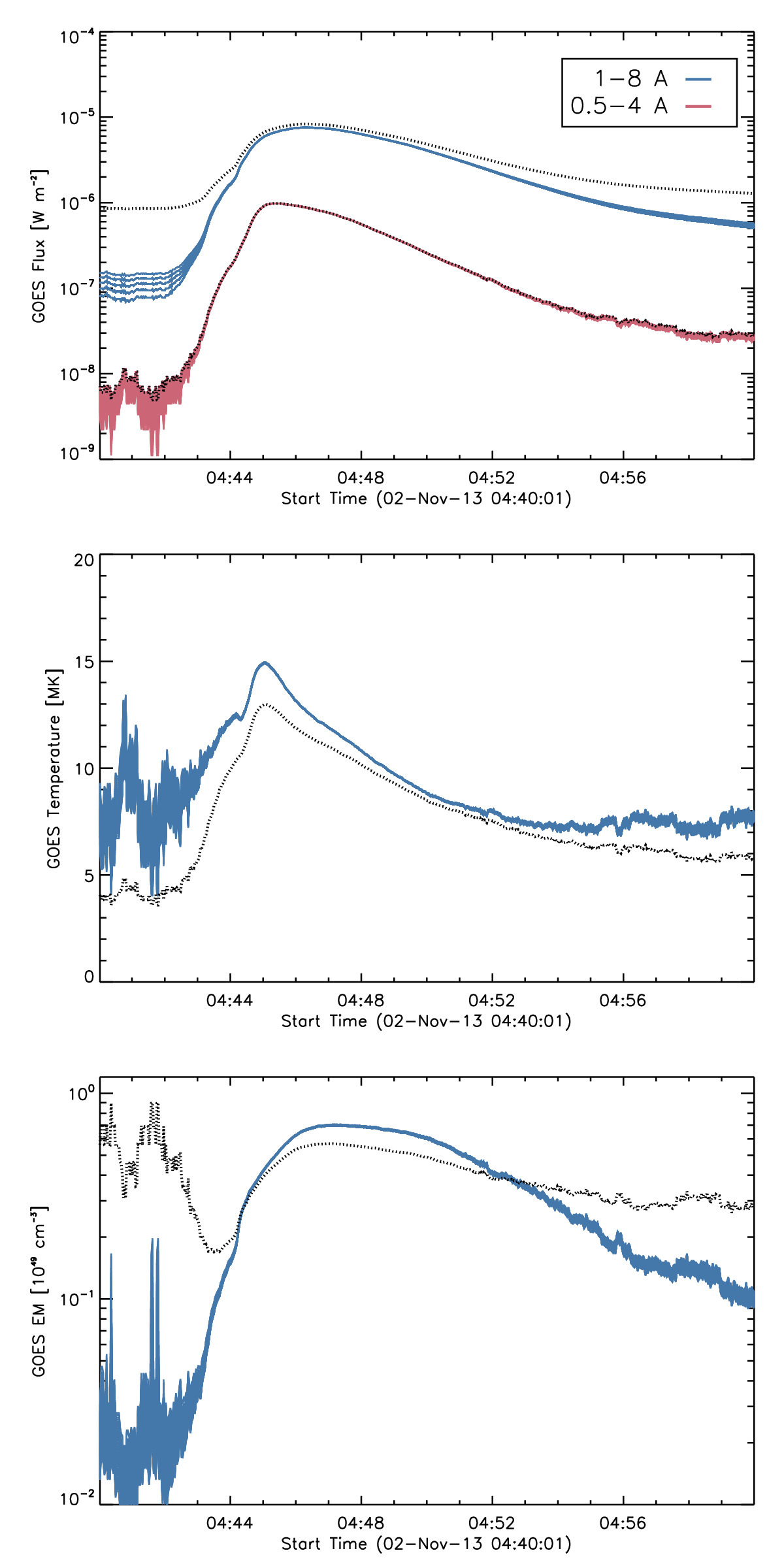}
    \caption{An example of the TEBBS algorithm \citep{ryan2012} for GOES background subtraction.  The plots show the 2 Nov. 2013 C8.2 flare.  The solid lines show the range of background subtracted values for the flux (top), temperature (middle), and EM (bottom), while the dotted black lines show the values without background subtracting.  Background subtraction can drastically alter the derived temperatures and EMs.}
    \label{fig:tebbs}
\end{figure}
We have implemented and used the TEBBS algorithm \citep{ryan2012} to perform background subtractions of the GOES data.  As explained by \citet{ryan2012}, estimates of the temperature and EM derived from the GOES flux ratio can be wildly inaccurate when either not background subtracting or subtracting with the pre-flare flux values.  The TEBBS algorithm gives a range of possible background levels that can be used to estimate the true fluxes, and therefore better approximate the temperature and EM.  For example, in Figure \ref{fig:tebbs}, we show the GOES light curves for the 2 November 2013 C8.2 flare with the range of possible fluxes, temperatures, and EMs derived with the TEBBS algorithm.  The dotted black curves show the original curves without background subtraction as a comparison.  It is clear that the derived EM and temperature, and their time profiles, change drastically with proper background subtraction.  In the scatter plots in Section \ref{sec:flux},  we use the median value of possible GOES light curves found with the TEBBS algorithm in order to estimate the peak temperature, EM, \textit{etc}.

Where pertinent, we have calculated both Pearson's correlation coefficient $r_{p}$, measuring the linear correlation between two variables, as well as Spearman's rank correlation coefficient $r_{s}$, measuring the monotonicity between two variables.  In cases where the data appear to be linearly correlated (in log-log space), we have used a Theil-Sen estimator to fit the linear correlation, and used a 95\% confidence interval to measure the uncertainties.  We indicate the slope and uncertainties on the plots in these cases.  In a few cases, where the data appear correlated but non-linear, we have fit a quadratic in log-log space to guide the eye.  The equation we fit is 
\begin{align}
\log{y} &= a + b \log{x} + c\ (\log{x})^{2} 
\label{eqn:quad}
\end{align}
\noindent which can be rewritten
\begin {align}
y = a x^{b + c \log{x}}
\end{align}
\noindent While the fit appears adequate in these cases, we do not expect that this function actually represents the physical relationship between the variables, so we do not list the fitted coefficients.  We avoid assumptions concerning correlations where possible, and have noted where there are clear biases or selection effects.  We begin by examining the geometry of the events as they relate to the flare size and duration.

\section{GOES Soft X-ray Flux}
\label{sec:flux}

\begin{figure}
\centering
\includegraphics[width=\linewidth]{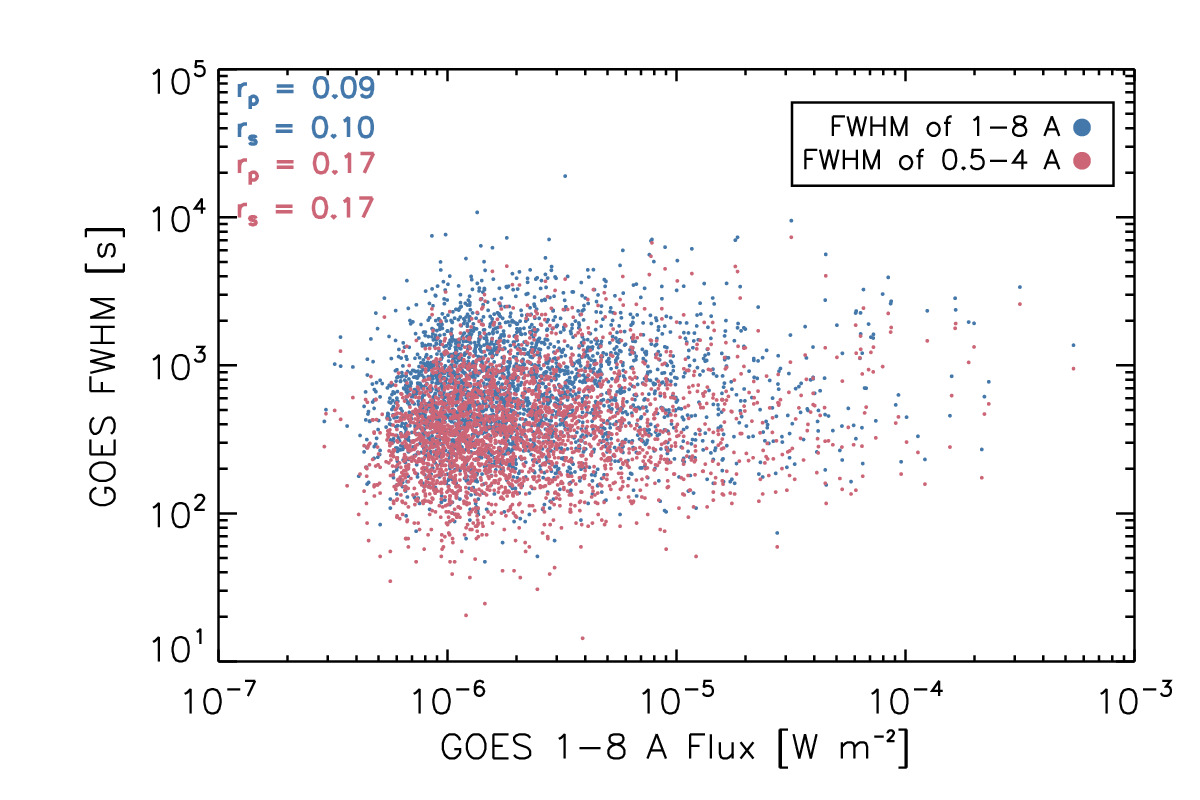}
\caption{A scatter plot showing the GOES class (peak flux in 1--8\,\AA\ channel) plotted against the FWHM of each channel, in blue (1--8\,\AA) and red (0.5--4\,\AA).  The Pearson correlation coefficient $r_{P}$ and Spearman rank correlation coefficient $r_{S}$ are indicated in the figure for each channel.  Since the values of the correlation coefficients are close to 0, we conclude that the two variables are not related.  In other words, flare class is independent of flare duration.  \label{fig:goesvsfwhm}}
\end{figure}

In this section, we first discuss the GOES flare classification as it relates to the basic properties of flares.  We will see that the GOES SXR flux depends intimately on the energy release, magnetic reconnection flux, temperature, EM, and ribbon area, as many previous studies have found \citep{feldman1996,veronig2002,emslie2005,warmuth2016a,warmuth2016b,kazachenko2017,sadykov2018}.  However, we find that there is no relation between the class of a flare and its SXR duration, as shown in Figure \ref{fig:goesvsfwhm}.  In this figure, we plot the GOES class, measured as the peak flux in the 1--8\,\AA\ channel against the FWHM of both channels for our data set of 2956 flares.  We show the 1--8\,\AA\ channel in blue, and the 0.5--4\,\AA\ channel in red, and use this convention for the rest of this paper.  We have indicated both the Pearson correlation coefficient and Spearman rank correlation coefficient for each channel.  The linear correlation was tested in log-log space, which is equivalent to fitting a power law in linear space.  The Pearson correlation coefficients are close to 0, indicating that there is no direct relation between the two variables, in either GOES channel.  By comparison, \citet{veronig2002} found a correlation coefficient $r_{p} = 0.25$ in their study of 1--8\,\AA\ GOES data.  This absence of correlation is surprising: shouldn't larger flares with a larger energy release last for a longer time?  In this work, we seek to examine this question in detail.

\begin{figure*}
\includegraphics[width=0.5\linewidth]{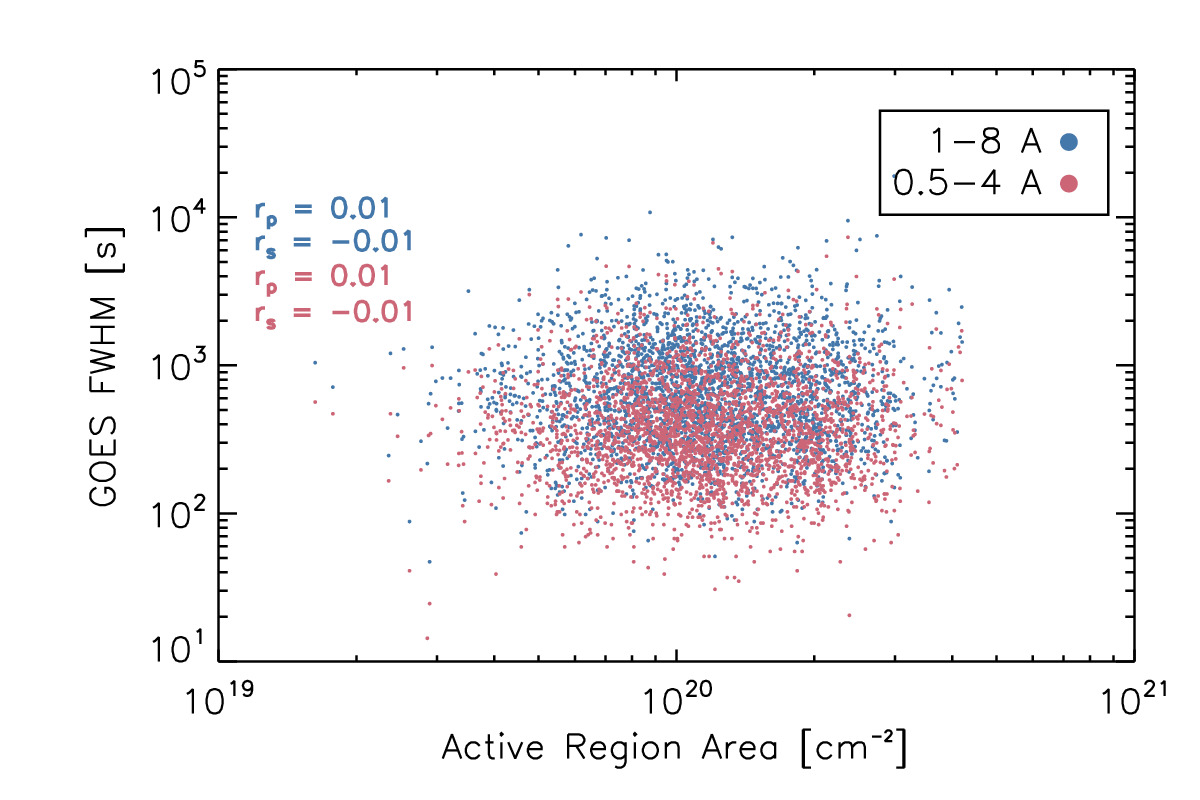}
\includegraphics[width=0.5\linewidth]{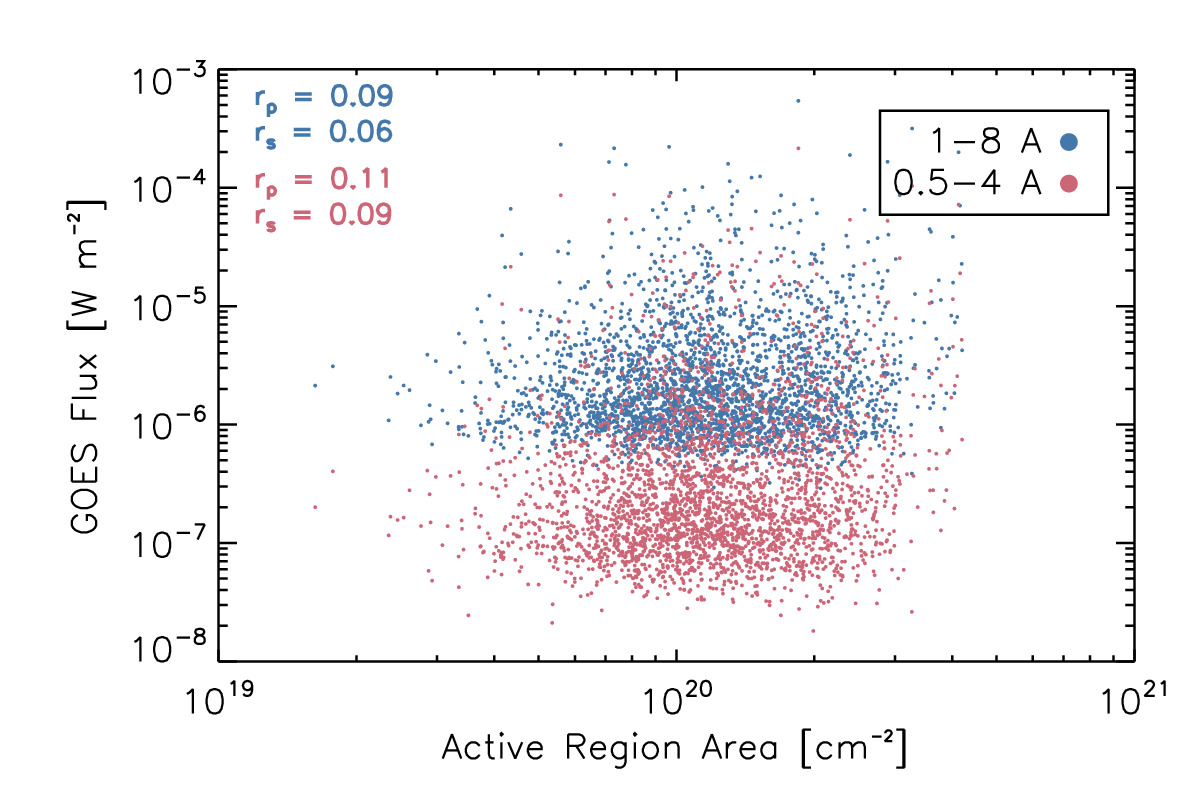}
\caption{The active region area plotted against the FWHM (left) and peak GOES flux (right) in both channels for the sample of flares.  There is no correlation between the active region area and flare duration or with the GOES flux (compare with \citealt{harra2016}). \label{fig:s_ar}}
\end{figure*}
We begin with the geometry of a flare.  We find that the area of the active region in which a flare occurs is not related to the size or duration of a flare.  Figure \ref{fig:s_ar} shows scatter plots of the AR area in which a flare occurred against the GOES FWHM (left) and peak GOES flux (right) in both channels.  The active region area was taken from the RibbonDB \citep{kazachenko2017}, measured with SDO/HMI continuum data.  From the scatter and the correlation coefficients, we find that there is no direct correlation.  In comparison, in a small set of X-class flares, \citet{harra2016} found a small correlation between the flare duration and sunspot area measured with SDO/HMI continuum data.  Those authors find no correlation between the SXR flux and sunspot area.  It is also worth noting here that we have not differentiated the ARs according to magnetic topology, and that different AR types could affect measured correlations (flares are more likely in certain magnetic configurations \textit{e.g.} \citealt{toriumi2017a,toriumi2017b}).  

\begin{figure*}
\includegraphics[width=0.5\linewidth]{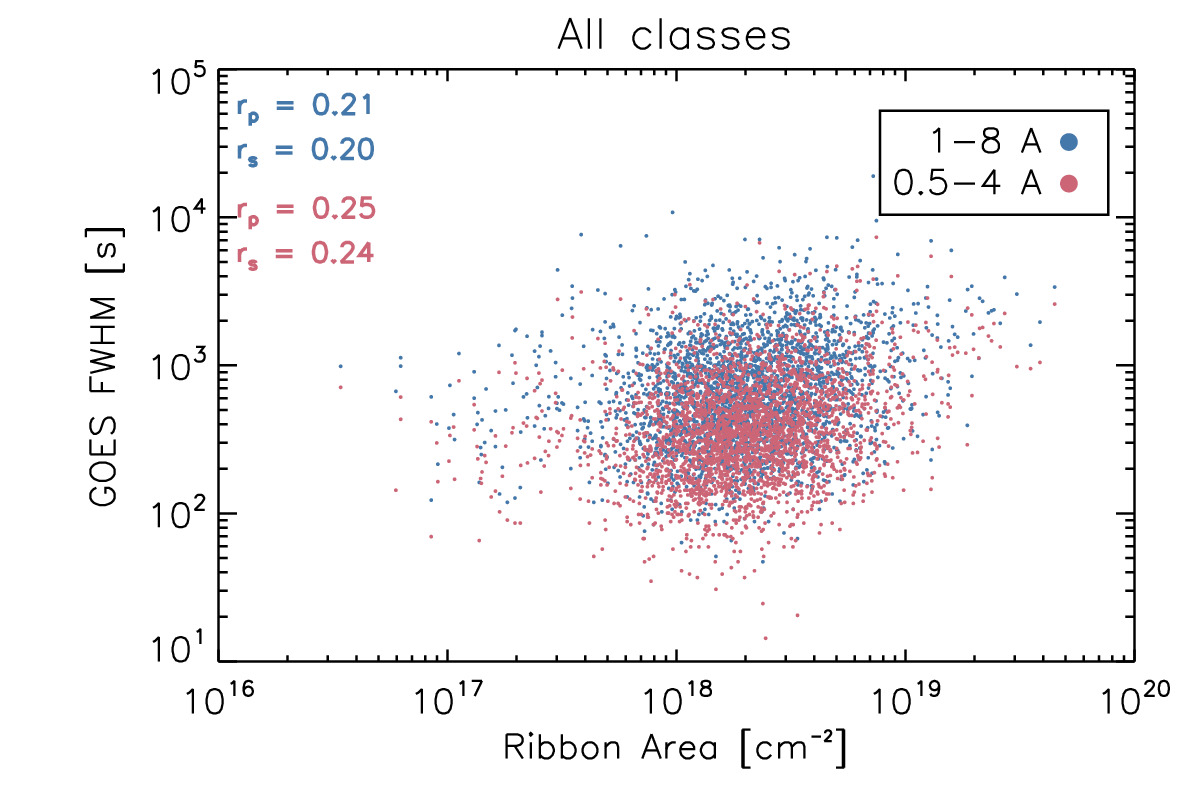}
\includegraphics[width=0.5\linewidth]{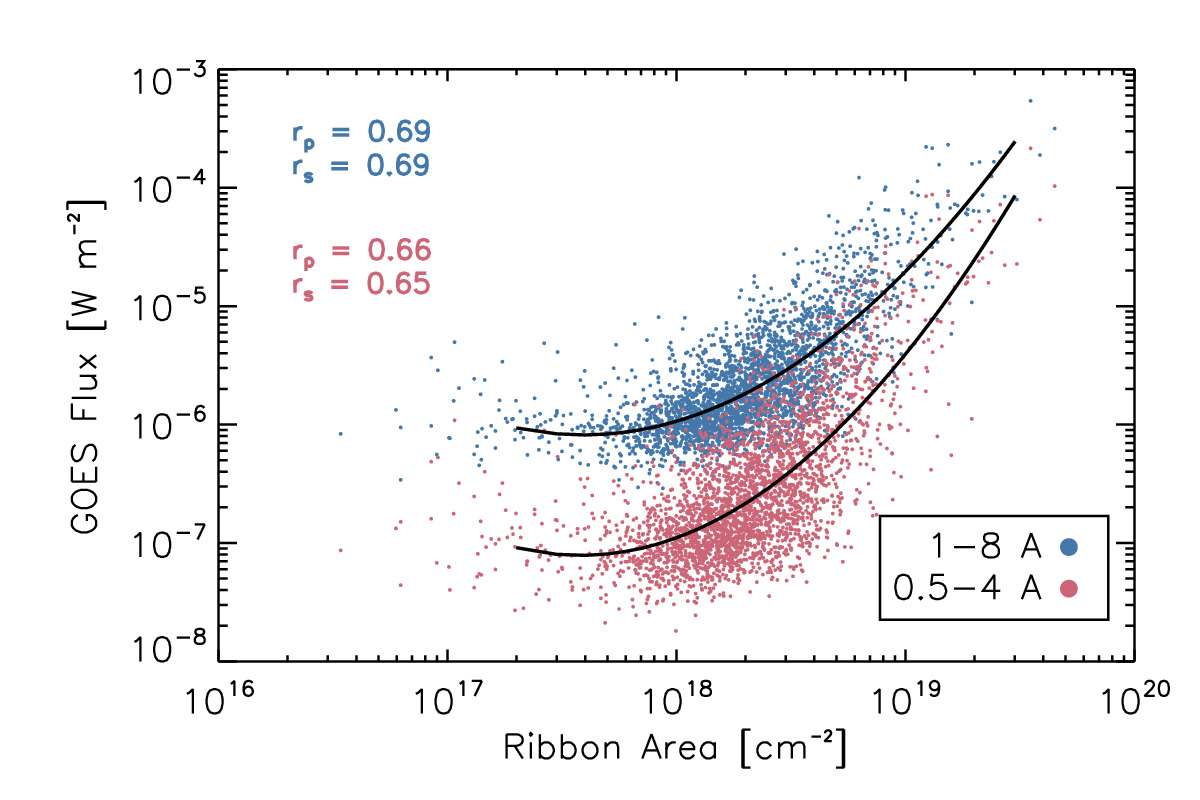}
\caption{The flare ribbon area plotted against the FWHM (left) and peak GOES flux (right) in both channels for the sample of flares.  There is no correlation between the flare ribbon area and flare duration, but there is a strong correlation with the GOES flux in both channels, which we indicate with a quadratic fit in log-log space, meant only to demonstrate the trend (see Equation \ref{eqn:quad}). \label{fig:s_rbn}}
\end{figure*}
In contrast, the area of the flare ribbons is related to the size of a flare, though not the duration (see also Section \ref{sec:duration}).  Figure \ref{fig:s_rbn} shows scatter plots relating these quantities.  The ribbon area is likely not correlated with the duration in either channel (though the lack of flares smaller than C1.0 represents a sampling bias).  In this case, however, we do find a strong and likely non-linear correlation between the GOES flux in both channels and the ribbon area.  We note that the slope appears to steepen for larger flares, additionally.  

\begin{figure*}
\includegraphics[width=0.5\linewidth]{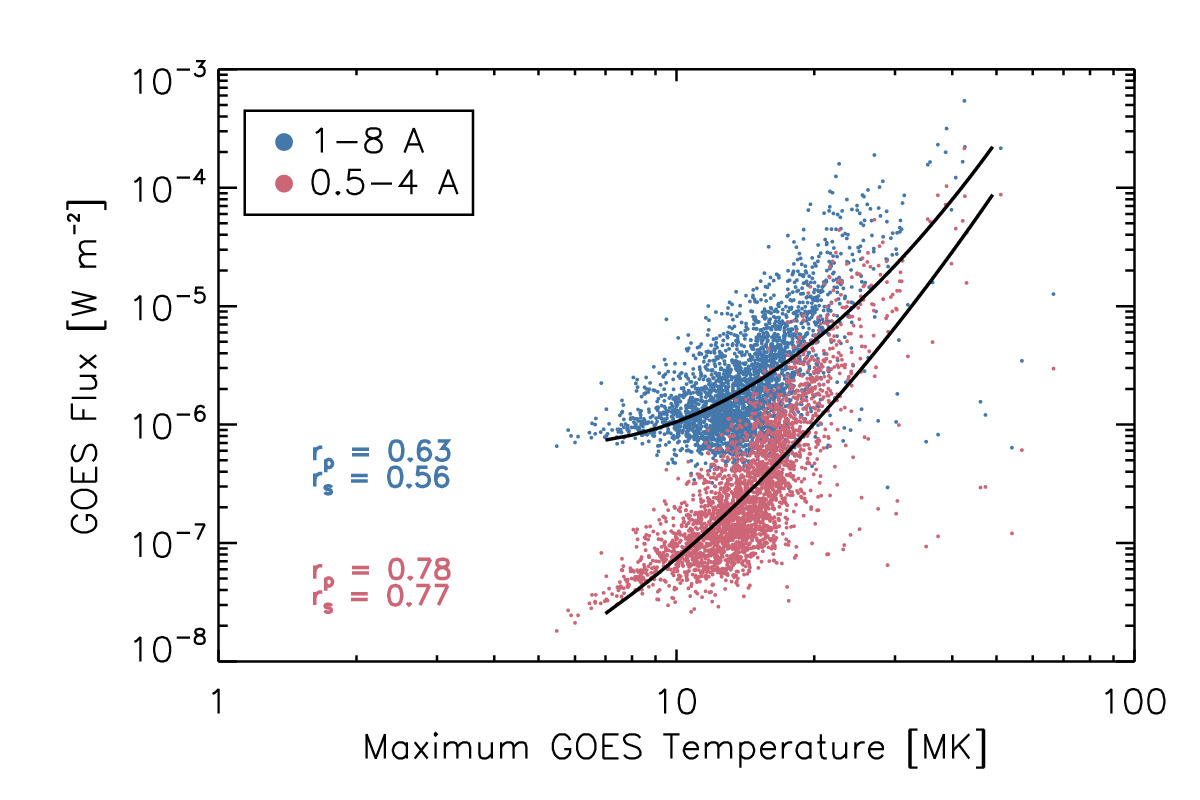}
\includegraphics[width=0.5\linewidth]{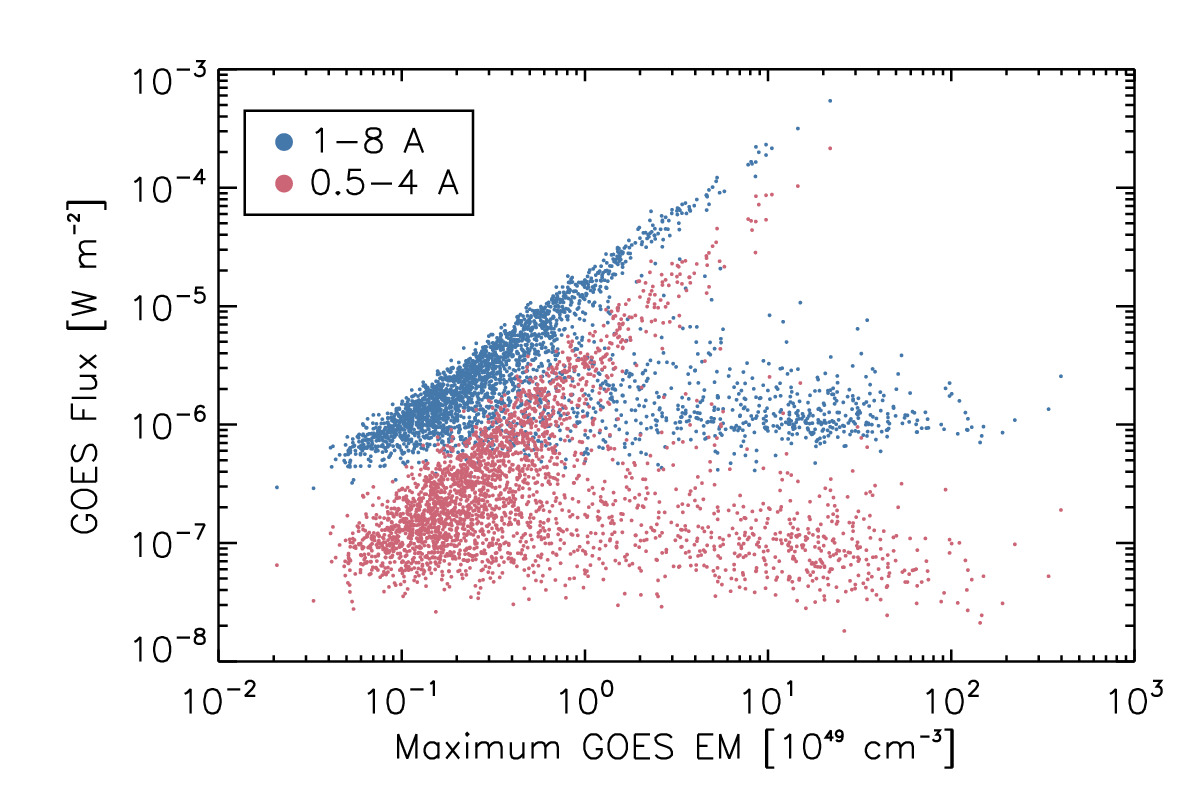}
\includegraphics[width=0.5\linewidth]{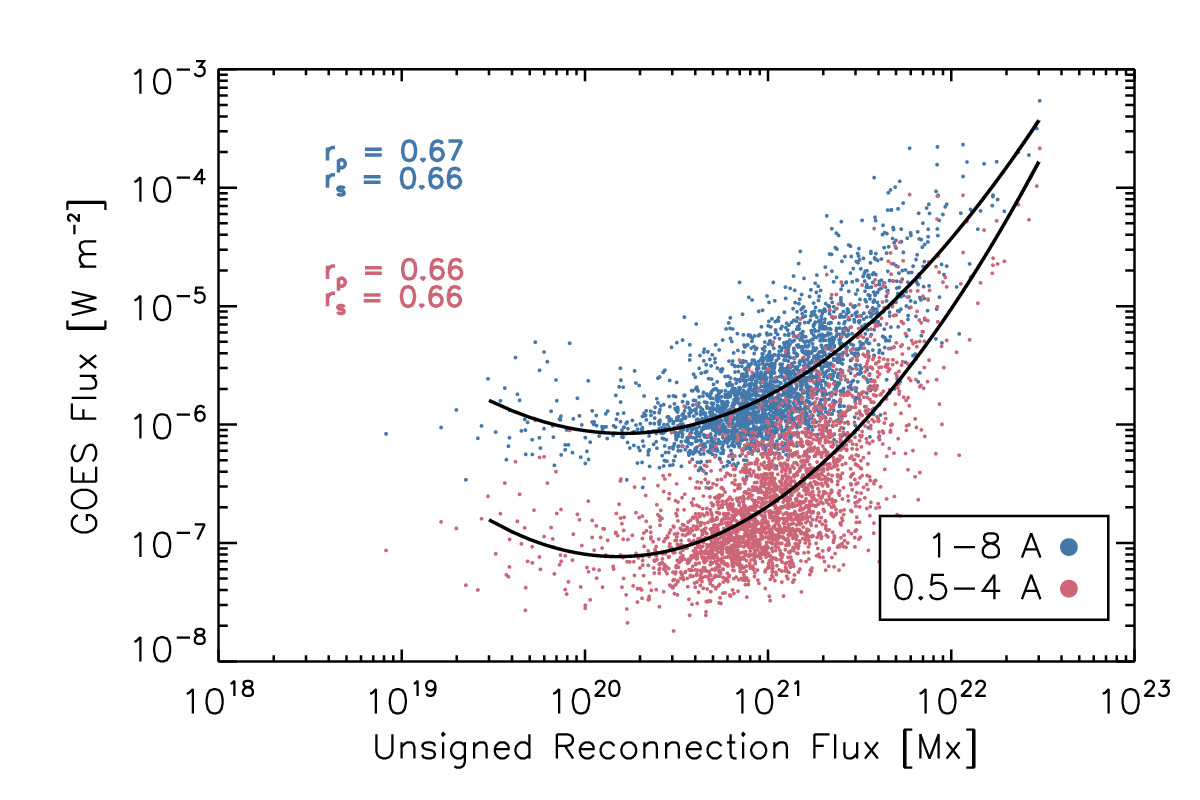}
\includegraphics[width=0.5\linewidth]{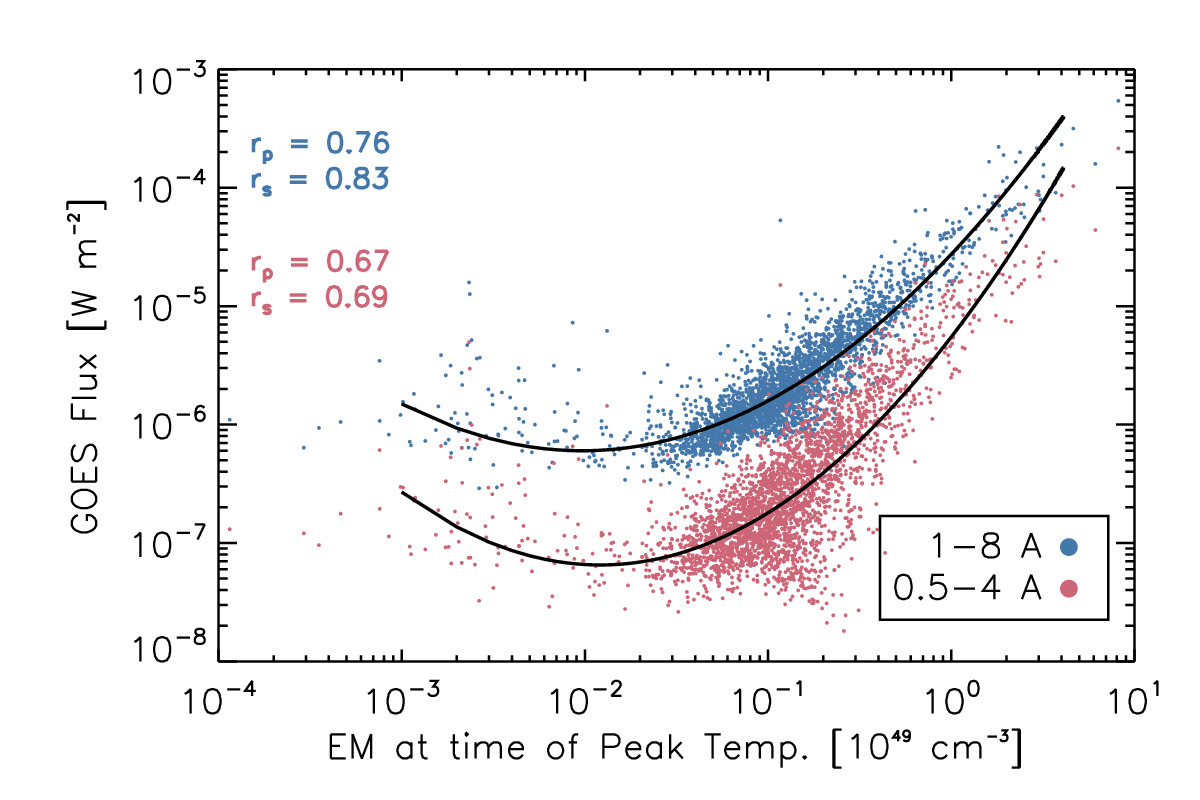}
\caption{The relationship between the GOES flux and the maximum GOES temperature (top left), maximum GOES EM (top right), unsigned reconnection flux (bottom left), and the EM at the time of the maximum temperature (bottom right).  The solid lines are quadratic fits, once again only to show the basic trend in the data (see Equation \ref{eqn:quad}).\label{fig:goes}}
\end{figure*}
The peak in the GOES flux scales directly with all of the basic properties of a flare.  In Figure \ref{fig:goes}, we show the relationship between the peak GOES flux in both channels with the maximum GOES temperature (top left), maximum GOES EM (top right), the unsigned magnetic reconnection flux (bottom left), and the GOES EM at the time of the maximum temperature (bottom right).  The GOES flux scales extremely strongly with the maximum temperature, with a sharper rise in the higher energy channel.  These trends can be compared against those found by \citet{feldman1996}, who measured flare temperatures with Yohkoh/BCS and found similar trends.  Those authors note that the GOES temperature and electron temperature measured by BCS do not agree well above $\approx 10$\,MK and they did not background subtract the GOES light curves, so their fits do not work well with the present data.  Since thermal bremsstrahlung emissions scale strongly with temperature (\textit{e.g.} \citealt{culhane1970}), and since the 0.5--4\,\AA\ channel is more sensitive to higher temperatures (\textit{e.g.} \citealt{warren2004}), we find that the scaling is strongest in the higher energy channel, but both scale rapidly with small changes in temperature.

The second plot in Figure \ref{fig:goes} shows how the flux in both channels scales with the maximum EM.  In this case, there are two populations of flares: those where the flux scales very strongly with the maximum EM, and those where the flux is independent of the maximum EM.  The first population of flares comprise the strongest flares in the data set, and in all of those flares, the temperature peaks prior to the peak of the EM, consistent with the standard chromospheric evaporation model \citep{hirayama1974,antiochos1978}.  In the second population of flares, the flux is independent of the maximum emission measure, and there are some flares where the temperature peaks first, and some where the EM peaks first.  In total, the EM peaks first in 10\% of the flares (297 out of 2956), while the temperature peaks first in 90\% of the flares (2659 out of 2956).  This should be contrasted with the recent study by \citet{sadykov2018}, who found that in 96\% of more than 15,000 flares the temperature peaked before the EM.\footnote{It is crucial that the background be properly subtracted in order to determine this number.  Without background subtracting, this 10\% rises to 30\% due to poor estimates of the EM.  See \citet{ryan2012} for a full discussion.}  

The third plot in Figure \ref{fig:goes} shows that there is a clear correlation between the unsigned magnetic reconnection flux $\Phi_{\text{ribbon}}$ and the SXR flux.  Unsurprisingly, larger flares sweep out more magnetic flux.  Note that the 1--8\,\AA\ plot here is identical to Figure 8b of \citet{kazachenko2017}, who find the relation $F_{1-8} \propto (\Phi_{\text{ribbon}})^{1.5}$ using the Levenberg-Marquardt algorithm to fit the function $y = ax^{b}$ (in linear space).

Finally, the fourth plot in Figure \ref{fig:goes} shows a clear correlation between the EM at the time of the peak temperature and the GOES flux.  The direct correlation here and absence of two distinct populations as in the top right plot of Figure \ref{fig:goes} suggests that the EM at the time of the peak temperature determines the peak flux, rather than the peak value of the EM.  This is because the response of GOES XRS increases monotonically with temperature in both channels (see \textit{e.g.} \citealt{warren2004}), and since the emission scales linearly with the EM.   

\begin{figure}
    \centering
    \includegraphics[width=\linewidth]{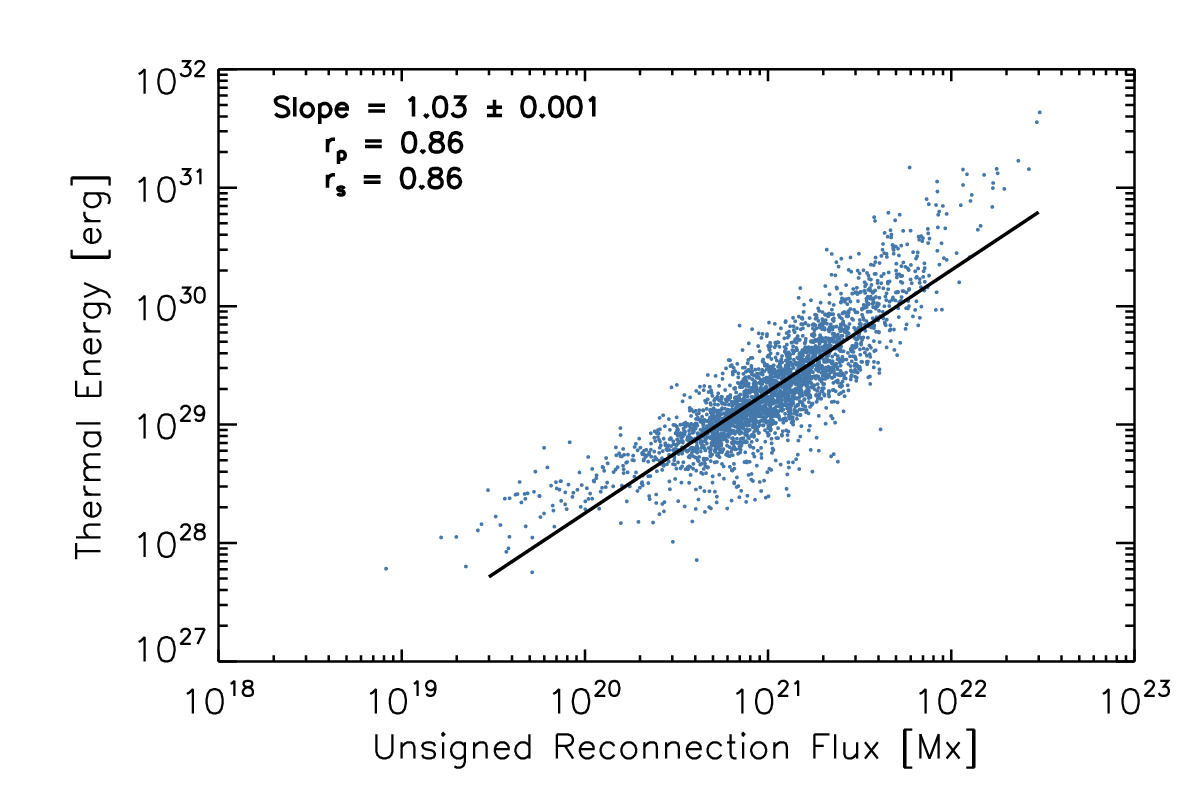}
    \caption{The correspondence between the unsigned magnetic reconnection flux and the total thermal energy content in a flare.  The two are linearly correlated, indicating that the reconnection flux is a good proxy for flare energy.}
    \label{fig:phithermal}
\end{figure}
How does the reconnection flux compare with the total energy release?  The thermal energy content of a flare can be written as $E_{th} = 3 n k_{B} T V$, which can be rewritten in terms of measured quantities.  The emission measure is $EM \approx n^{2} V$, which implies $n \approx \sqrt{\frac{EM}{V}}$.  Approximating the volume $V$ as $(\frac{S_{\text{ribbon}}}{2})^{3/2}$, we then find $E_{th} = \frac{3}{8^{1/4}} k_{B} (EM)^{1/2} S_{\text{ribbon}}^{3/4} T$.  We use the EM at the time of the peak temperature in this approximation, as this is what determines the GOES flux (as opposed to the maximum EM).  In Figure \ref{fig:phithermal}, we show that this scales linearly with the reconnection flux $\Phi_{\text{ribbon}}$, indicating that $\Phi_{\text{ribbon}}$ is a valid proxy for the flare's energy.  Note that this approximation assumes an isothermal fit, which generally under-estimates the total energy content compared to a multithermal fit (see \textit{e.g.} \citealt{aschwanden2015}).

\begin{figure*}
\includegraphics[width=0.5\linewidth]{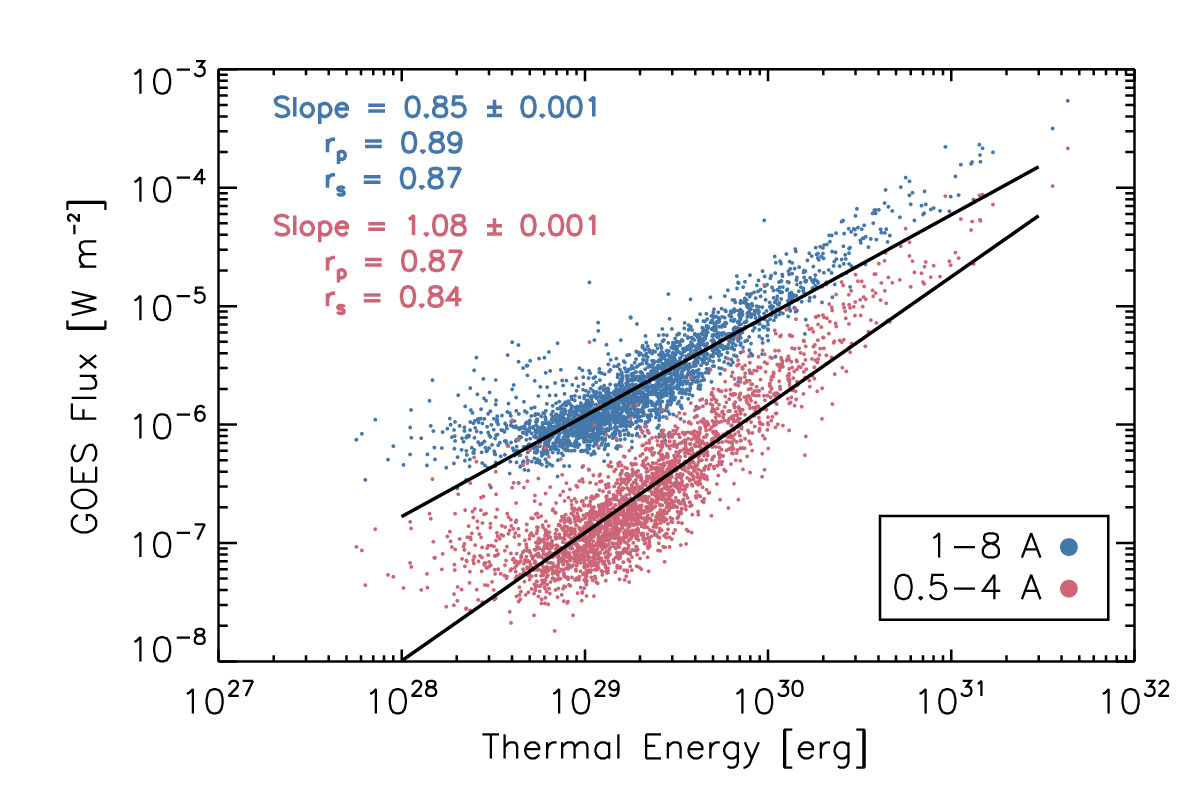}
\includegraphics[width=0.5\linewidth]{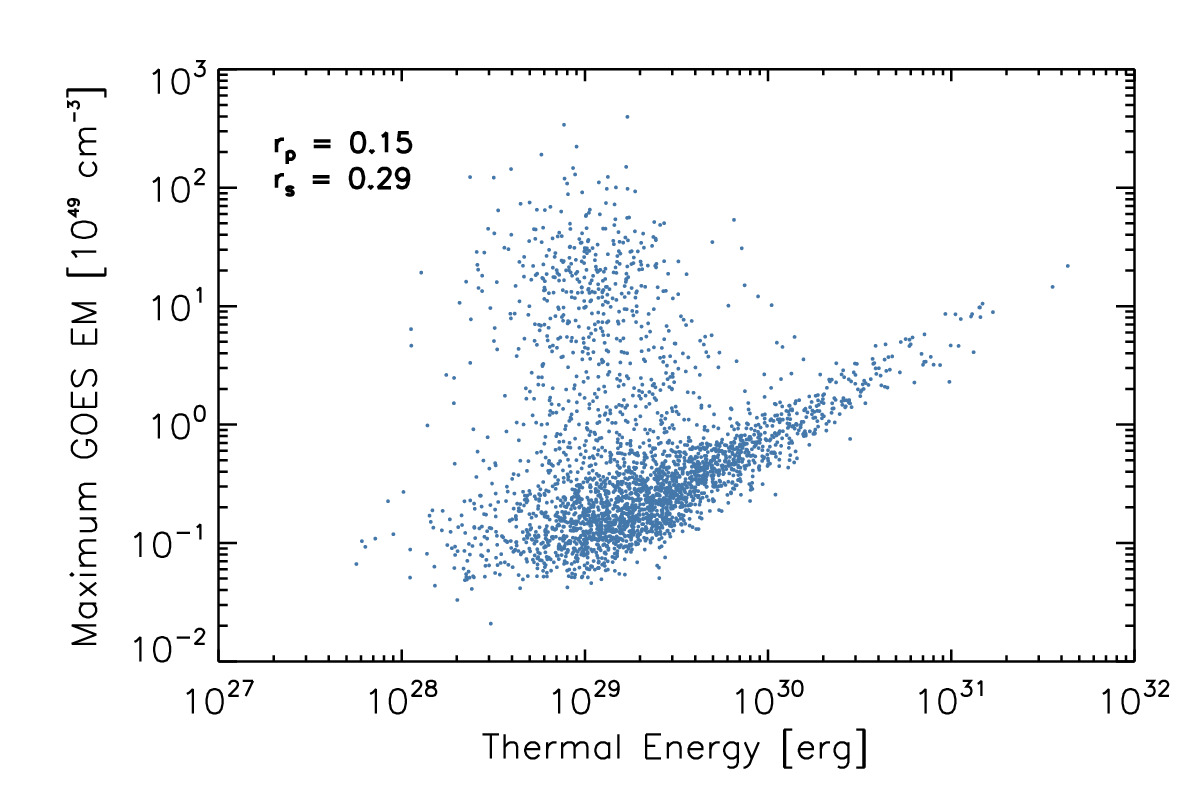}
\includegraphics[width=0.5\linewidth]{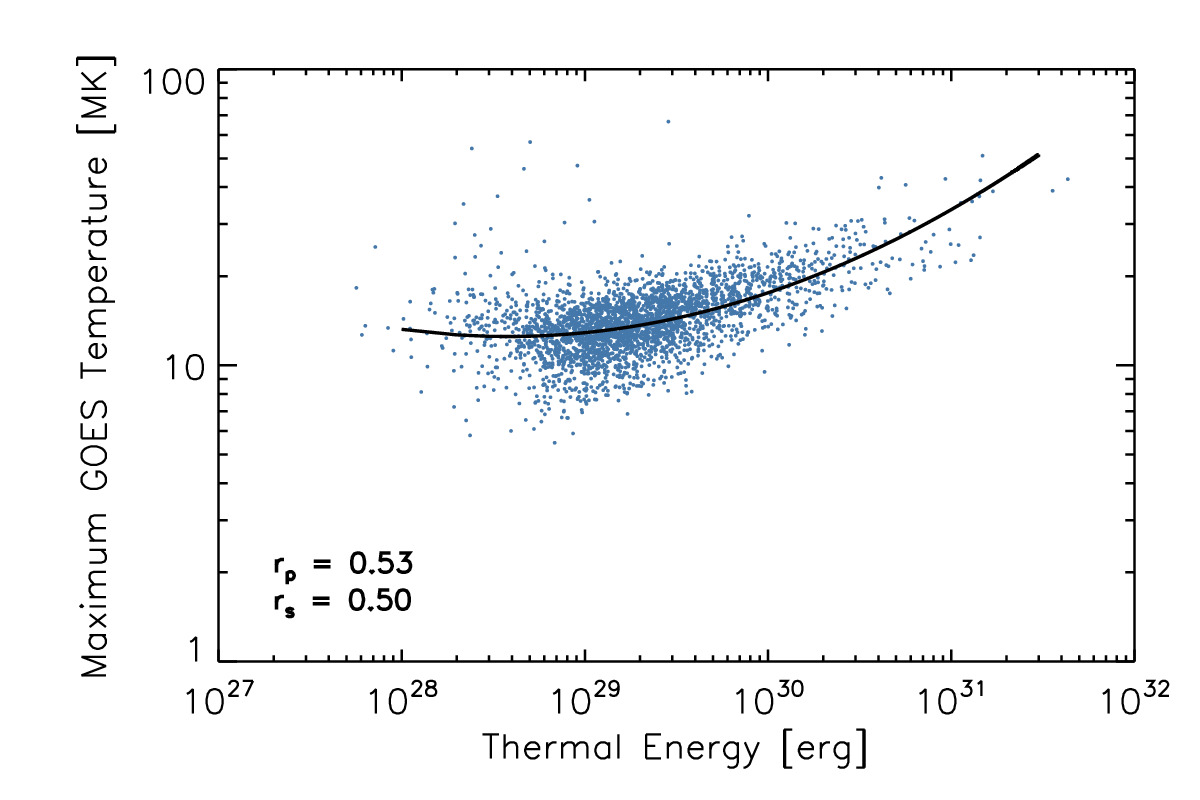}
\includegraphics[width=0.5\linewidth]{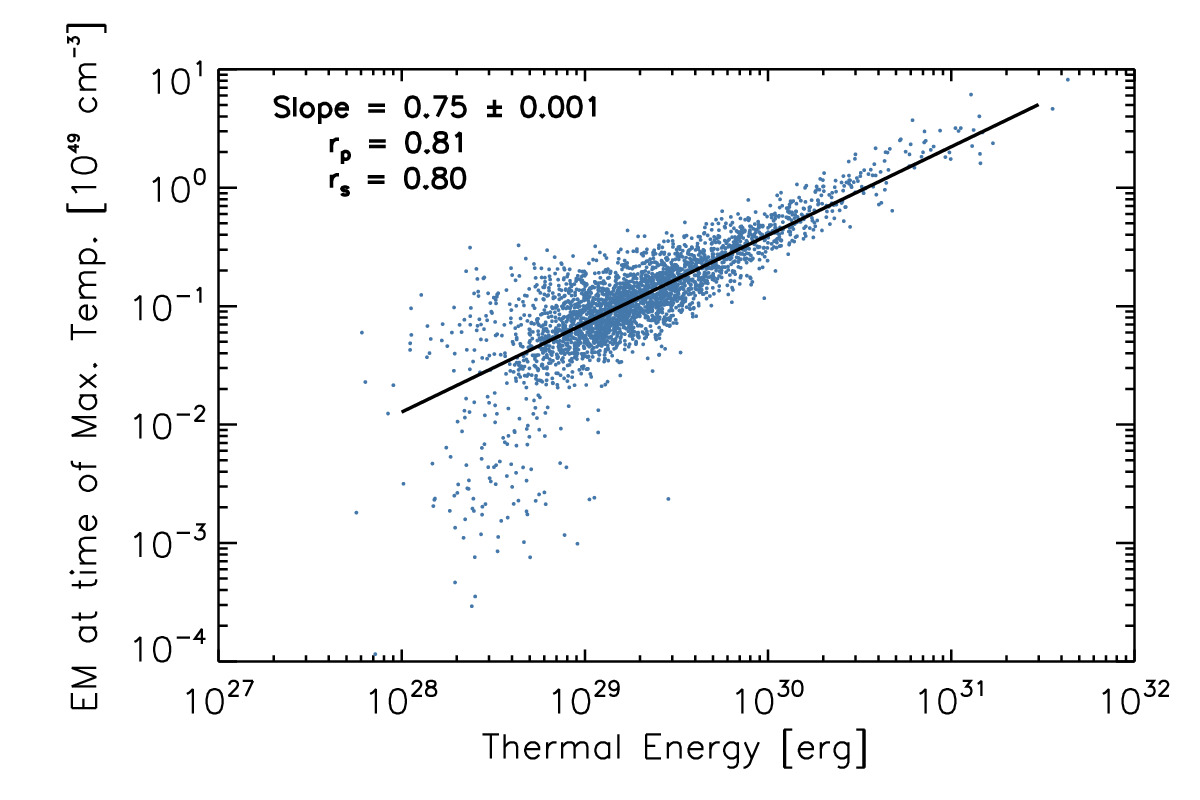}
\caption{The relationships between the flare thermal energy and and the GOES flux (top left), maximum EM (top right), maximum temperature (bottom left), and EM at the time of the maximum temperature (bottom left).  The SXR flux depends strongly on the total energy.  \label{fig:thermal}}
\end{figure*}
Using this approximation to the thermal energy content, we show how the basic properties of flares scale in Figure \ref{fig:thermal}.  The top left plot shows that the SXR flux scales strongly with the energy content:
\begin{align}
    F_{1-8} &\propto E_{th}^{0.85 \pm 0.001} \\ \nonumber
    F_{0.5-4} &\propto E_{th}^{1.08 \pm 0.001}
\end{align}
\noindent There appears to be a linear relationship with the energy content, suggesting that it could be used to approximate flare energies.  The top right plot shows that the maximum EM does not generally scale with the thermal energy.\footnote{\textit{N.B.}  We note again that we fit the thermal energy using the EM at the time of the peak temperature, rather than the maximum EM.}  As with the GOES flux, there are two populations of flares: those that do and those that do not scale with the maximum EM.  This may be due to what \citet{sadykov2018} refer to as ``temperature-dominated'' and ``EM-dominated'' flares.  The maximum temperature is correlated with the thermal energy content, though the increase is rather shallow.  In over three orders of magnitude in energy, the temperature varies by only a factor of about 3.  Finally, the EM at the time of the peak temperature is strongly correlated with the energy release, increasing by a factor of around 100 over three orders of magnitude in energy.    

Interestingly, these fittings may be related to the Neupert effect. The Neupert effect \citep{neupert1968,dennis1993} describes a linear correlation between the hard X-ray (HXR) flux and the time derivative of the soft X-rays (SXRs), or, equivalently, the HXR fluence and SXR flux.  It is thought to be caused as a result of electron beam heating of the chromosphere, where the non-thermal HXR emission is produced by non-thermal \textit{bremsstrahlung} emitted by the beam \citep{brown1971}, which begins to heat the plasma and therefore ablate material into the corona as the pressure rises, in turn causing a brightening of the SXRs in the corona.  \citet{lee1995} showed that this implies the following relation between the SXR flux and the total energy:
\begin{align}
E \propto \frac{F_{\text{SXR}} T^{3/2}}{n\ I(T)}
\end{align}
\noindent where $I(T)$ is effectively the detector response.  We can simplify this to determine the scaling relation that characterizes the Neupert effect.  From Figure \ref{fig:thermal}, we have roughly $T \propto E^{1/8}$ and $EM \propto E^{3/4}$.  The density can be written in terms of the EM and volume: $n \approx \sqrt{\frac{EM}{V}}$, and by definition the energy content scales linearly with the volume $E \propto V$.  Combining these expressions, we therefore find
\begin{align}
F_{\text{SXR}} &\propto \frac{E n I(T)}{T^{3/2}}   \\  \nonumber
			&\propto E (EM)^{1/2} V^{-1/2} T^{-3/2} I(T) \\ \nonumber
			&\propto E^{11/16} I(T)
\end{align} 
\noindent Following \citet{warren2004}, the detector responses scale with temperature as $I_{\text{1--8\,\AA}} \propto T^{5/4}$ and $I_{\text{0.5--4\,\AA}} \propto T^{11/4}$ (in the range of approximately 8--30\,MK).  We therefore find the scaling relations for GOES under the assumption of the Neupert effect's validity:
\begin{align}
F_{\text{1--8\,\AA}}    &\propto E^{27/32} \approx E^{0.84} \\ \nonumber
F_{\text{0.5--4\,\AA}} &\propto E^{33/32} \approx E^{1.03} 
\end{align}
\noindent Both of these scaling laws are close to the observed distributions found in Figure \ref{fig:thermal}, so the predictions of the Neupert effect appear consistent with this data set.

Previous solar studies, using various methodologies, have found a super-linear correlation between the flux and the energy release.  \citet{warren2004} used an analytic estimation based on combining the RTV scaling laws \citep{rosner1978} with the GOES temperature response functions, showing that this implies $F_{\text{1--8\,\AA}} \propto E^{1.75}$ and $F_{\text{0.5--4\,\AA}} \propto E^{2.24}$.  \citet{reep2013} used hydrodynamic modeling to synthesize GOES emissions for various energy inputs, finding that the GOES flux increased according to energy as $F_{\text{1--8\,\AA}} \propto E^{1.7}$ and $F_{\text{0.5--4\,\AA}} \propto E^{1.6}$.  \citet{kazachenko2017}, using a Levenberg-Marquardt fitting algorithm, found a relation between the peak flux in the 1--8\,\AA\ channel and the reconnection flux $F_{\text{1--8\,\AA}} \propto \Phi_{\text{ribbon}}^{1.5}$.  It is not clear why there is a discrepancy between the data here and these previous results.  One possibility is that the estimate of thermal energy has a number of uncertainties, such as the approximation of volume as $(S_{\text{ribbon}}/2)^{3/2}$ or the GOES-derived temperature and EM.  

\section{Flare Duration}
\label{sec:duration}

As we have shown, a flare's soft X-ray flux scales directly with the basic properties of the flare.  What about the duration?  In this section, we show that the duration of flares does not correlate well with any of the basic variables, including thermal energy, peak temperature, peak EM, peak flux, ribbon area, or magnetic flux.  When the flares are sorted by their GOES classes, however, trends develop in larger flares that are absent in smaller flares.  

\begin{figure*}
\includegraphics[width=0.5\linewidth]{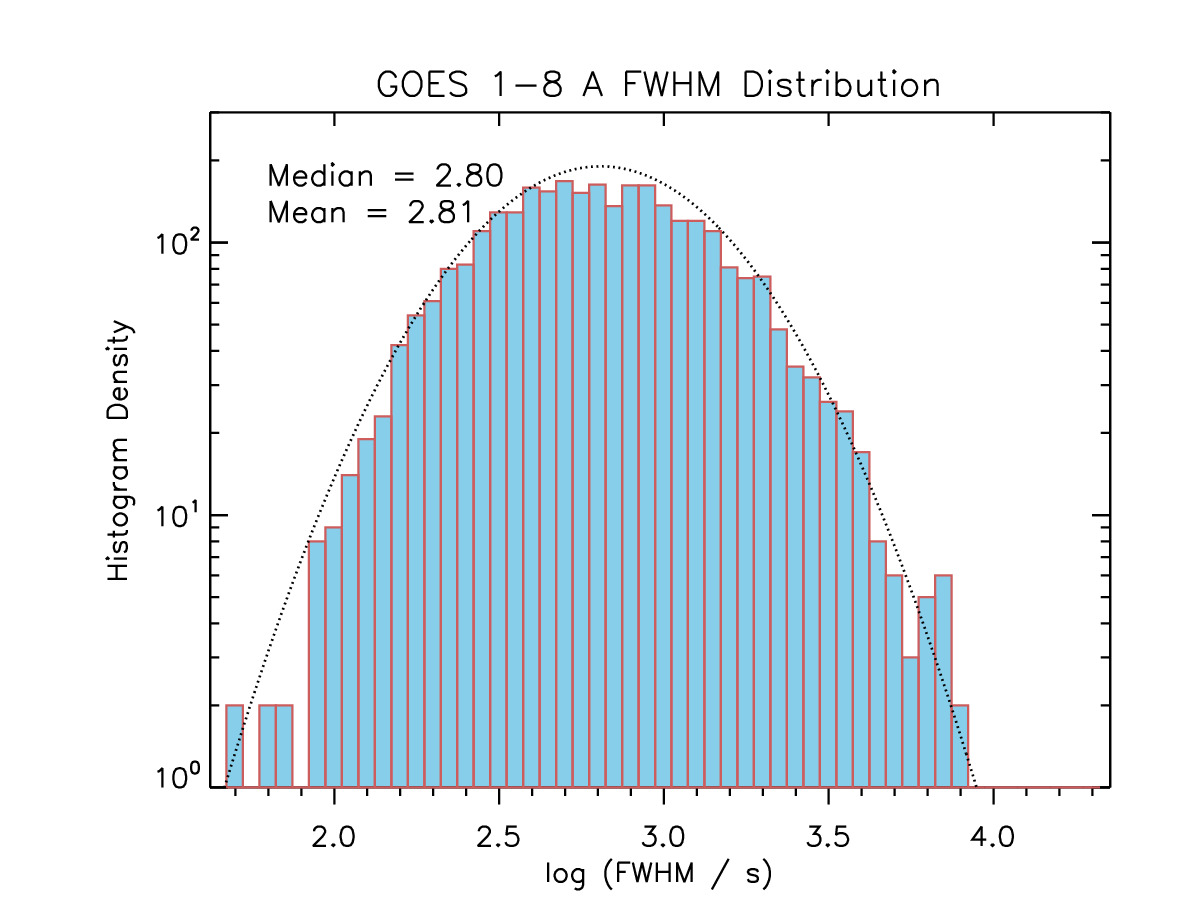}
\includegraphics[width=0.5\linewidth]{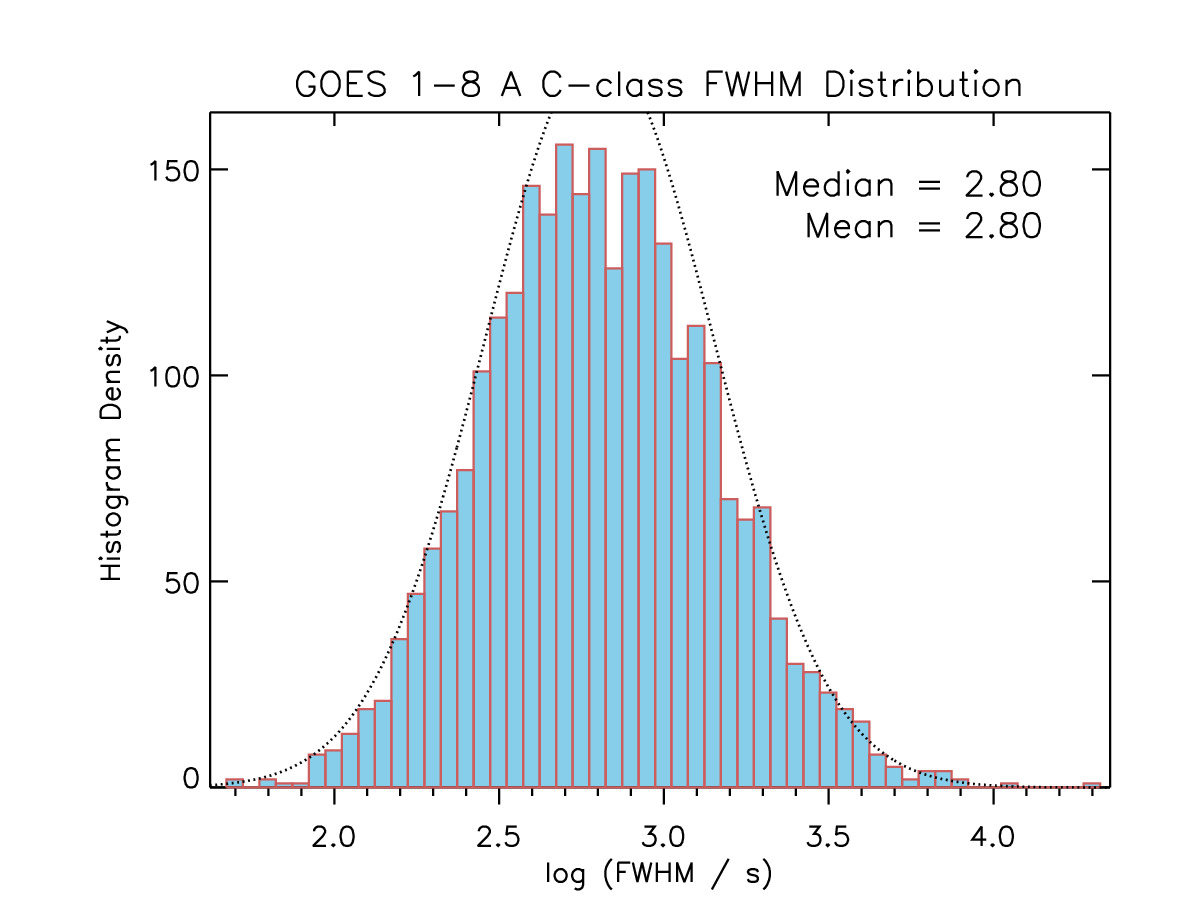}
\includegraphics[width=0.5\linewidth]{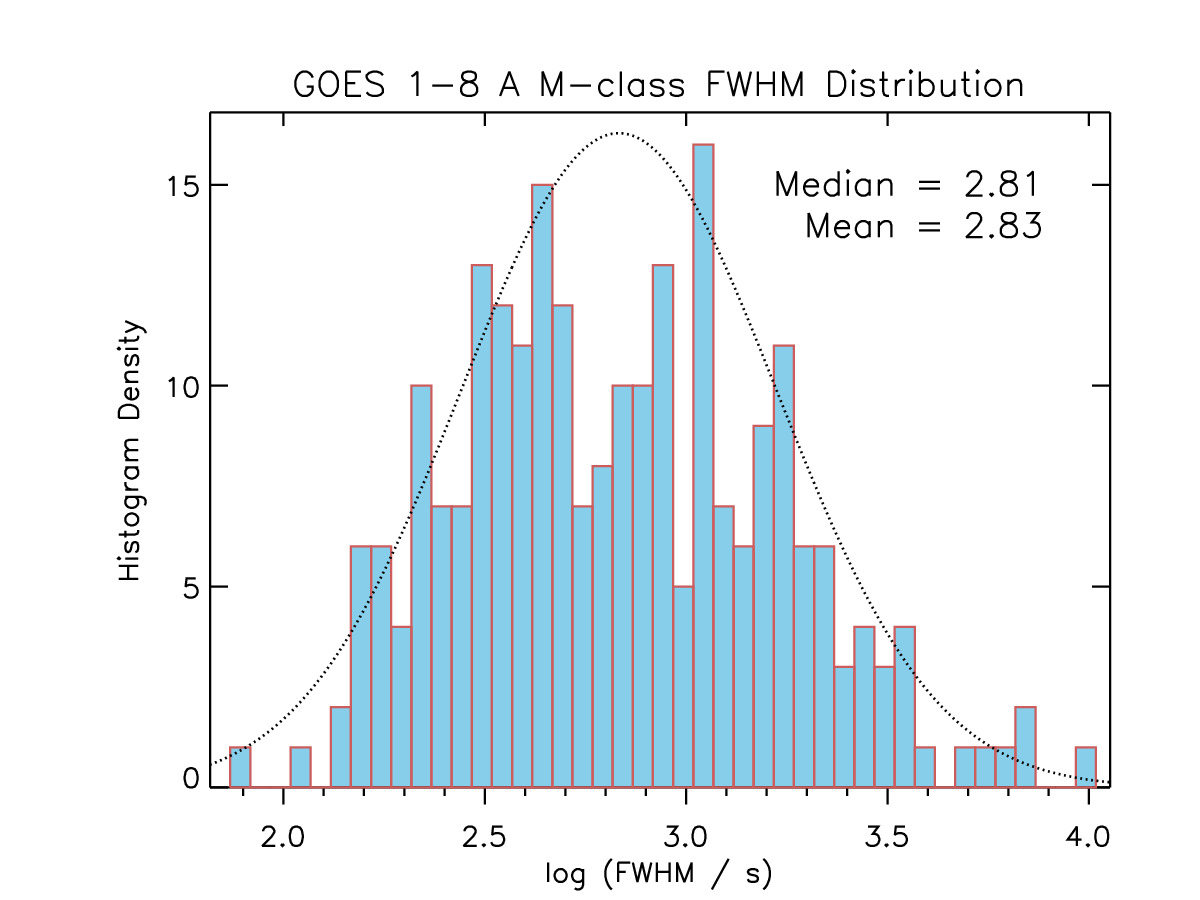}
\includegraphics[width=0.5\linewidth]{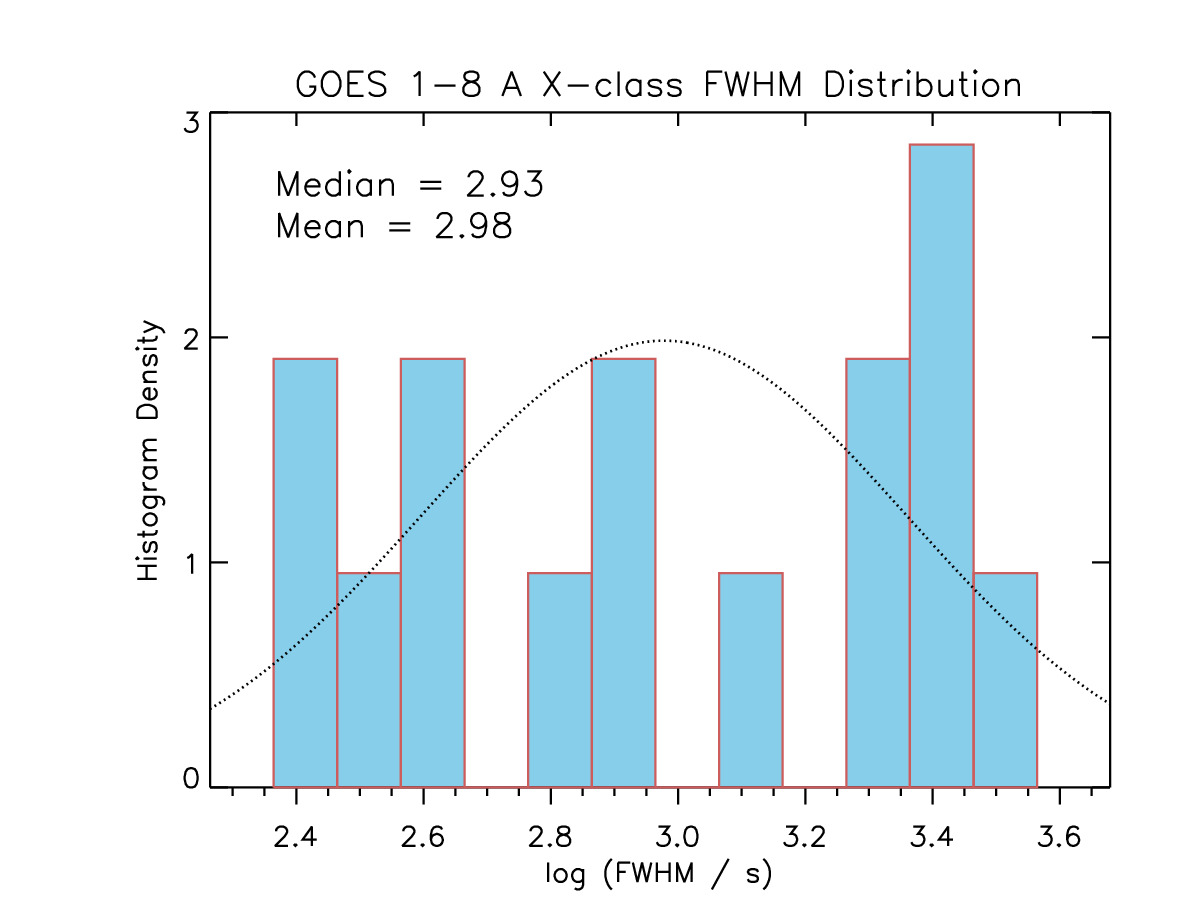}
\caption{The distribution of flare durations in 1--8\,\AA.  There are four histograms, showing the total distribution (top left), C-class flare distribution (top right), M-class flare distribution (bottom left), and X-class flare distribution (bottom right).  We find that the distributions are consistent with log-normal (see text), with median values of around 11 minutes.  The dotted curves show the best-fit Gaussian derived from a maximum likelihood estimate for each case.  Note that the total distribution is shown on a log-scale, the rest on a linear scale.  \label{fig:fwhm18_dist}}
\end{figure*}
\begin{figure*}
\includegraphics[width=0.5\linewidth]{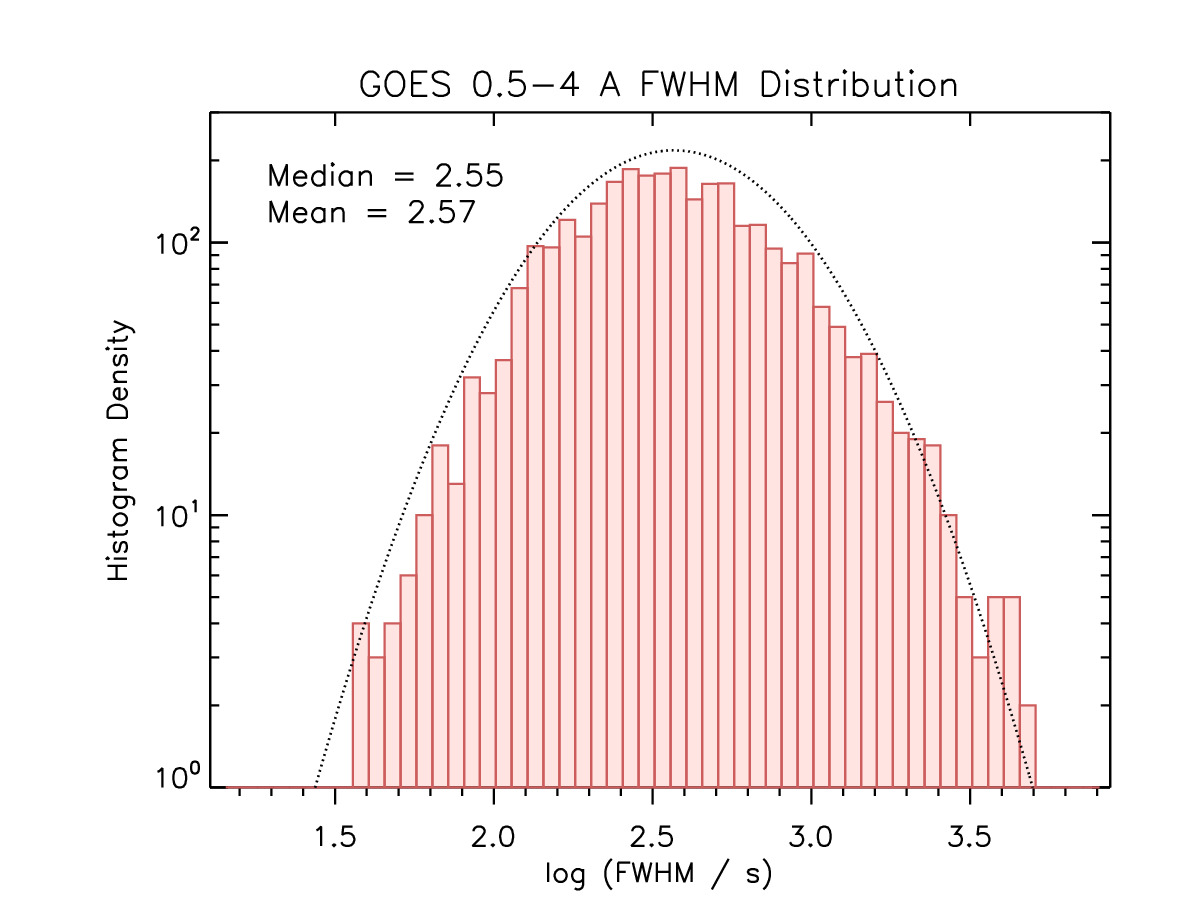}
\includegraphics[width=0.5\linewidth]{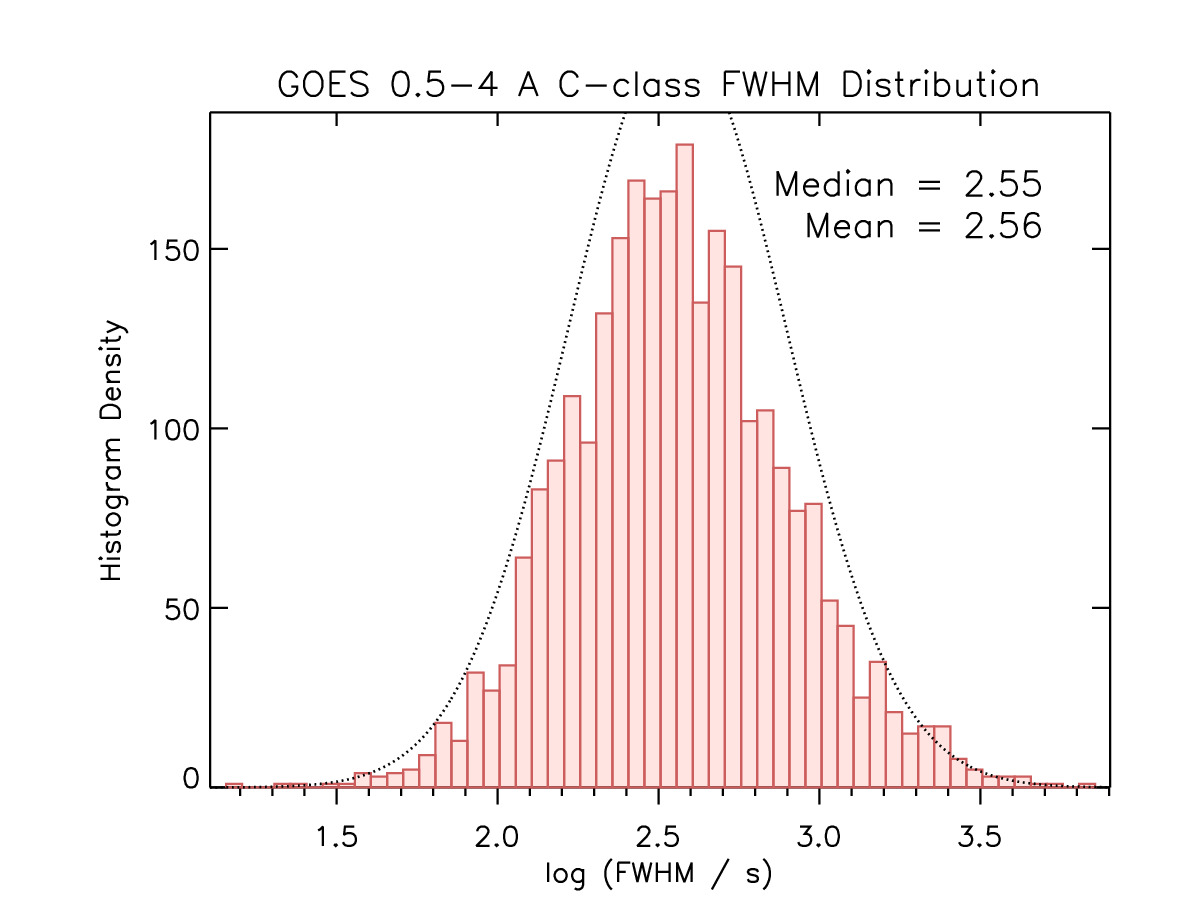}
\includegraphics[width=0.5\linewidth]{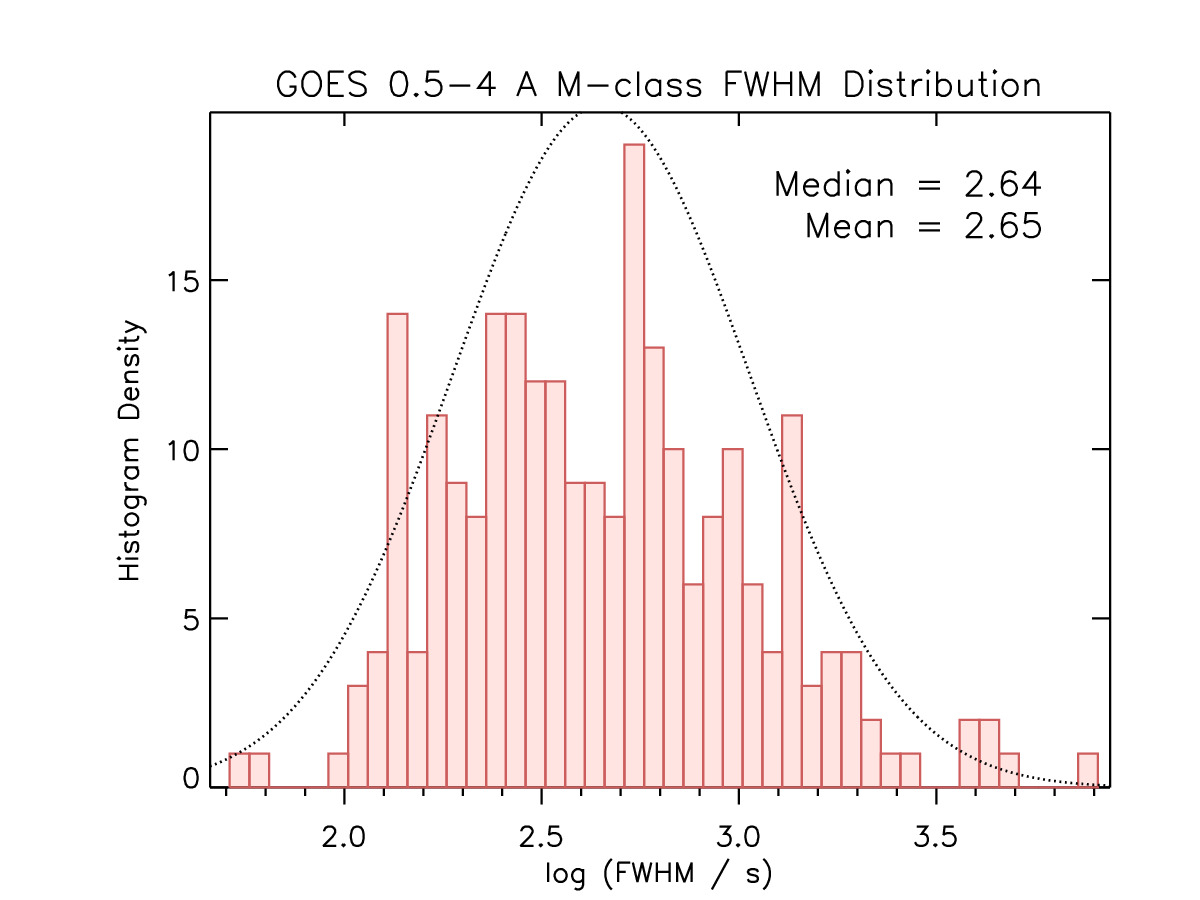}
\includegraphics[width=0.5\linewidth]{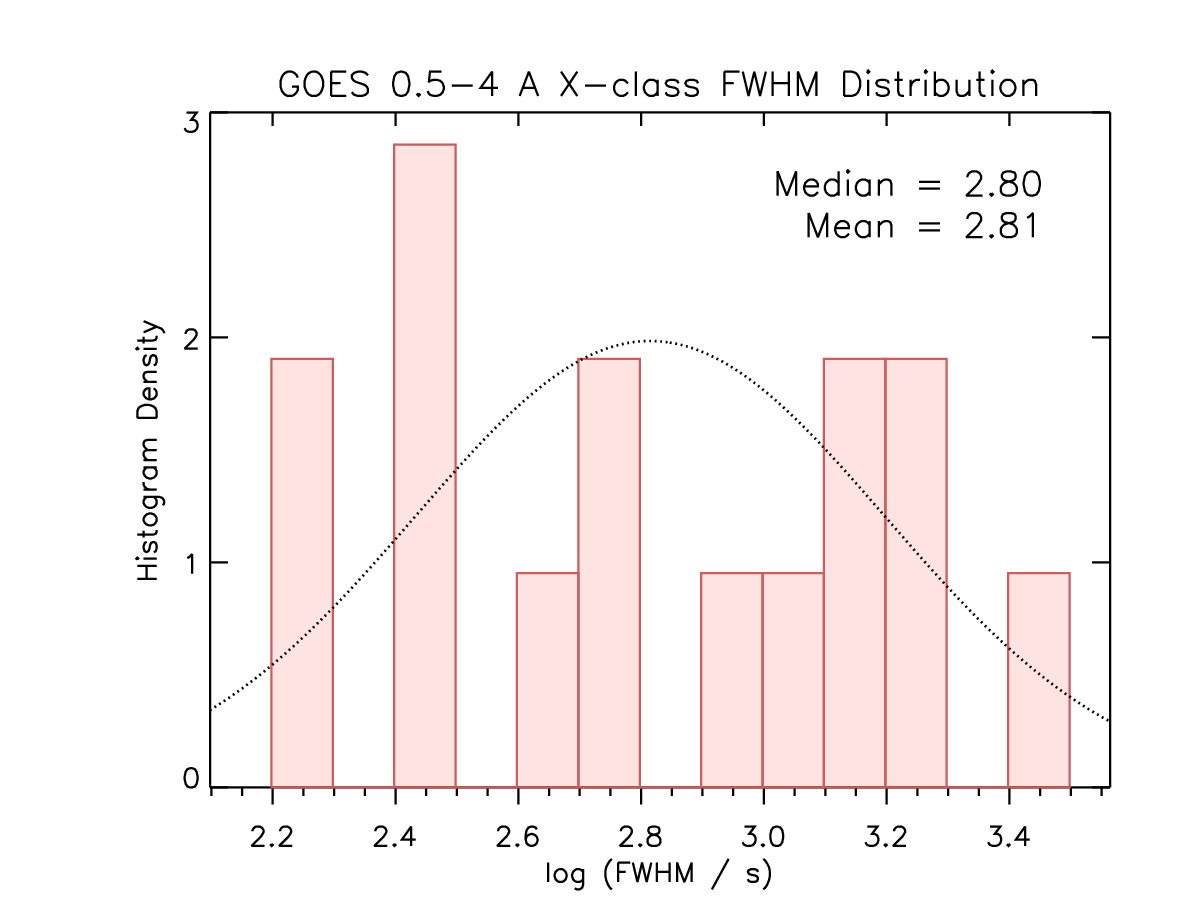}
\caption{Similar to Figure \ref{fig:fwhm18_dist}, but for the 0.5--4\,\AA\ channel.  We find that the three sub-divided distributions are consistent with log-normal, while the total distribution may be inconsistent with a log-normal distribution (see text).  The dotted curves show the best-fit Gaussian derived from a maximum likelihood estimate for each case.  Note that the total distribution is shown on a log-scale, the rest on a linear scale.  \label{fig:fwhm054_dist}}
\end{figure*}
We first show the distribution of flare SXR durations in Figures \ref{fig:fwhm18_dist} and \ref{fig:fwhm054_dist}, which show histograms of the FWHM in the 1--8\,\AA\ channel and 0.5--4\,\AA\ channel.  For each channel, there are four histograms, showing the total distribution (top left), C-class flare distribution (top right), M-class flare distribution (bottom left), and X-class flare distribution (bottom right).  The total distribution is shown on a log-scale, while the rest are on a linear scale.  

\begin{table}[t]
\caption{The mean values $\log{\mu}$ and standard deviations $\log{\sigma}$ for each of the FWHM distributions shown in Figures \ref{fig:fwhm18_dist} and \ref{fig:fwhm054_dist}, calculated using a maximum likelihood estimation.  The 1--8\,\AA\ channel averages around 10--11\,min FWHM, while the 0.5--4\,\AA\ channel is around 6\,min average FWHM.  There appears to be a slight tendency for larger flares to last longer.}
\begin{tabular}{ c | c | c  }
 & $\log{\mu}$ & $\log{\sigma}$ \\ \hline
 All, 1--8\,\AA & 2.81 $\pm$ 0.01 & 0.35 $\pm$ 0.01 \\ 
C-class, 1--8\,\AA & 2.81 $\pm$ 0.01 & 0.35 $\pm$ 0.01 \\ 
M-class, 1--8\,\AA & 2.83 $\pm$ 0.03 & 0.39 $\pm$ 0.02 \\ 
X-class, 1--8\,\AA & 2.98 $\pm$ 0.14 & $0.38^{+0.12}_{-0.07}$ \\ 
All, 0.5--4\,\AA & 2.57 $\pm$ 0.01 & 0.34 $\pm$ 0.01 \\ 
C-class, 0.5--4\,\AA & 2.56 $\pm$ 0.01 & 0.34 $\pm$ 0.01 \\
M-class, 0.5--4\,\AA & 2.65 $\pm$ 0.03 & 0.38 $\pm$ 0.02 \\ 
X-class, 0.5--4\,\AA & 2.81 $\pm$ 0.14 & $0.38^{+0.12}_{-0.08}$ 
\label{table:estimates}
\end{tabular}
\end{table}
The distributions appear to be log-normal, which we now test.  We first fit each of the 8 distributions using a maximum likelihood estimation to determine the mean and standard deviation, shown in Table \ref{table:estimates}.  Using those estimates, we then perform a Kolmogorov-Smirnov test to examine the null hypothesis $H_{0}$ that the distributions are consistent with log-normal.  The results are shown in Table \ref{table:kstest}, where we list the test statistic $D_{max}$ along with the confidence levels at 1, 5, and 10\%, and the conclusion.  We find that all of the distributions are consistent with log-normal at the 1\% level, and 7 of the 8 are additionally consistent with log-normal at the 5 and 10\% levels.  The distribution of the FWHM in the 0.5--4\,\AA\ channel for all flares, however, is inconsistent with log-normal at the 5 and 10\% levels.  The median values are approximately 10 and 5--6 minutes for the low and high energy channels, respectively.  In the high energy channel, there may be a slight tendency for larger flares to last longer (compare M and C-class median FWHM), which may suggest that there are physical differences in the heating, but this hypothesis would need to be stringently checked with spectrally-resolved observations.  This may be the reason why the total distribution appears inconsistent with log-normal, due to a skew related to flare size.
\begin{table*}[t]
\centering
\caption{Kolmogorov-Smirnov test to determine whether the FWHM distributions in Figures \ref{fig:fwhm18_dist} and \ref{fig:fwhm054_dist} are consistent with a log-normal distribution.  Each row shows the test statistic $D_{max}$, the confidence levels at 1, 5, and 10\%, and the conclusion.  We accept the null hypothesis $H_{0}$ that the distribution is log-normal in all cases at the 1\% level, and accept at the 5 and 10\% levels for all but one case, the FWHM of the 0.5--4\,\AA\ channel for the entire set of flares.}
\begin{tabular}{ c | c | c | c | c | c }
 & $D_{max}$ & $\alpha_{0.01}$ & $\alpha_{0.05}$ & $\alpha_{0.10}$ & Conclusion \\ \hline
All, 1--8\,\AA & 0.021 & 0.030 & 0.025 & 0.023 & Accept $H_{0}$ \\ 
C-class, 1--8\,\AA & 0.017 & 0.031 & 0.026 & 0.024 & Accept $H_{0}$ \\ 
M-class, 1--8\,\AA & 0.059 & 0.105 & 0.087 & 0.079 & Accept $H_{0}$ \\ 
X-class, 1--8\,\AA & 0.188 & 0.420 & 0.351 & 0.316 & Accept $H_{0}$ \\ 
All, 0.5--4\,\AA & 0.027 & 0.030 & 0.025 & 0.023 & Accept $H_{0}$ at 1\%, reject at 5 and 10\% \\ 
C-class, 0.5--4\,\AA & 0.023 & 0.031 & 0.026 & 0.024 & Accept $H_{0}$ \\ 
M-class, 0.5--4\,\AA & 0.051 & 0.105 & 0.087 & 0.079 & Accept $H_{0}$ \\ 
X-class, 0.5--4\,\AA & 0.132 & 0.420 & 0.351 & 0.316 & Accept $H_{0}$ 
\label{table:kstest}
\end{tabular}
\end{table*}

Many studies in the literature have discussed the distribution of flare durations.  In the hard X-ray range ($> 25$\,keV), \citet{crosby1993}, \citet{lu1993}, \citet{bromund1995}, and \citet{aschwanden2016} have found that there is a power law distribution of flare durations with slopes of $\approx -2$.  At slightly lower energies ($> 12$\,keV), \citet{christe2008} fit a similar power law to the longer durations of their distribution, but show an approximately log-normal distribution of durations of microflares with a mean value of about 6 minutes, consistent with our data.  In the soft X-ray range ($2-12$\,\AA\ $\approx 1-6$\,keV), \citet{drake1971} found a distribution peaked at around 15 minutes, skewed towards longer durations (though it is not clear what fit would be appropriate).  Finally, using GOES 1--8\,\AA\ data, \citet{veronig2002} found a similarly skewed distribution (that appears to be log-normal) with median value of 12 minutes.  In comparison, a recent study using magnetohydrodynamic modeling of reconnection event durations found probability distributions slopes of $-1.93 \pm 0.11$ (using current density as a proxy for reconnection, \citealt{knizhnik2018}).  It is possible that there are different distributions for the HXR and SXRs, however, since the emission in each range is produced by different physical mechanisms (non-thermal \textit{bremsstrahlung} and a combination of thermal line emission and thermal \textit{bremsstrahlung}, respectively).

\begin{figure*}
\includegraphics[width=0.5\linewidth]{srbn_vs_fwhm.jpeg}
\includegraphics[width=0.5\linewidth]{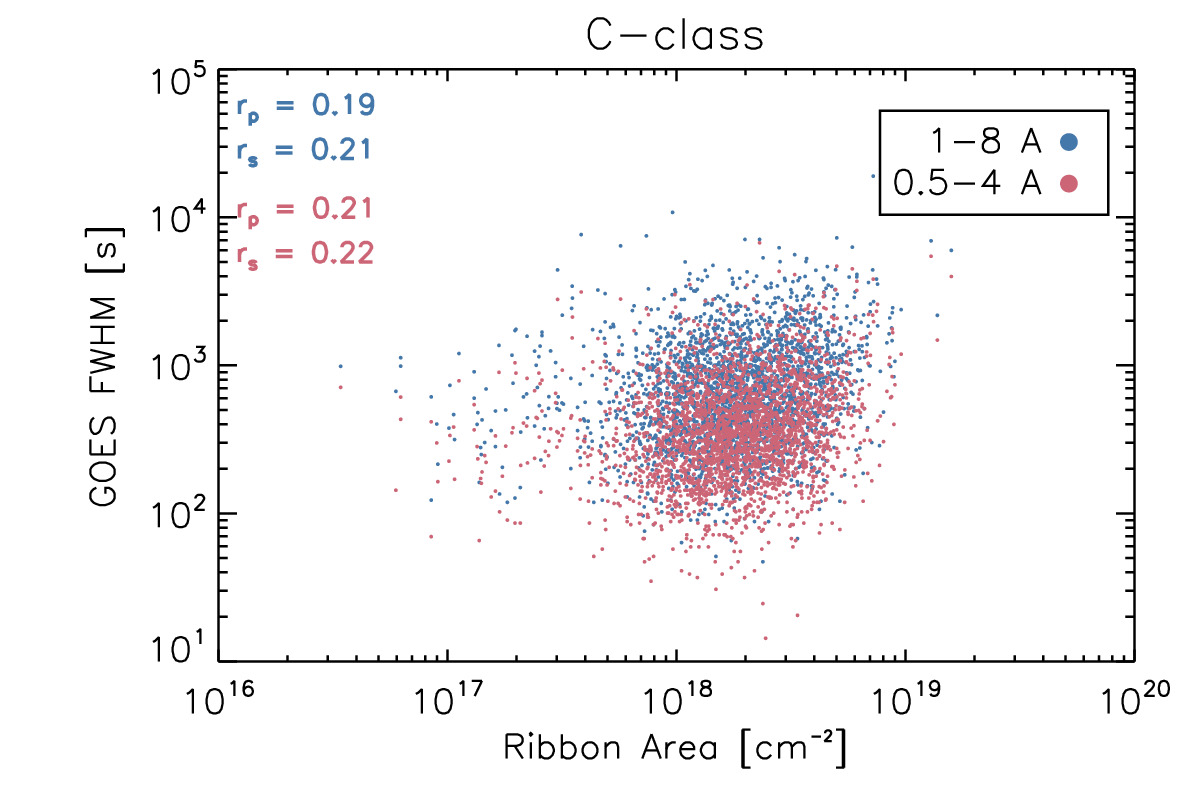}
\includegraphics[width=0.5\linewidth]{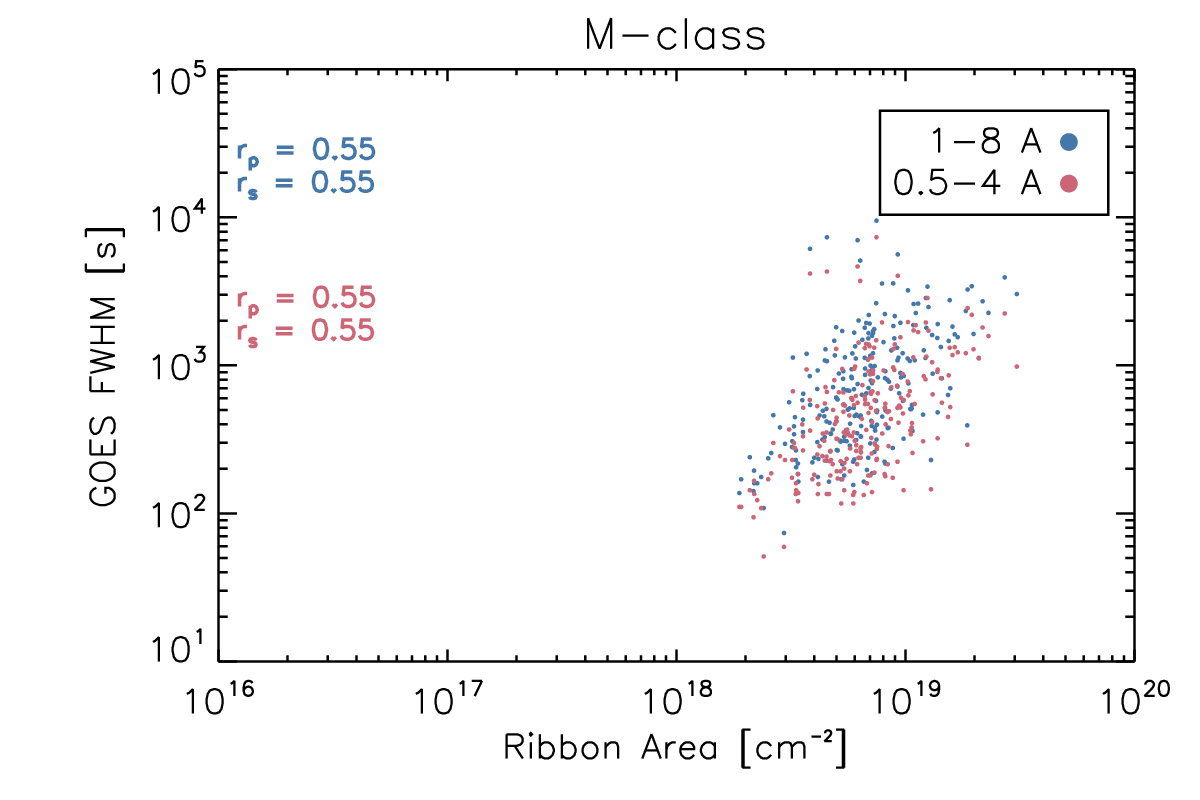}
\includegraphics[width=0.5\linewidth]{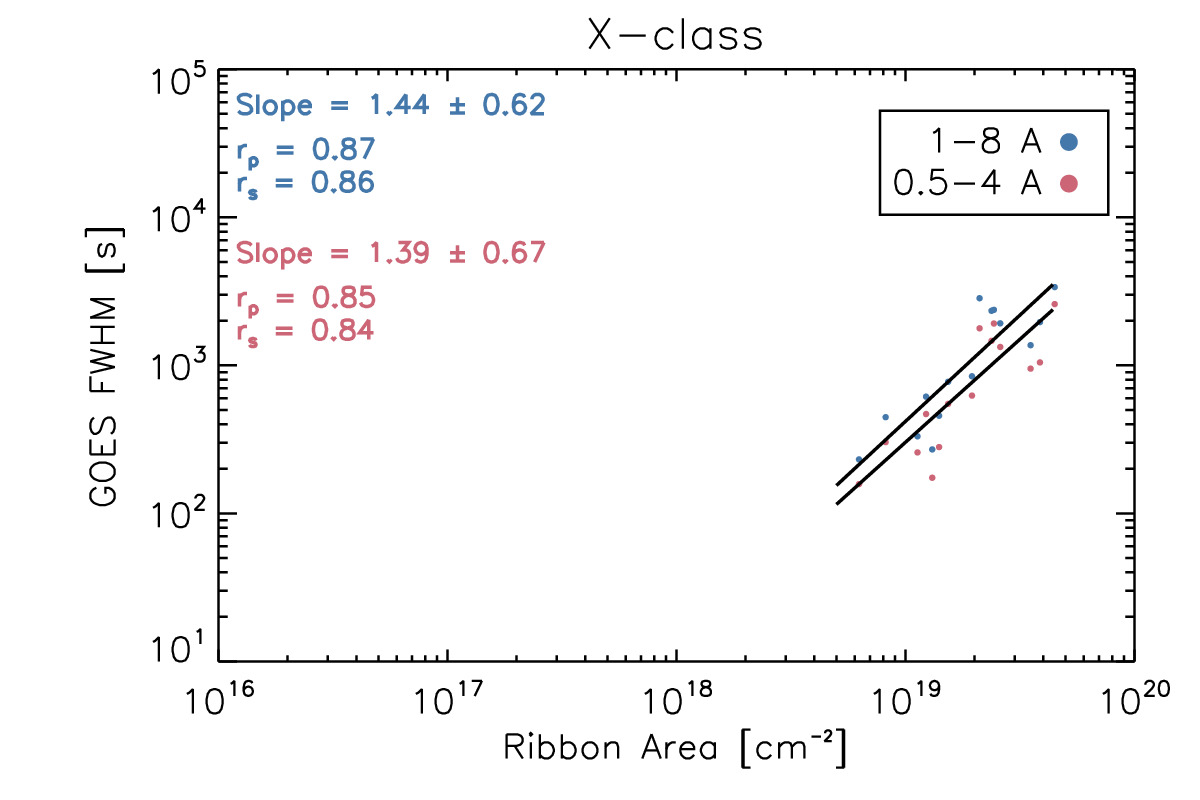}
\caption{The relationships between the ribbon area and and the GOES FWHM, when unsorted (top left), and sorted into C-class (top right), M-class (bottom left), and X-class flares (bottom right).  There is no clear trend between the ribbon area and flare duration, in general, though in the largest flares there appears to be a correlation (\textit{i.e.} the longest X-class flares have the largest ribbons).  \label{fig:duration_srbn}}
\end{figure*}
In order to better understand what causes these distributions, we now turn to the relations with the basic properties of flares.  We begin by showing the relation between the ribbon area and flare duration in Figure \ref{fig:duration_srbn}.  When examining all of the flares, there does not appear to be a clear correlation between the area of the flare ribbon and the duration of the flare.  In X-class flares, a trend appears to develop.  The longest X-class flares appear to sweep out the largest area.  We caution, however, that X-class flares are comparatively rare, so there may be an observational bias due to the limited data.  In contrast, the C-class flares, which have much better statistics, show absolutely no relation between duration and ribbon area.  

\begin{figure*}
\includegraphics[width=0.5\linewidth]{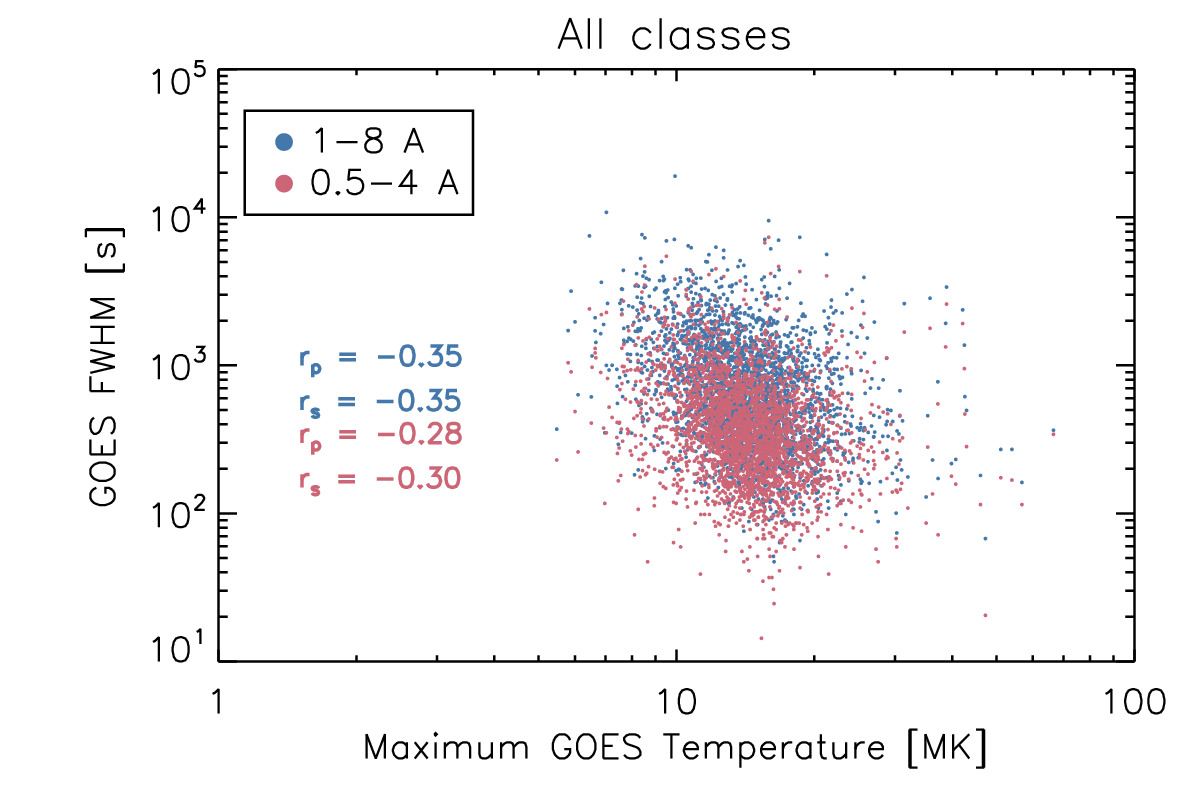}
\includegraphics[width=0.5\linewidth]{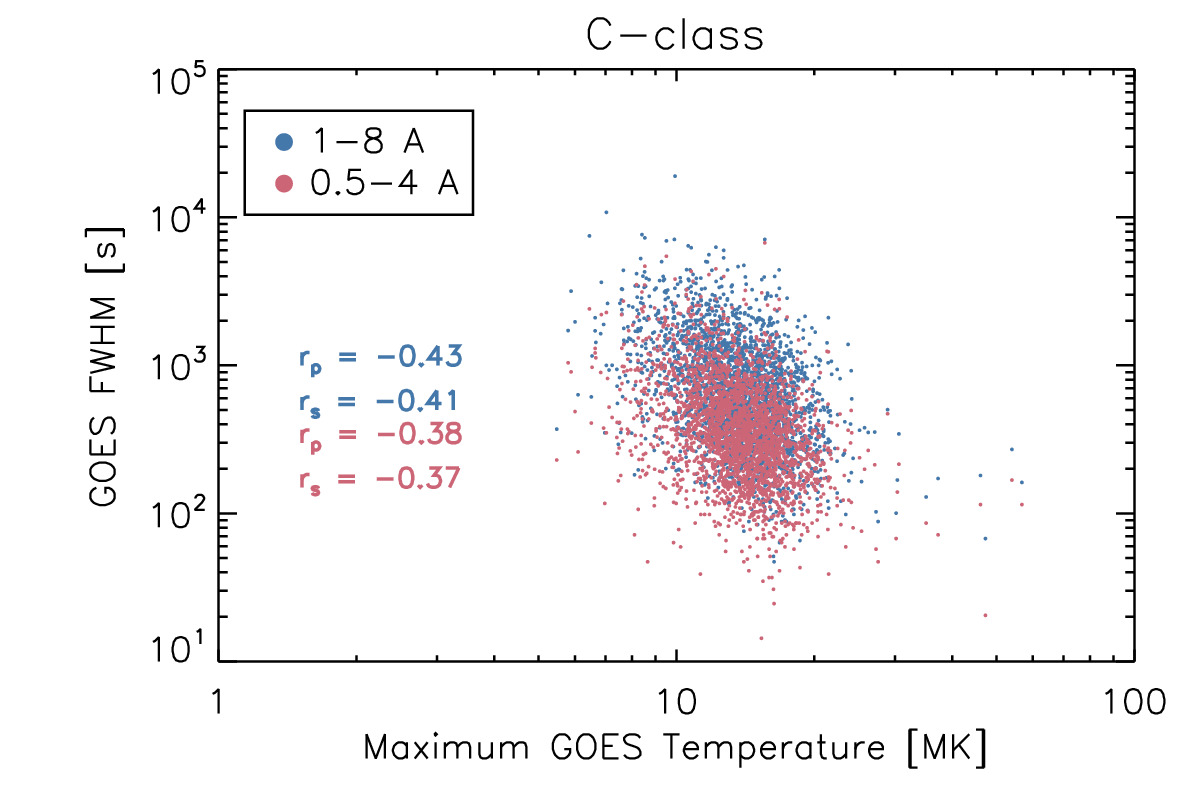}
\includegraphics[width=0.5\linewidth]{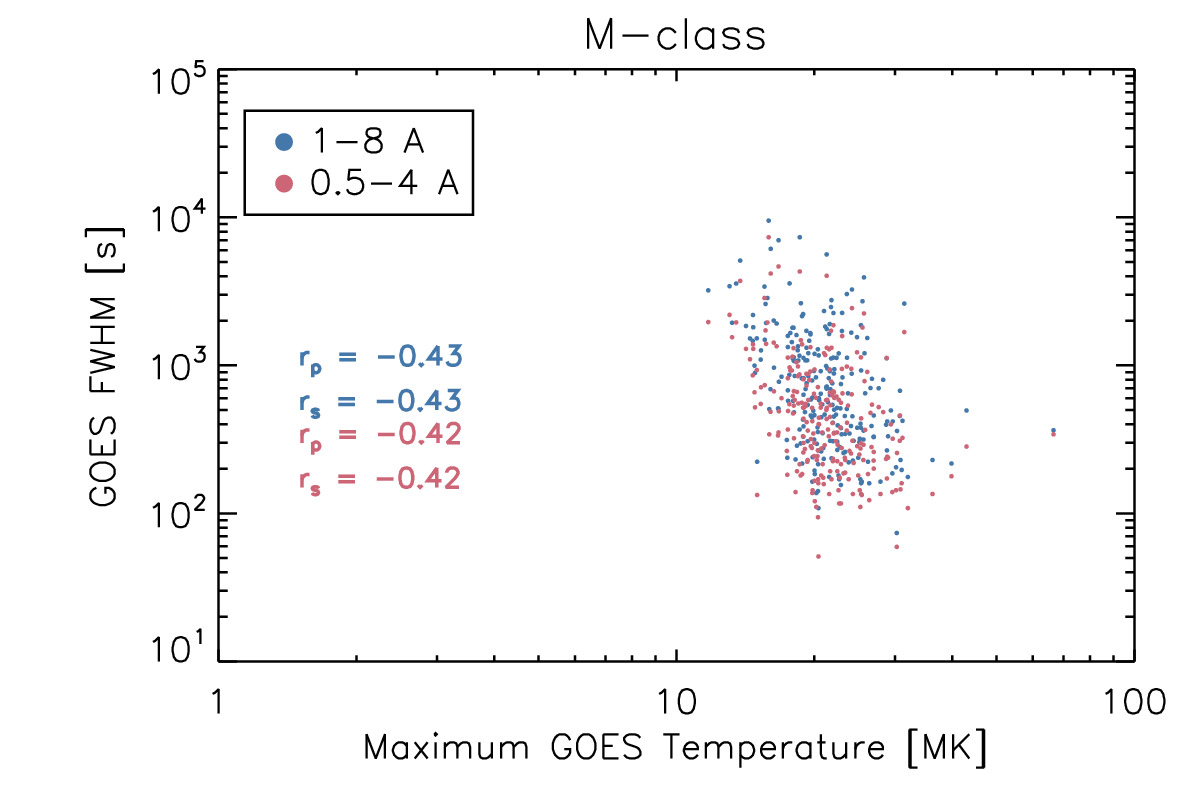}
\includegraphics[width=0.5\linewidth]{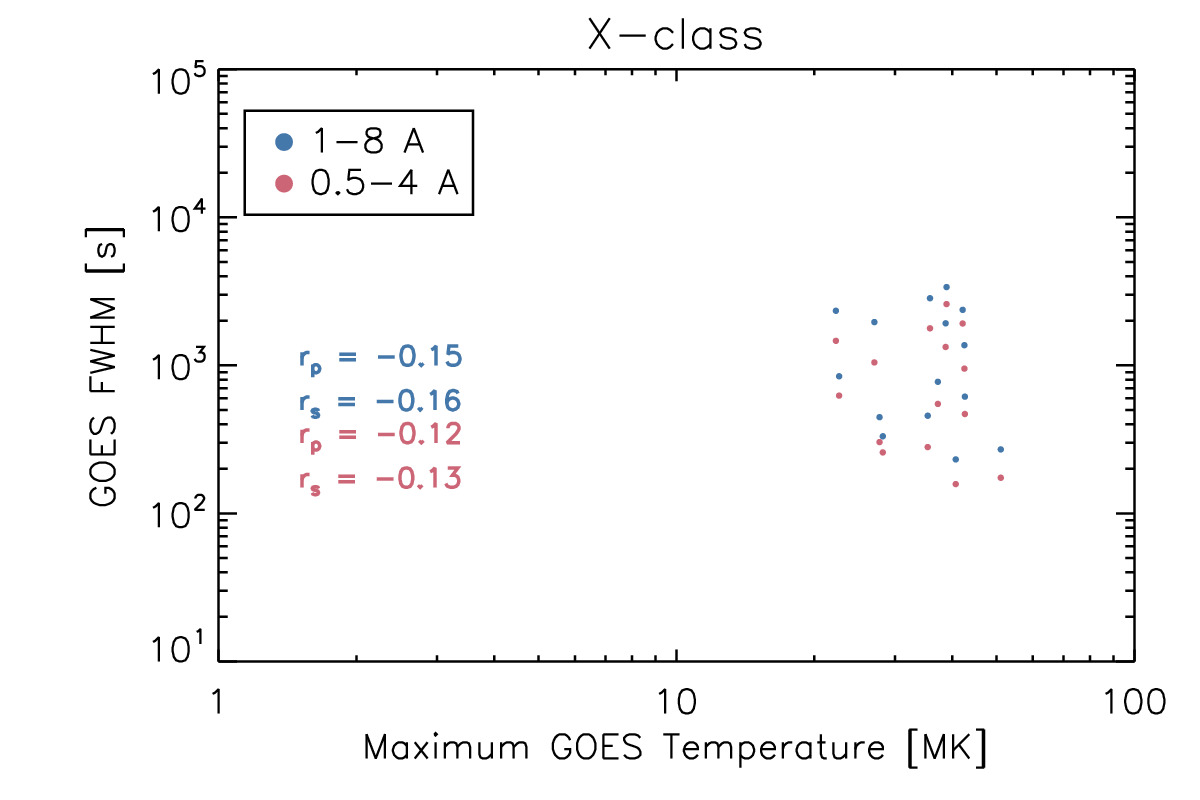}
\caption{Similar to Figure \ref{fig:duration_srbn}, except showing the relation between flare duration and maximum temperature.  There is no clear relation between the two.  \label{fig:duration_tmax}}
\end{figure*}
The maximum temperature does not appear to be related to the flare duration, as shown by Figure \ref{fig:duration_tmax}.  In this case, at all sizes, there is no strong correlation ($r_{P} < 0.5$), though it is possible that a weak, negative correlation exists.  Since the scatter is large, however, we refrain from performing fits to the data.  From a hydrodynamic standpoint, shorter, impulsive bursts of heating with the same amount of energy as a more gradual event would lead to more rapid increases in temperature since more energy is deposited into a lower density plasma, so it may be reasonable to expect a negative correlation.  Due to the multi-stranded nature of flares and the isothermal assumption in the temperature fitting here, though, this is far from clear in the current data.

\begin{figure*}
\includegraphics[width=0.5\linewidth]{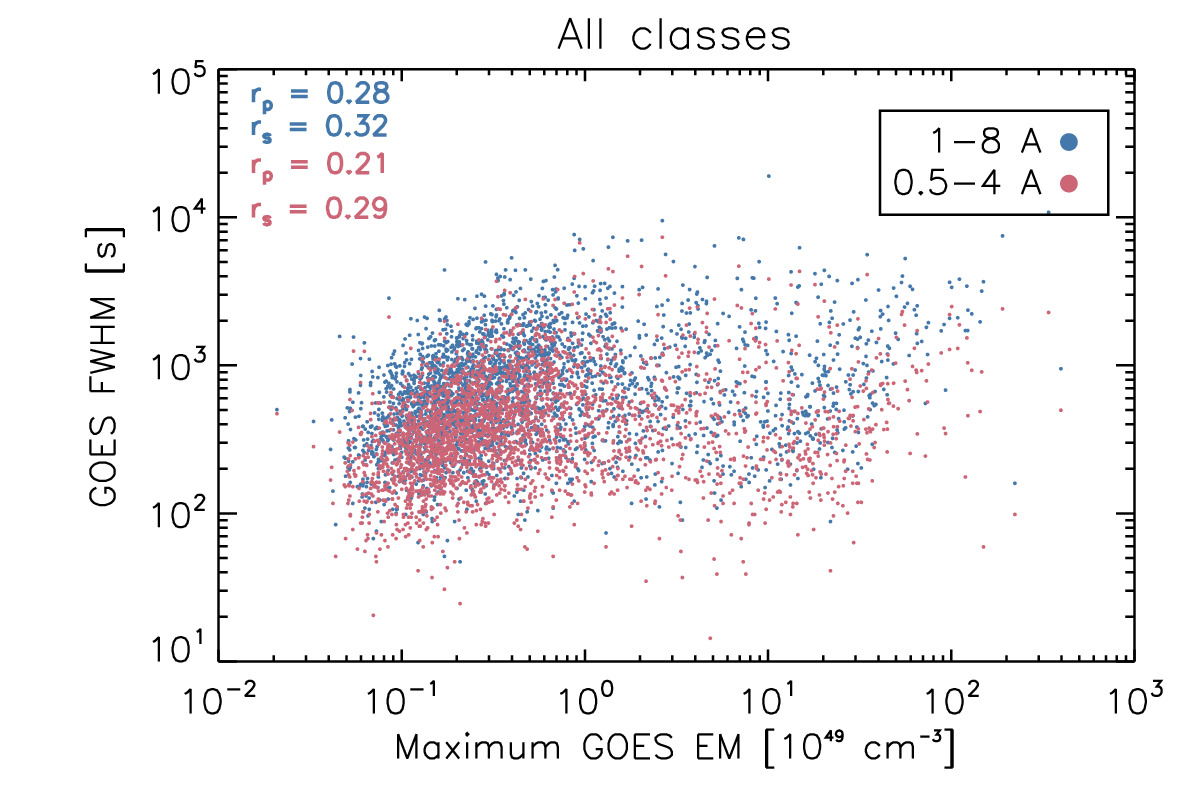}
\includegraphics[width=0.5\linewidth]{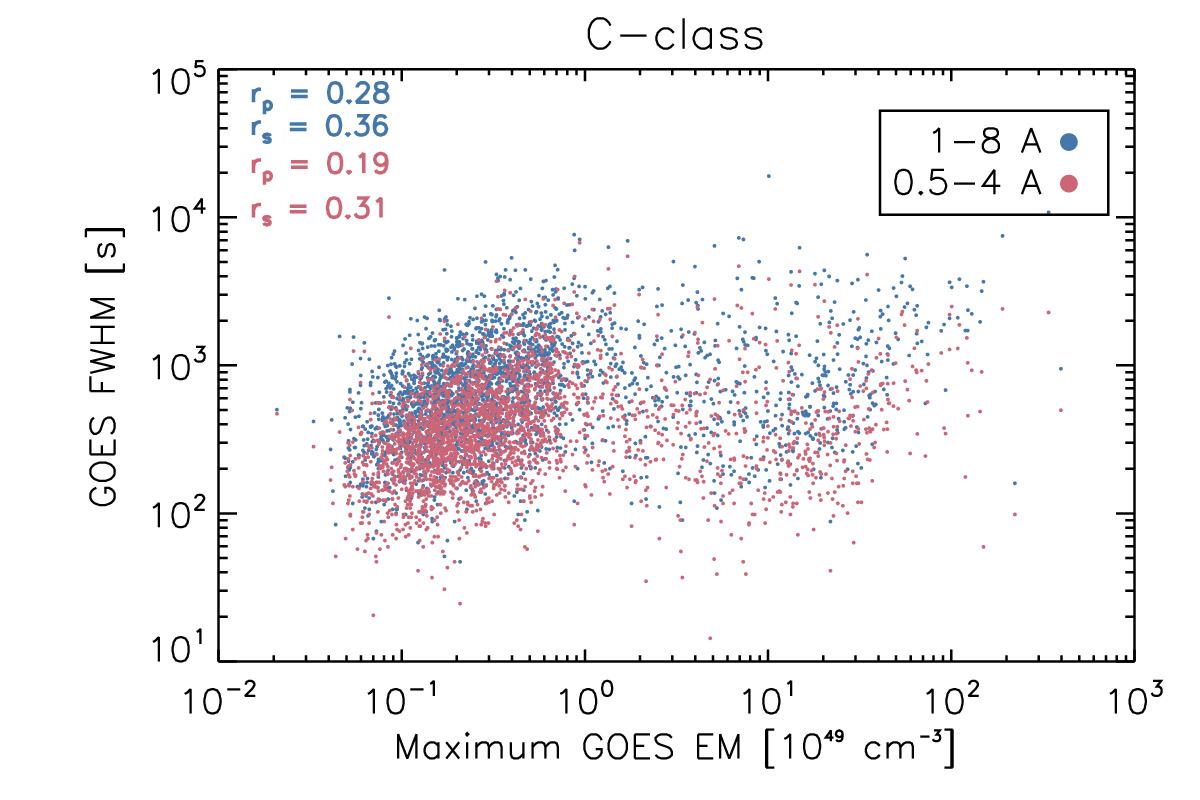}
\includegraphics[width=0.5\linewidth]{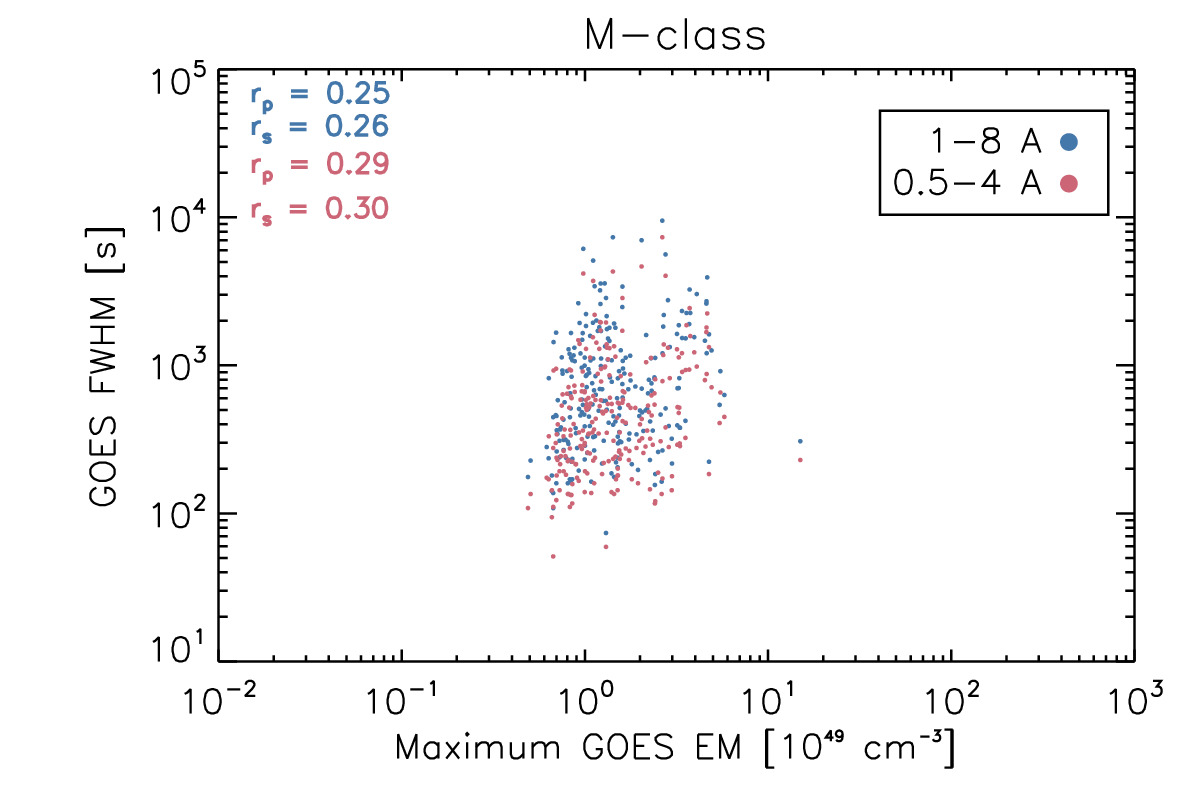}
\includegraphics[width=0.5\linewidth]{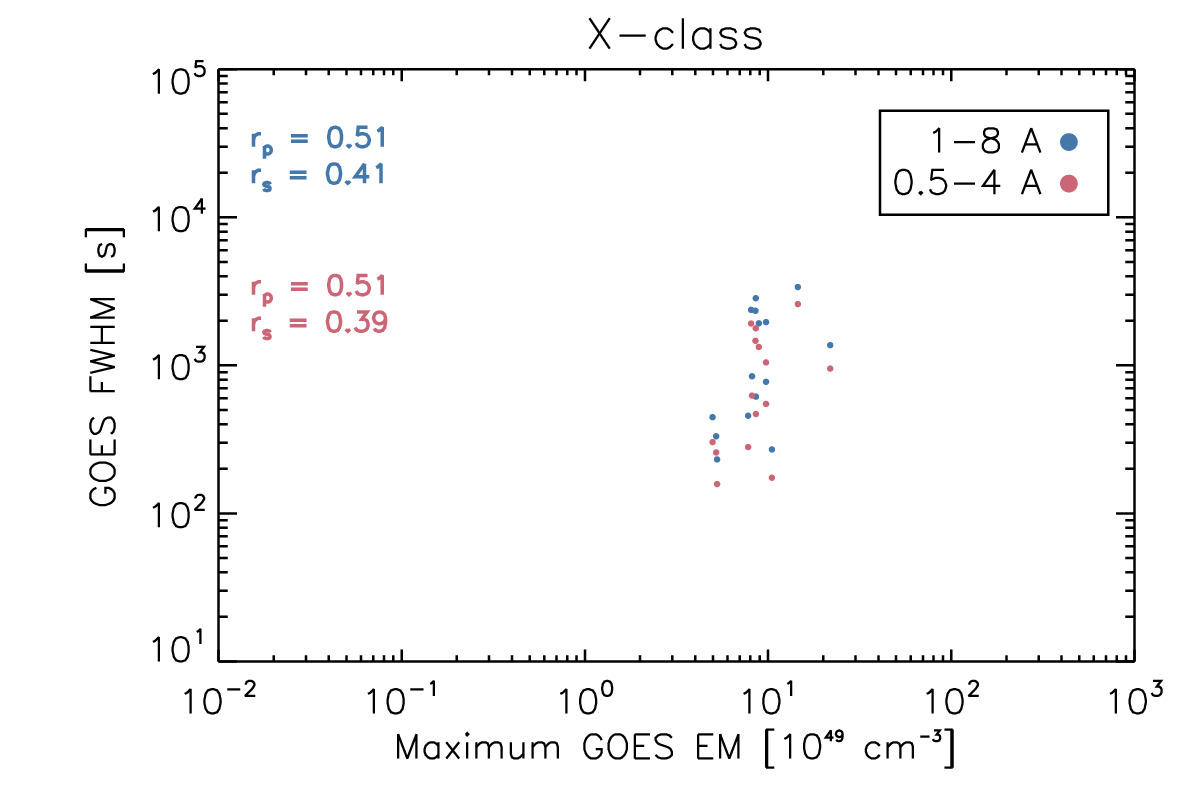}
\caption{Similar to Figure \ref{fig:duration_srbn}, except showing the relation between the flare duration and maximum EM.  There is no relation between the two in general.  \label{fig:duration_emmax}}
\end{figure*}
\begin{figure*}
\includegraphics[width=0.5\linewidth]{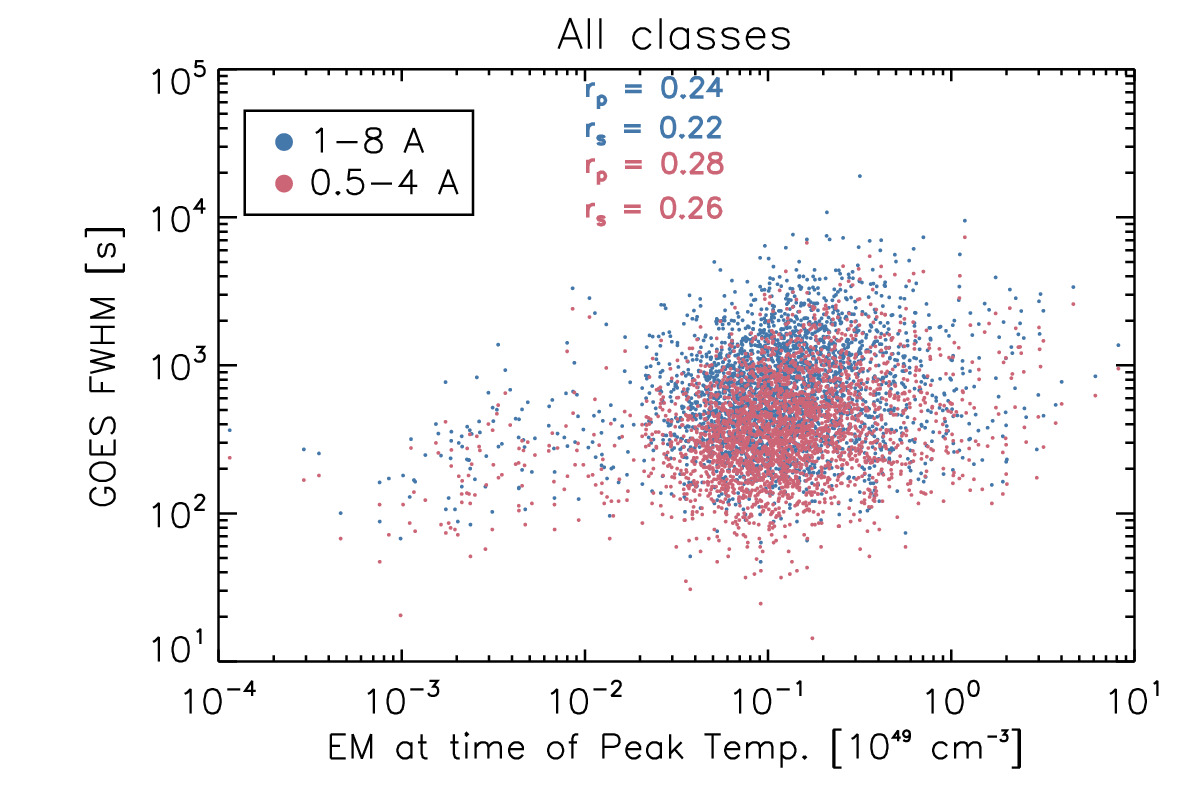}
\includegraphics[width=0.5\linewidth]{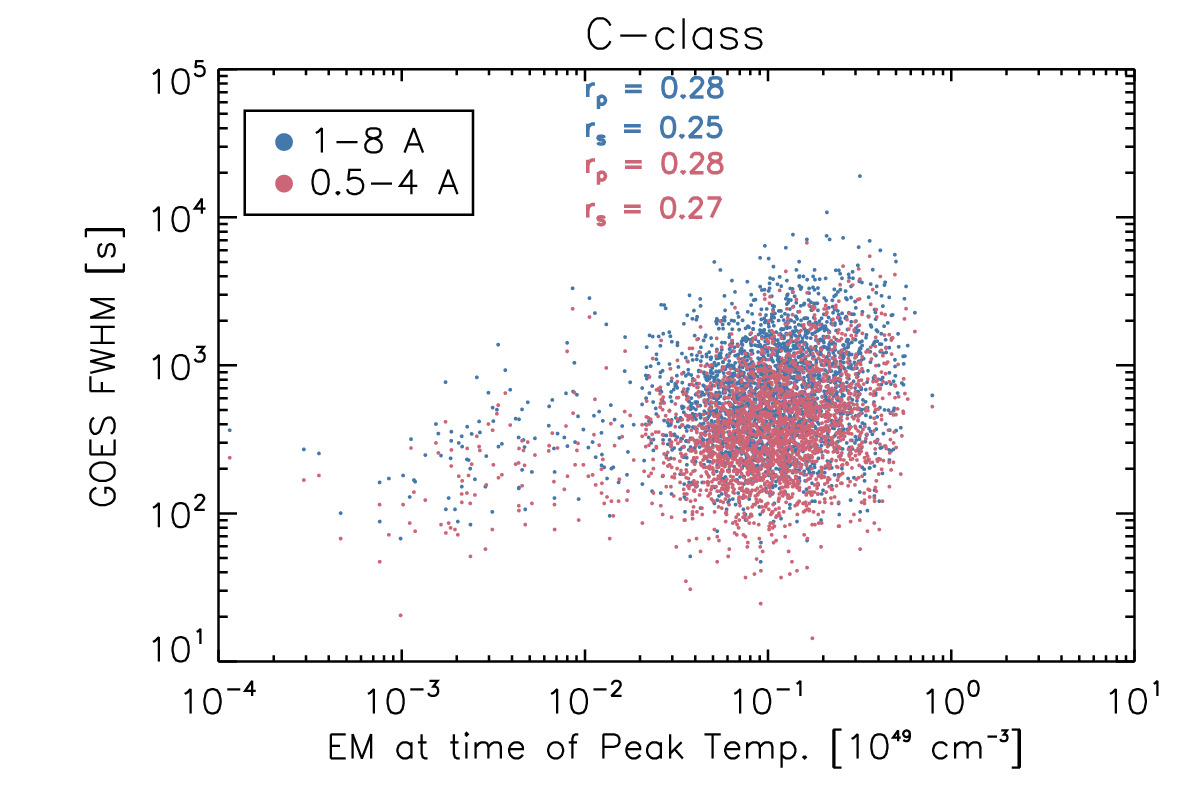}
\includegraphics[width=0.5\linewidth]{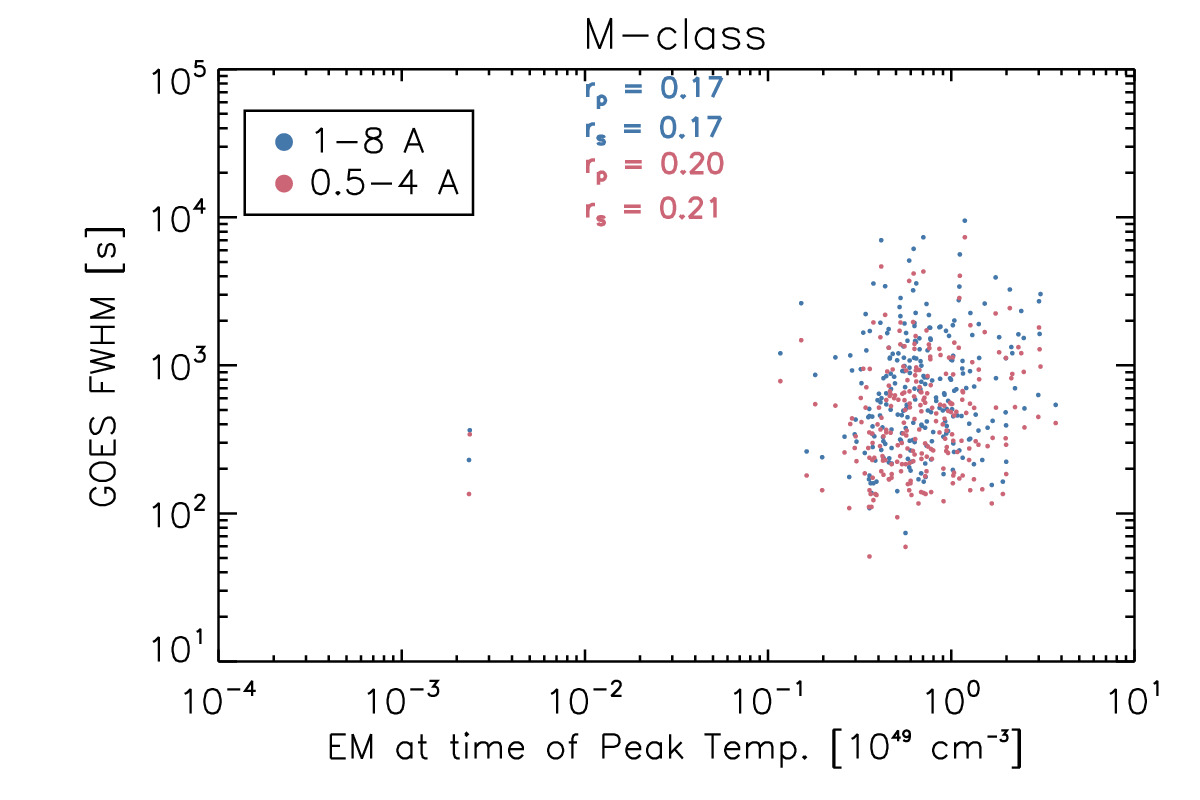}
\includegraphics[width=0.5\linewidth]{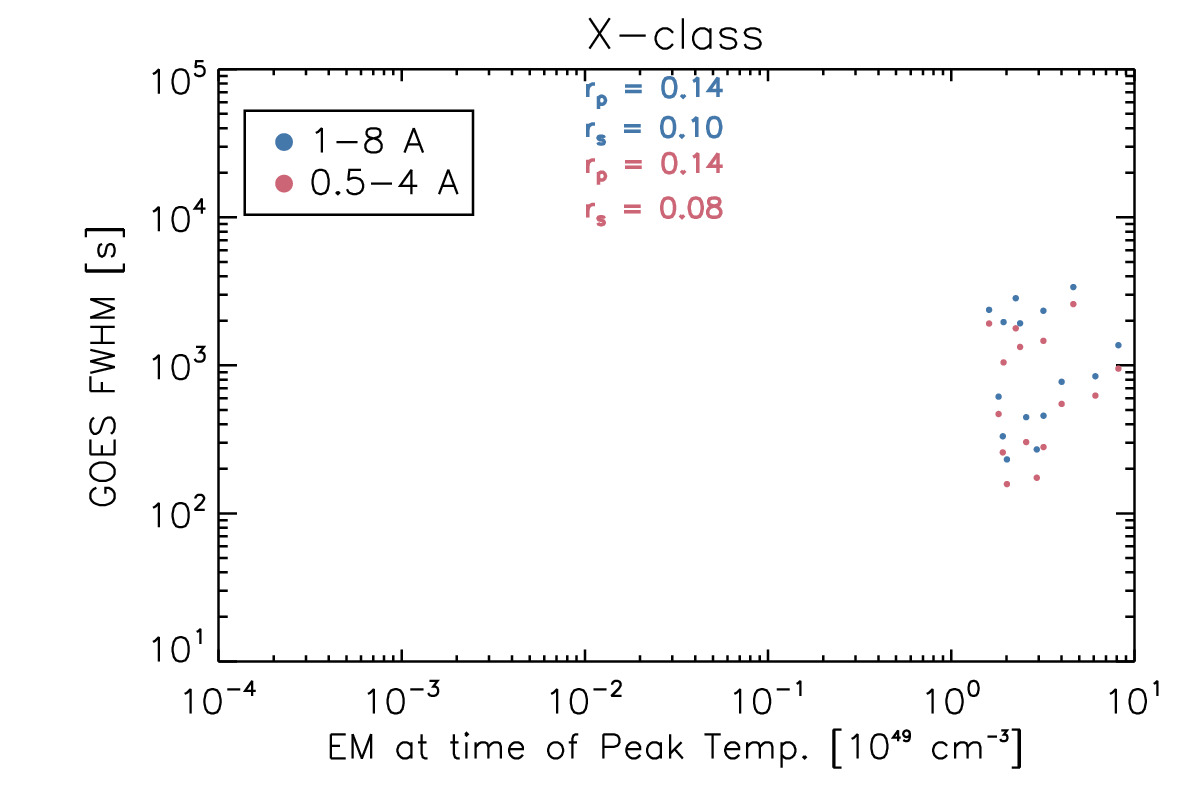}
\caption{Similar to Figure \ref{fig:duration_srbn}, except showing the relation between the flare duration and the EM at the time of the peak temperature.  There is no relation between the two.  \label{fig:duration_emattmax}}
\end{figure*}

The maximum EM does not appear to be related to the FWHM in general, as demonstrated by Figure \ref{fig:duration_emmax}.  In C-class flares, the duration ranges over two orders of magnitude, while the peak EM ranges over three orders of magnitude, showing no relation between the two.  The EM is generally larger in M-class flares, but there is no obvious connection with the FWHM.  In comparison, the EM at the time of the peak temperature does not appear to be related to the flare duration, for any size flare, as shown by Figure \ref{fig:duration_emattmax}.  While there is a trend that larger flares have a larger EM at the peak temperature, this does not correlate with the flare's duration.

\begin{figure*}
\includegraphics[width=0.5\linewidth]{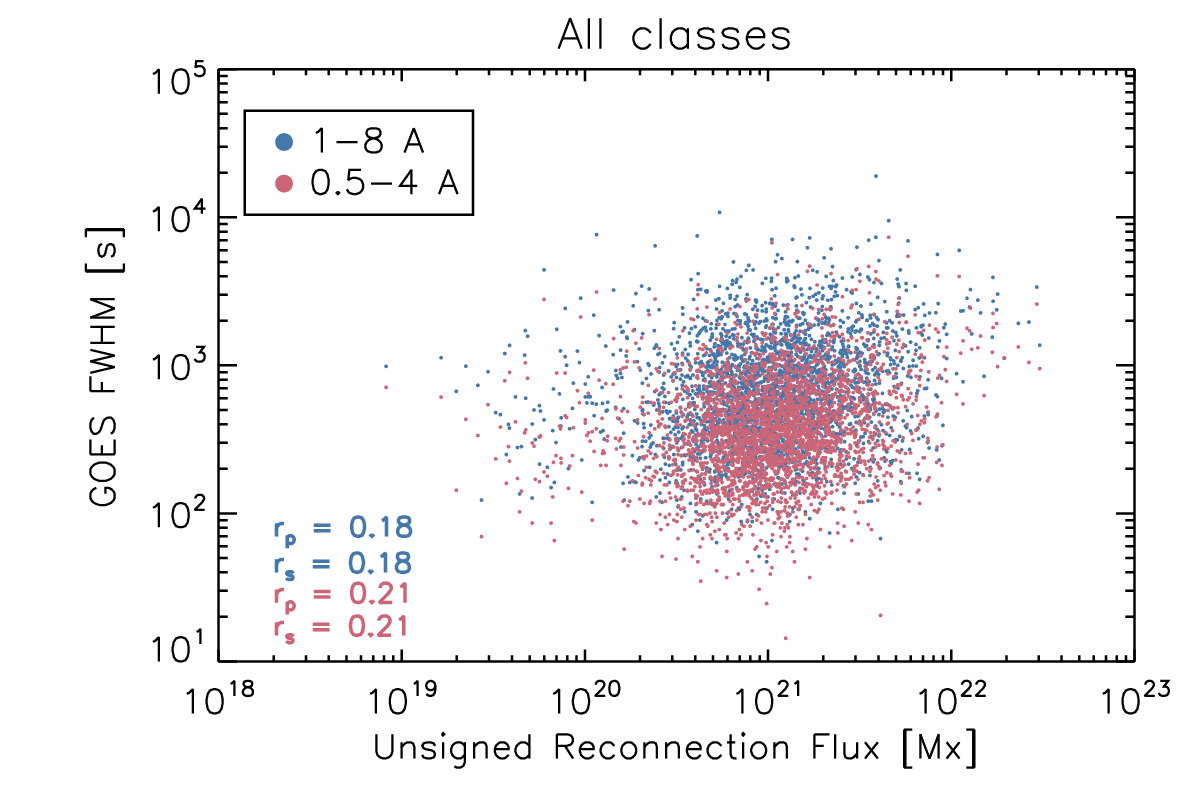}
\includegraphics[width=0.5\linewidth]{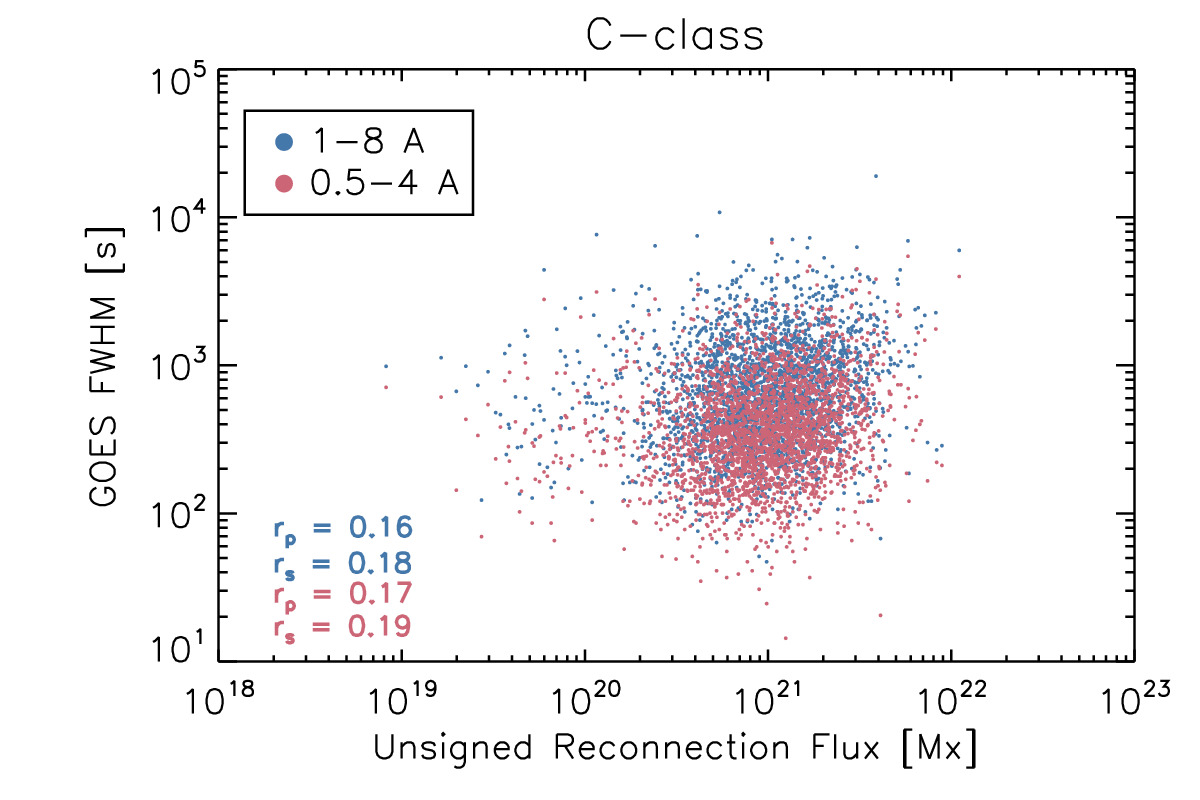}
\includegraphics[width=0.5\linewidth]{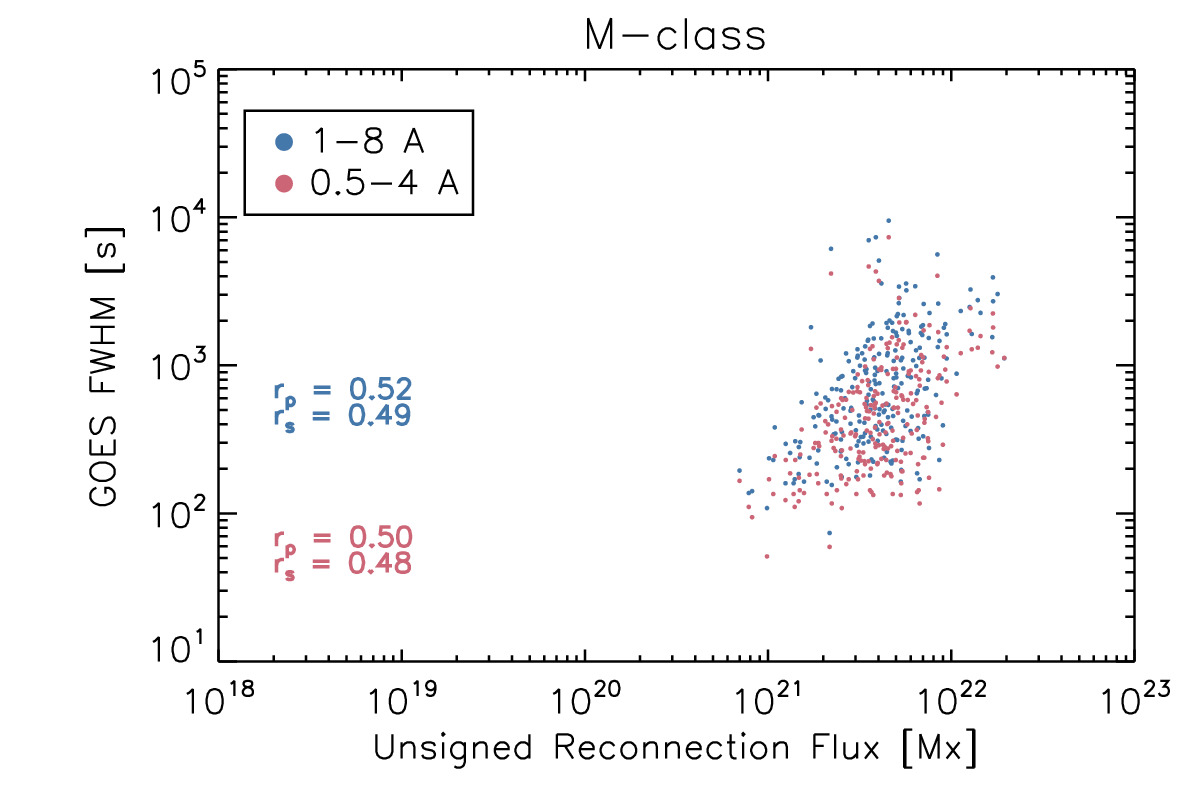}
\includegraphics[width=0.5\linewidth]{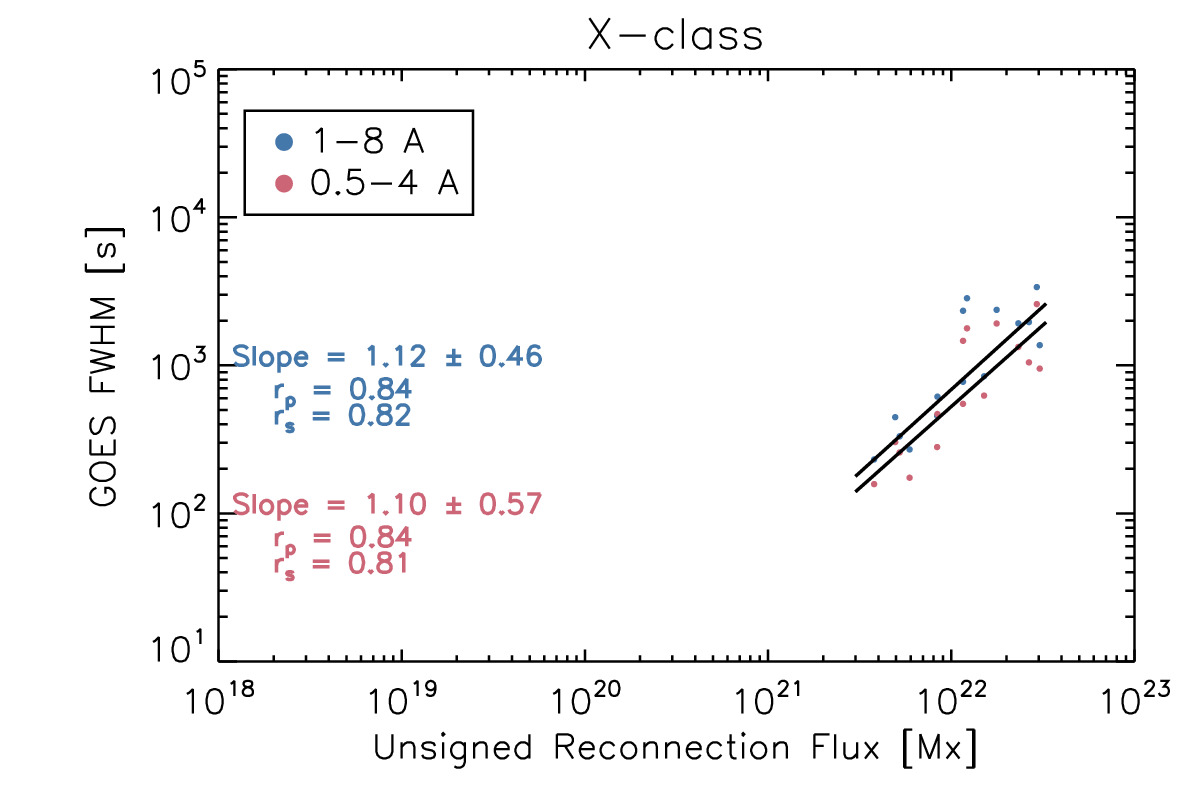}
\caption{Similar to Figure \ref{fig:duration_srbn}, except showing the relation between the flare duration and the reconnection flux.  The largest flares are approximately linearly related to the flux, but there is considerable scatter. \label{fig:duration_phirbn}}
\end{figure*}
The magnetic flux similarly does not appear to have an obvious relation across all flare sizes, but a trend does develop in larger flares, as shown in Figure \ref{fig:duration_phirbn}.  We naively expect that the duration should depend on how much flux is swept out by the flare ribbon, but the rate may vary due to changes in the ribbon speed, for example \citep{asai2004}.  In C-class flares, there is no relation between the flux and duration, in either GOES channel.  In X-class flares, there does appear to be a linear trend that longer flares sweep out more magnetic flux, but we again caution that there are too few of these events to strongly conclude anything.   

\begin{figure*}
\includegraphics[width=0.5\linewidth]{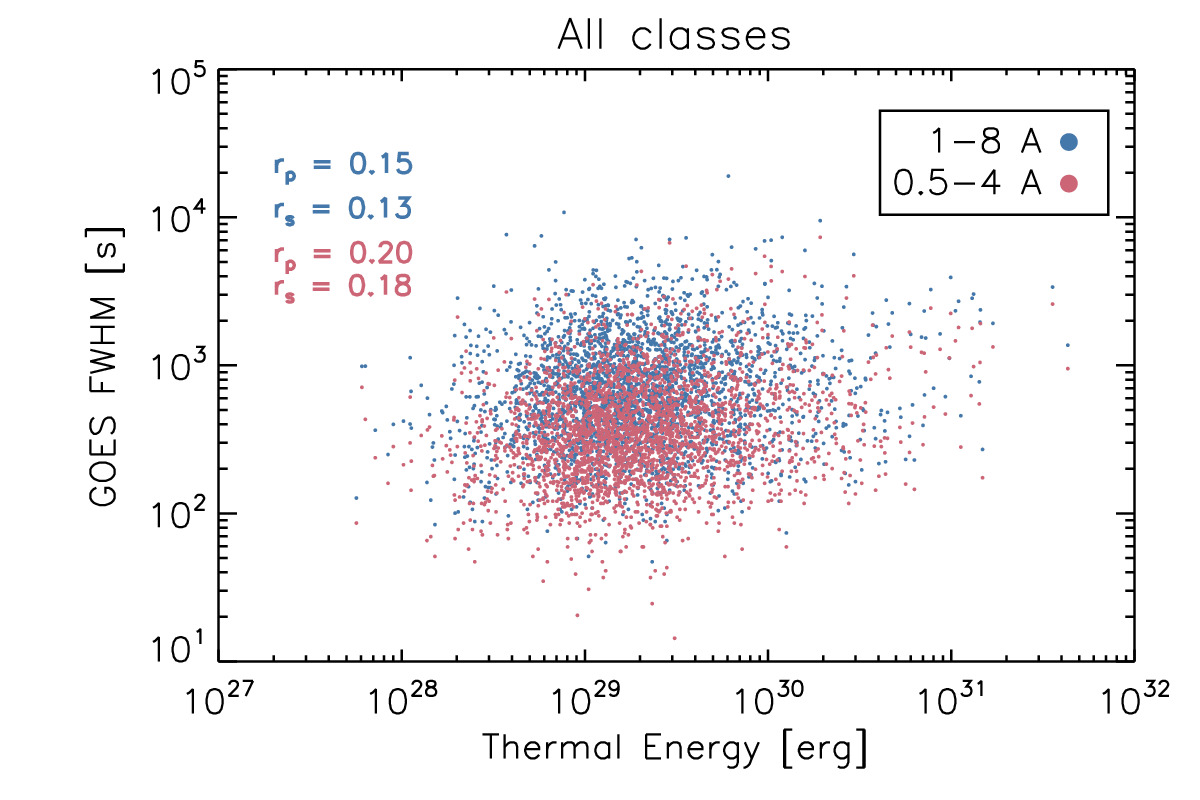}
\includegraphics[width=0.5\linewidth]{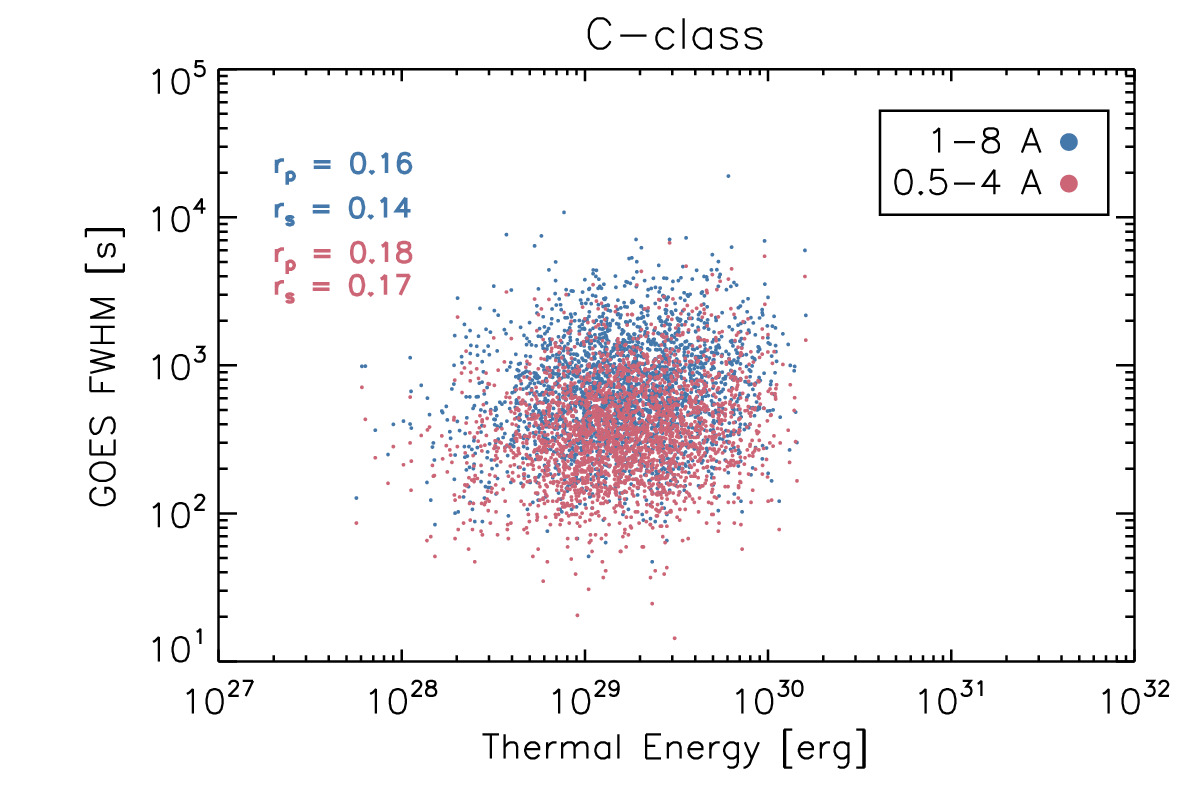}
\includegraphics[width=0.5\linewidth]{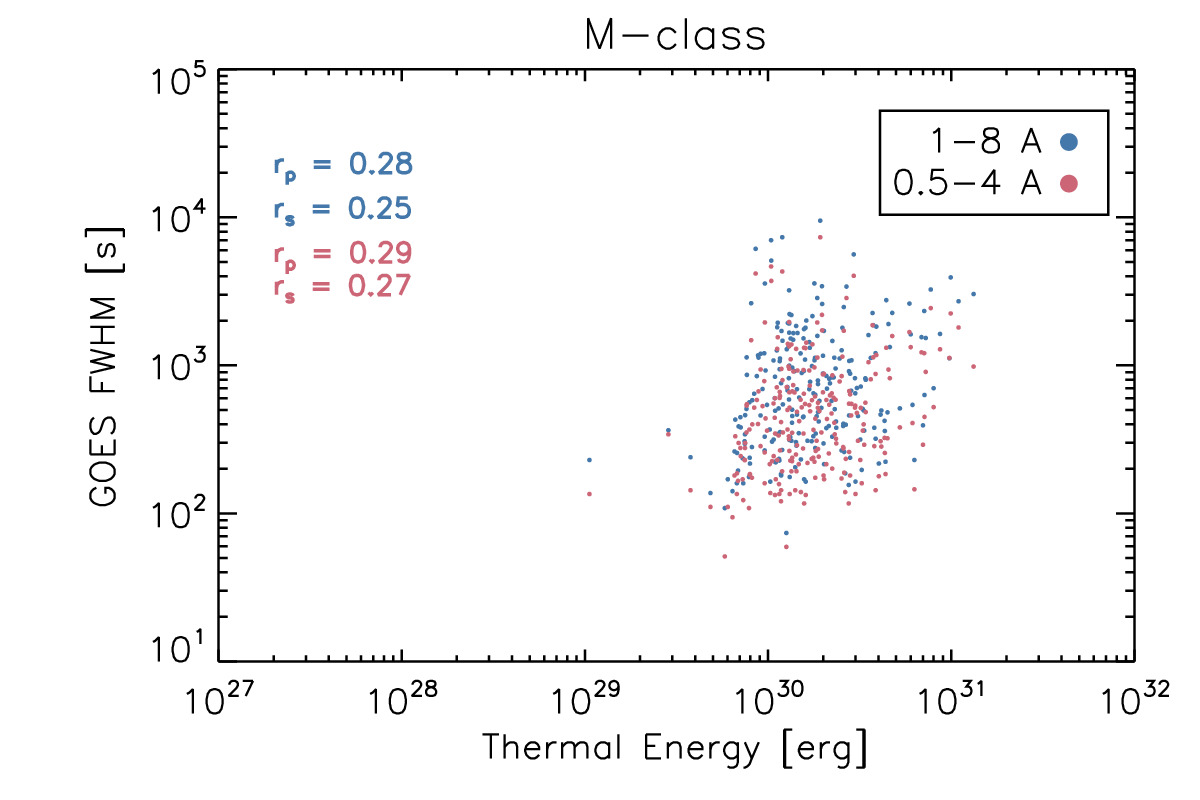}
\includegraphics[width=0.5\linewidth]{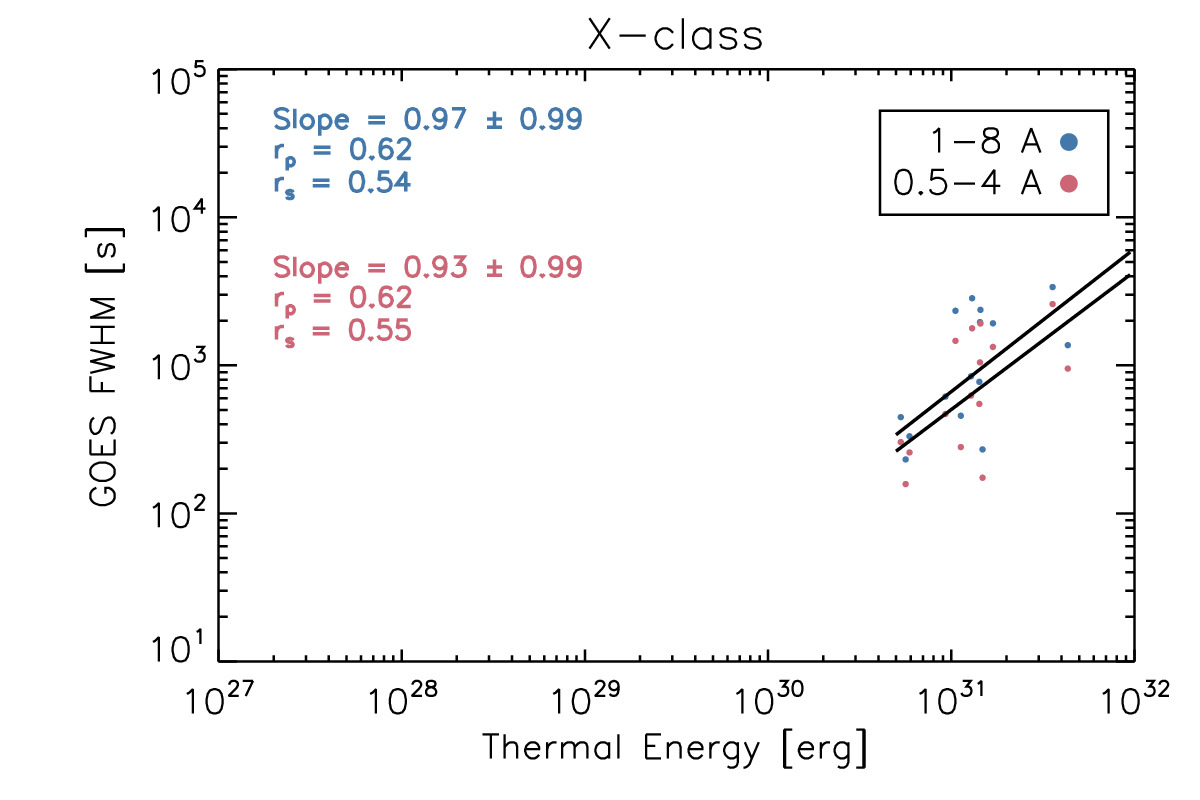}
\caption{The relationships between the flare thermal energy and and the GOES FWHM, when unsorted (top left), and sorted into C-class (top right), M-class (bottom left), and X-class flares (bottom right).  The duration of smaller flares appears to be independent of the thermal energy content, while the X-class flares appear to show a linear relation with the energy. \label{fig:duration_thermal}}
\end{figure*}
\begin{figure*}
\includegraphics[width=0.5\linewidth]{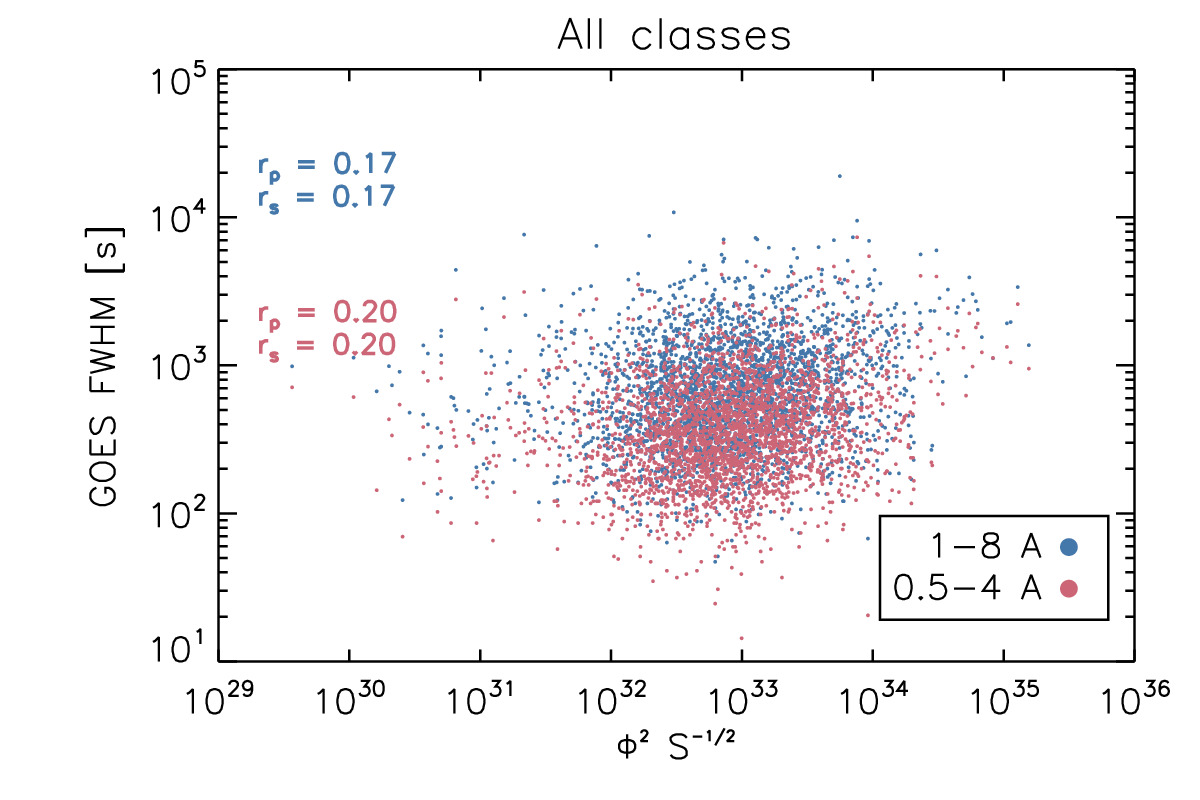}
\includegraphics[width=0.5\linewidth]{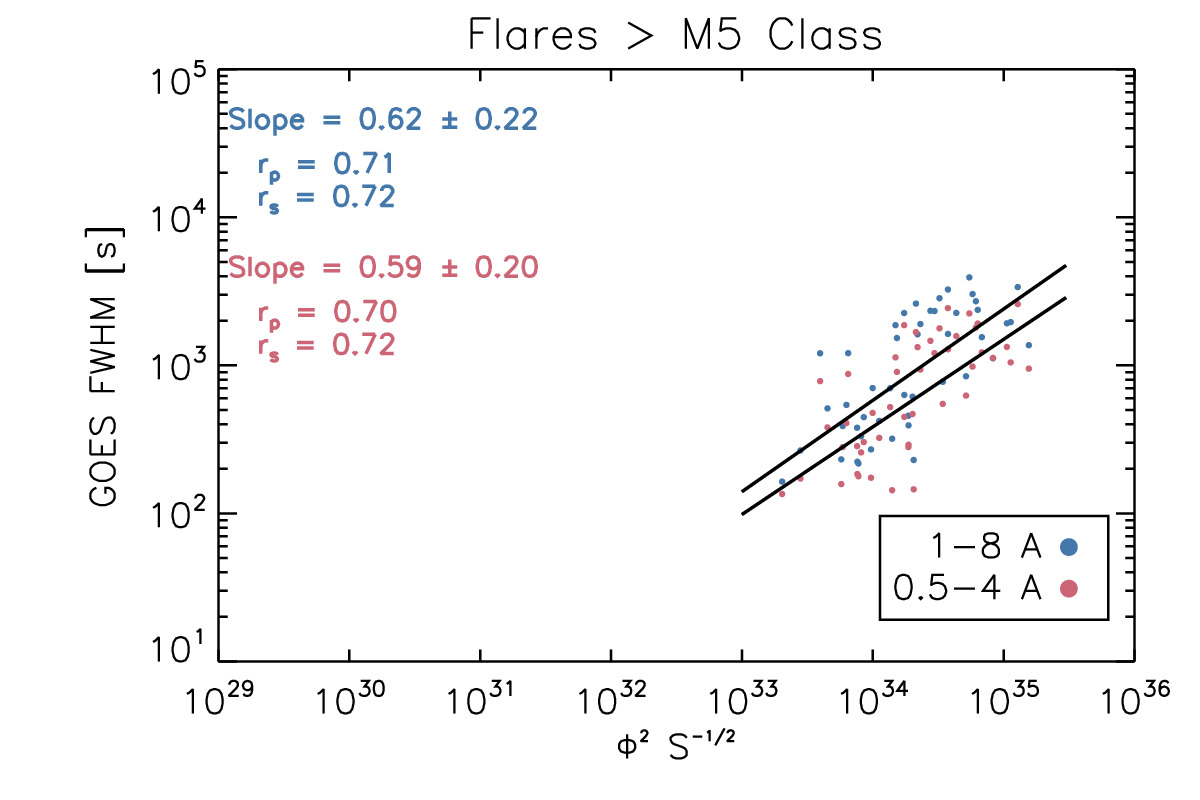}
\caption{The relationships between the flare magnetic energy $E_{\text{mag}} \approx \Phi_{\text{ribbon}}^{2} S_{\text{ribbon}}^{-1/2}$ (see text) and and the GOES FWHM, when unsorted (left), and sorted into flares above M5 (right) as way of comparison to \citet{toriumi2017a}.  The full set of flares is independent of the magnetic energy content, while the larger flares appear to show a sub-linear dependence on the energy ($\tau_{\text{FWHM}} \propto E_{\text{mag}}^{3/5}$).  \label{fig:duration_magnetic}}
\end{figure*}

Next, we show the duration plotted against the thermal energy content in Figure \ref{fig:duration_thermal}.  When unsorted, there does not appear to be any relation between the duration and energy content of a flare.  For X-class flares, there visually appears to be a linear relation between the two variables, but the uncertainties are so large that we cannot conclude that this is real.  Additionally, due to the relative rarity of X-class flares, however, it is not clear if there are sufficiently many data points to conclude that the linear dependence is real, especially since the M-class flares do not show the same dependence.  For the C- and M-class flares, in fact, the FWHM ranges from around a minute to a few hours, without any obvious relation to the energy content.  Interestingly, \citet{toriumi2017a} found a correlation between the FWHM (or decay time) of the 1--8\,\AA\ channel and the magnetic energy.  Expressing the magnetic energy as:
\begin{align}
\label{eqn:emag}
E_{\text{mag}} &= \frac{B^{2}}{8\pi} V \\ \nonumber
			&\propto B^{2} S_{\text{ribbon}} d_{\text{ribbon}}
\end{align}
\noindent where $V \approx S_{\text{ribbon}} d_{\text{ribbon}}$ is the flare's volume.  In their set of flares, they find that $\tau_{\text{FWHM}} \propto E_{\text{mag}}^{0.45}$ with a Pearson correlation coefficient of 0.81.  They similarly find a correlation with the e-folding decay time.  In our current set of flares, we do not have $B_{\text{ribbon}}$ or $d_{\text{ribbon}}$, but we can rewrite Equation \ref{eqn:emag} by using the magnetic flux in place of the field strength, $\Phi_{\text{ribbon}} = B_{\text{ribbon}} S_{\text{ribbon}}$, and approximating the volume less accurately with $V \approx S_{\text{ribbon}}^{3/2}$.
\begin{align}
E_{\text{mag}} \propto \Phi_{\text{ribbon}}^{2} S_{\text{ribbon}}^{-1/2}
\end{align}
\noindent We plot this approximation to the magnetic energy against the FWHM in Figure \ref{fig:duration_magnetic}.  On the left, we show the full set of flares without restriction, while on the right, we restrict the data to flares greater than M5, which is the same restriction used by \citet{toriumi2017a}.  As with many of the other variables, we find that the magnetic energy is independent of the GOES FWHM when plotted against all of the flares, but there may be a correlation when restricted to larger flares.  The slope is found to be $0.62 \pm 0.22$ in the 1--8\,\AA\ channel, approximately consistent with \citet{toriumi2017a}.  

This lack of correlation between the reconnection flux or energy and the FWHM is particularly surprising, as studies of stellar flares have claimed to find a relation between the energy release and duration of flares (\textit{e.g.} \citealt{hawley2014,maehara2015,namekata2017}).  A major difference between solar and stellar flares is that the size, geometry, emergence rate, and decay rate of star spots can vary quite significantly for different stars \citep{namekata2018}, though this may not explain the differences we have found.  In stellar flares, scaling laws have been proposed that relate flare duration to the energy release.  We reiterate the basic derivation here, and show that they do \textit{not} hold for GOES SXR light curves (at least with the current data set).

\begin{figure*}
\centering
\includegraphics[width=0.32\linewidth]{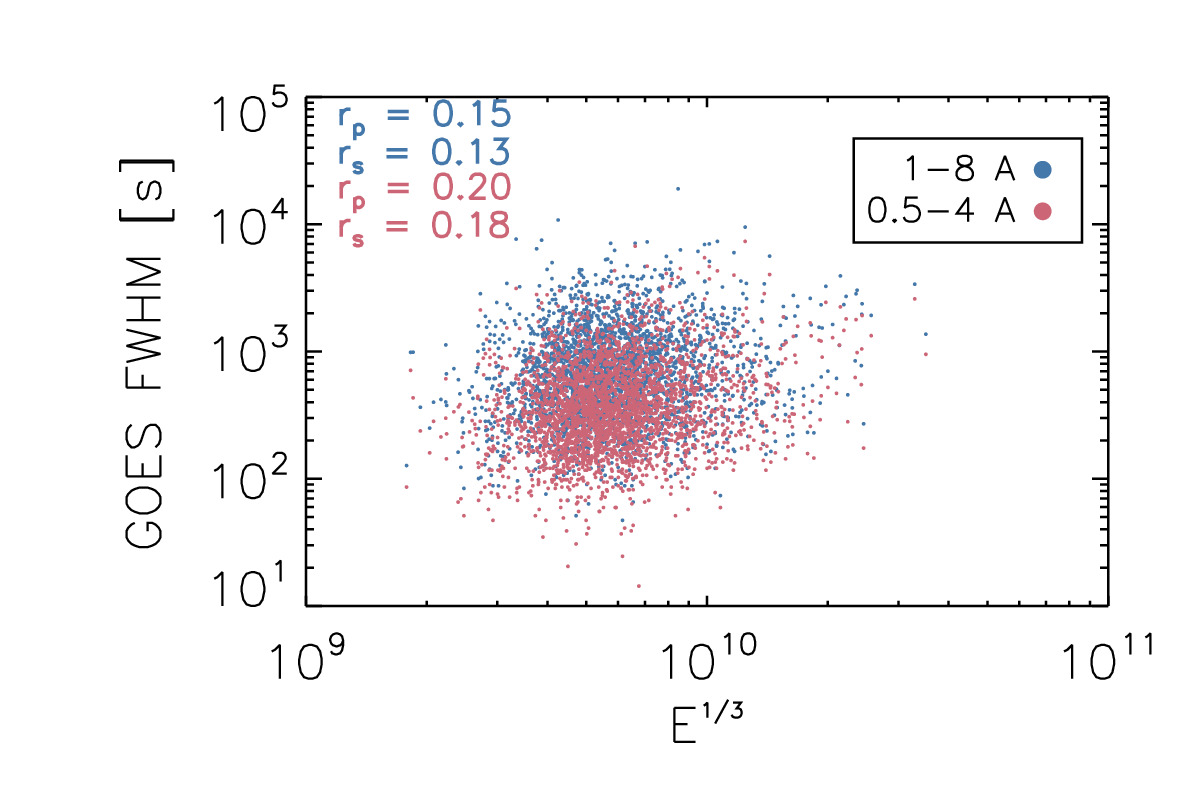}
\includegraphics[width=0.32\linewidth]{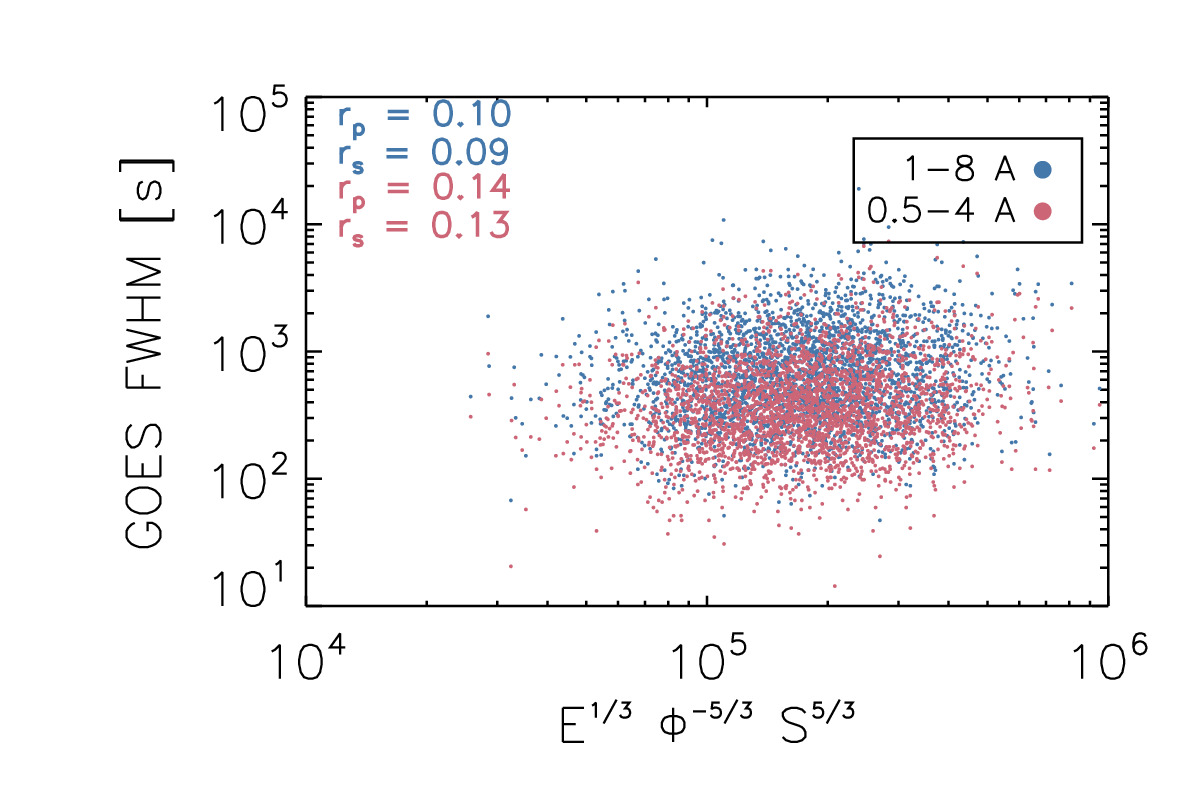}
\includegraphics[width=0.32\linewidth]{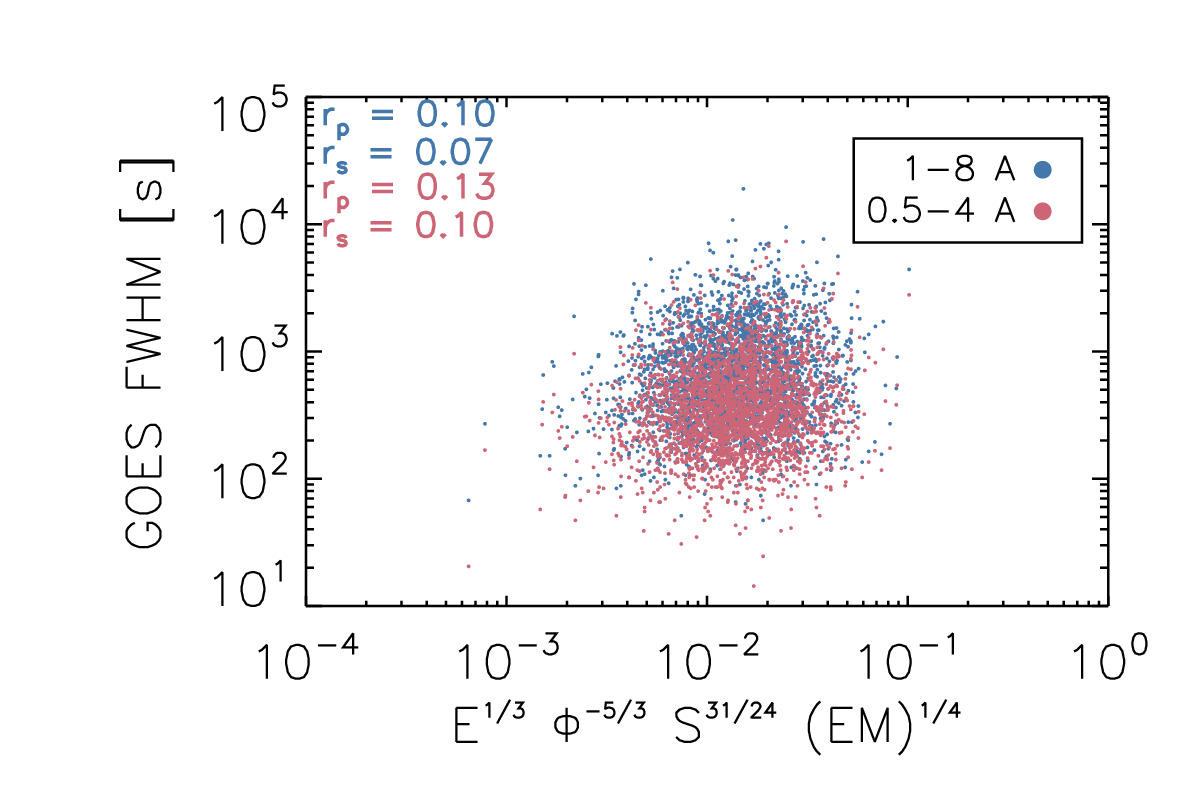}
\caption{Comparison of the derived scaling law for the duration of flares against the thermal energy, magnetic field, and density.  There is no evidence that the scaling laws hold in this data set.  \label{fig:scaling1}}
\end{figure*}

Following the derivation of \citet{maehara2015}, we define a magnetic reconnection time-scale $t$:
\begin{equation}
    t = \frac{L}{v_{A} R}
\end{equation}
\noindent where $L$ is the loop length, $v_{A}$ the Alfv\'en speed, and $R$ the magnetic reconnection rate.  We also express the magnetic energy $E$ simply:
\begin{equation}
    E = \frac{B^{2}}{4 \pi} L^{3} 
\end{equation}
\noindent where $B$ is the field strength.  This is an upper limit, that assumes that all of the magnetic energy is converted to thermal energy, but the fraction of converted energy certainly would vary from flare to flare.  In the usual approximation, the reconnection rate $R$ is taken to be constant $\approx 0.1$ (\textit{e.g.} \citealt{liu2018}), which allows one to solve for $L = v_{A} t$, and substitute:
\begin{align}
    E &\propto B^{2} (v_{A} t)^{3} \\ \nonumber
      &\propto B^{2} \frac{B^{3}}{\rho^{3/2}} t^{3}
\end{align}
\noindent where $\rho$ is the mass density.  Finally, solving for $t$, we find the scaling law:
\begin{equation}
    t \propto E^{1/3} B^{-5/3} \rho^{1/2}
    \label{scalinglaw}
\end{equation}
\noindent This scaling law was derived by \citet{maehara2015}, who found a similar trend in superflare observations on G-type stars seen by Kepler.  They argue that the magnetic field and density should be comparable on all G-stars, so that $t \propto E^{1/3}$ should hold, and they find a linear fit with slope close to 1/3.  In Figure \ref{fig:scaling1}, we show that the scaling law does \textit{not} hold in GOES flares.  Writing the magnetic field $B = \Phi / S_{\text{ribbon}}$ and $\rho \approx n \approx \sqrt{\frac{EM}{S_{\text{ribbon}}^{3/2}}}$, we can rewrite Equation \ref{scalinglaw}:
\begin{align}
	t \propto E^{1/3} \Phi^{-5/3} (EM)^{1/4} (S_{\text{ribbon}})^{31/24} 
\end{align}
\noindent In Figure \ref{fig:scaling1}, using the measured variables, we show the duration plotted (left) against $E^{1/3}$, (center) against $E^{1/3} B^{-5/3} \propto E^{1/3} \Phi^{-5/3} (S_{\text{ribbon}})^{5/3}$, and (right) against $E^{1/3} B^{-5/3} \rho^{1/2} \propto E^{1/3} \Phi^{-5/3} (EM)^{1/4} (S_{\text{ribbon}})^{31/24}$.  There is no relation between the FWHM and these combination of variables, so we conclude that the scaling laws do not hold for this set of flares.  Although we do not show the plots here, the scalings also do not hold when limited to larger flares only (in fact, the X-class flares have $\tau_{\text{FWHM}} \propto E$ -- see Figure \ref{fig:duration_thermal}).  

The scaling laws may not hold here for a few reasons.  First, the flare volume is not well approximated by $L^{3}$, but should more accurately be written $S_{\text{ribbon}} \times L$ (perhaps with a filling factor), and we have used the poor approximation $L \propto S^{1/2}$.  Second, the reconnection rate $R$ is not necessarily constant from flare to flare, since it depends on both the Alfv\'en speed and density.  Third, the definition of ``duration'' needs to be stringently defined, and depends on what physical processes produce a given light curve, which importantly means that different wavelengths could have rather different durations.  Finally, and most importantly, the definition of the reconnection time-scale as $L / v_{A}$ is perhaps incorrect.  The duration $\tau$ scales linearly with the ribbon separation $\tau \propto d_{\text{ribbon}} \propto L$ \citep{toriumi2017a}, but the time-scale for the ribbon to spread apart is essentially determined by the velocity of the ribbon (on the order of $\approx 20$\,km\,s$^{-1}$, \citealt{asai2004}), which is likely unrelated to the Alfv\'en speed.  Using a multi-threaded hydrodynamic model that only varied the loop lengths and the ribbon speed, \citet{reep2017} showed that this reproduces the linear relation between the ribbon separation and the GOES FWHM (or e-folding decay time).  We consider it likely that determining the cause of the ribbon speed would lead to a better scaling law, though it is not clear how the ribbon speed relates to a flare's energy (or even if it does).  Interestingly, \citet{krucker2005} found that there is a clear correlation between the time variation of the ribbon speed and the HXR flux observed by RHESSI.  

\section{Discussion}
\label{sec:discussion}

We summarize the relationships we have found in Table \ref{table:corr}.  We indicate whether a variable is related to the GOES flux of a flare and whether it is related to the GOES FWHM of a flare.  We first note that the flare's duration and SXR flux are not related.  
\begin{table}[h]
\caption{Table showing whether a pair of variables is correlated. `---' indicates no correlation, a check mark indicates a correlation, and the letter $X$ near the check mark indicates that the correlation is only for X-class flares. \label{table:corr}}
\begin{tabular}{lll}
                   & FWHM$_{GOES}$  & Flux$_{GOES}$           \\
                   \hline
FWHM$_{GOES}$        &    &   ---              \\
Flux$_{GOES}$        &      ---   &           \\
$\Phi$ &                \checkmark (X)          &    \checkmark            \\
$EM_{max}$            &     ---                     &      \checkmark \footnote{There are clearly two different distributions, see Figure 5.}         \\
Therm. En.    &      ---                    &       \checkmark                      \\
AR area            &  ---                        &   ---                       \\
Ribbon area        &      \checkmark (X)                  &      \checkmark               \\
$T_{max}$          &     ---                    &              \checkmark           \\
EM ($t_{peak}$)        &    ---                      &        \checkmark            
\end{tabular}

\end{table}
The duration of the flares, measured here with the FWHM in both GOES channels, is not correlated with \textit{any} of the measured parameters when considering the whole data set.  Flare duration is independent of the flare class, the peak temperature, the peak EM, the thermal energy content, the ribbon area, the active region area, and the reconnection flux.  The same range of durations is found for X-class flares as for smaller flares.\footnote{A larger sample of flares that is not restricted to those in the RibbonDB returns the same results: the SXR duration of a flare is completely independent of the flare class.}   When restricting the durations within the different classes, however, some trends do appear to develop in larger flares, and the cause of the discrepancy is discussed below.  The total distribution of flare SXR durations is well-described by a log-normal function, with median values of around 11 and 6 minutes in the low and high energy GOES channels, respectively.  These distributions are comparable to those found by other studies (\textit{e.g.} \citealt{veronig2002}), though many authors have concluded that the distribution is a power law \citep{christe2008}.  

 \citet{toriumi2017a} presented compelling evidence that flare durations scale with the separation of the two ribbons in a sample of 51 flares larger than M5, finding a linear scaling between the duration and ribbon separation.  \citet{reep2017} showed that this implies that there is a direct connection between the duration of magnetic reconnection and the duration of the GOES light curves, which is due to the increasing length of the loops that span the ribbon separation.  There are two important parts to this.  First, the cooling time of a coronal loop scales directly with the length $\propto L^{5/6}$ \citep{cargill1995}, so longer lengths result in SXR light curves that take longer to fade.  Second, as long as reconnection continues, new loops form with longer and longer lengths as the ribbons separate, thus extending the fading time.  
 
It is not clear, however, how the energy release changes with time in a reconnection event -- is it gradual or impulsive?  Because the HXR burst is extremely impulsive, lasting only a few minutes at most, and electron beams are thought to be the primary heating mechanism in flares, it has traditionally been thought that flare heating must be impulsive.  However, there have been many indications that show that there is persistent heating well into the gradual phase of flares, ranging from high temperatures \citep{svestka1982}, high densities \citep{moore1980}, to on-going evaporation flows \citep{czaykowska1999,czaykowska2001}.  In fact, a recent study by \citet{kuhar2017} has shown that the energy released during the gradual phase is at least an order of magnitude \textit{larger} than the energy released during the impulsive phase.  Models of on-going reconnection have been shown to be compatible with the late-phase properties of flares \citep{cargill1983}, and multi-threaded modeling of post-flare loops have been able to reproduce long duration light curves observed with GOES and SDO/AIA \citep{li2014,qiu2016,zhu2018}.  

The results of the present work imply that the duration of reconnection may not be directly related to the total energy release of that reconnection.  In the total data set, there is no correlation between the SXR durations and the temperatures, EMs, or reconnection flux, all of which are found to be proportional to the total energy release in a given flare.  In contrast, studies of stellar flares have found a relation between the flare duration and the energy release (\textit{e.g.} \citealt{maehara2015}): $\tau \propto E^{1/3}$, which does not appear to hold in GOES events.  Different wavelength emission, produced by different physical mechanisms (\textit{e.g.} HXRs produced by non-thermal \textit{bremsstrahlung}), would certainly have different durations and may therefore have a different relation to the energy release.  

\begin{figure*}
\centering
\includegraphics[width=0.49\linewidth]{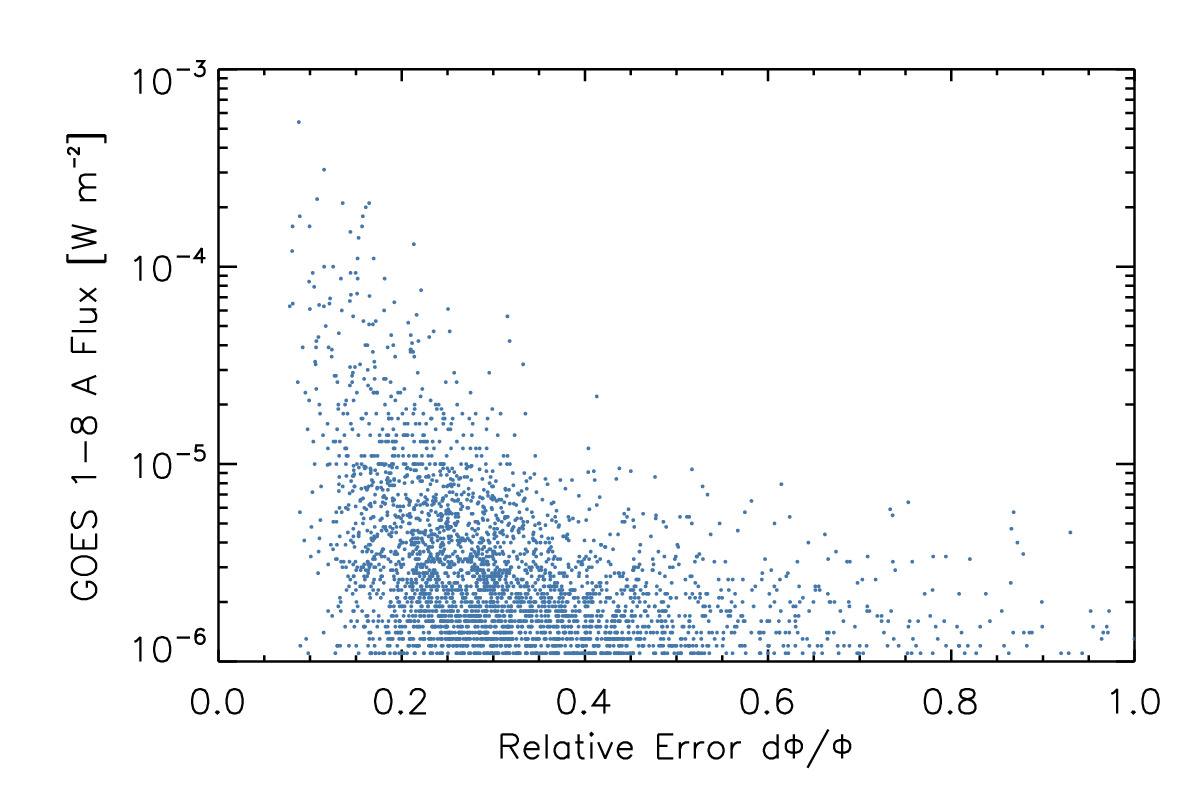}
\includegraphics[width=0.49\linewidth]{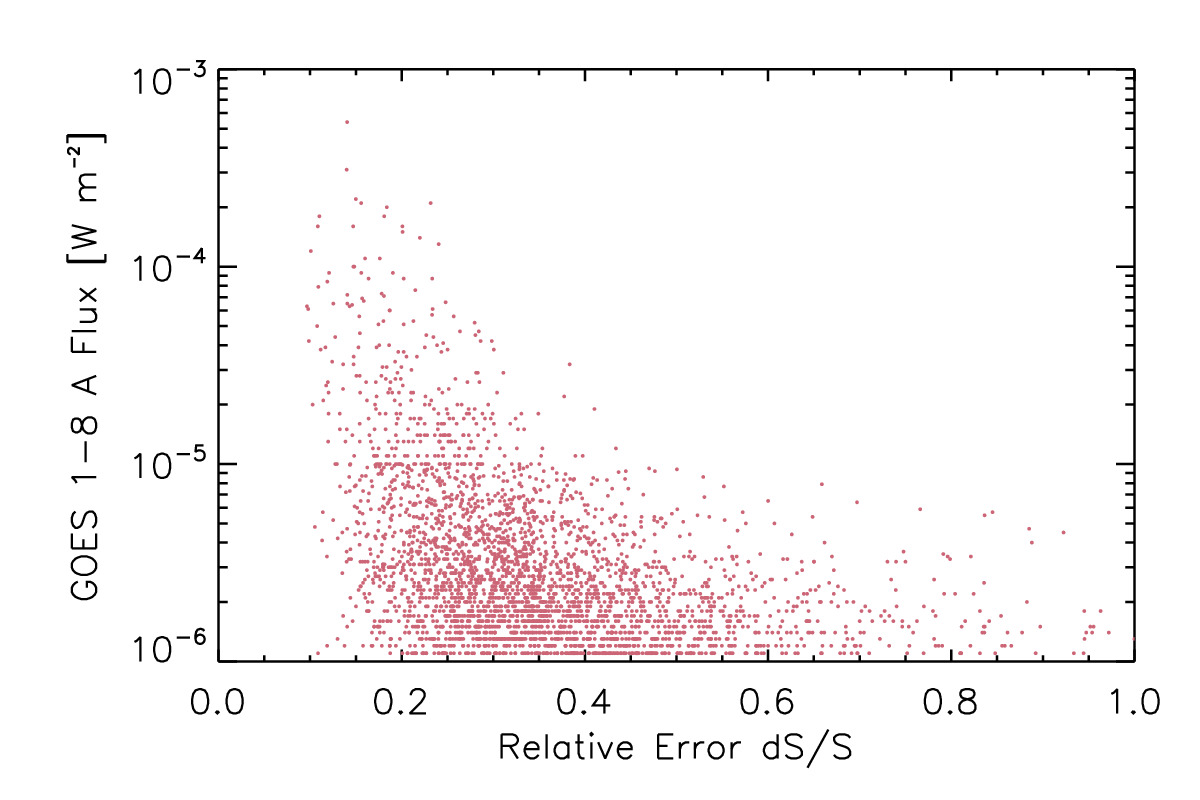}
\caption{The relative errors of the magnetic reconnection flux $\frac{d\Phi}{\Phi}$ (left) and of the ribbon area $\frac{dS_{\text{ribbon}}}{S_{\text{ribbon}}}$ (right) compared against the GOES class (not background subtracted).  The largest flares have the smallest relative errors.  \label{fig:error}}
\end{figure*}

Why, then, are there correlations between the duration and some of these variables for larger flares but not across the whole data set?  We consider three issues.  First, the relative errors associated with the ribbon area and magnetic flux are larger in smaller flares.  That is, the measurements of $\Phi_{\text{ribbon}}$ and $S_{\text{ribbon}}$ are less accurate in smaller flares (and therefore derived parameters like the thermal energy are also less accurate).  Additionally, the GOES-derived EM and temperatures are much more sensitive to accurate background subtraction in smaller flares than larger ones, which could skew those values.  For example, in Figure \ref{fig:error}, we show the relative errors of the magnetic reconnection flux $\frac{d\Phi}{\Phi}$ and of the ribbon area $\frac{dS_{\text{ribbon}}}{S_{\text{ribbon}}}$ compared with the GOES class in the RibbonDB.  The X-class flares have errors of around 15--20\%, while the C-class flares have errors of 15--100\% (though 87\% of the flares have errors of less than 50\%).  It is clear that the signal-to-noise ratio can impact the observations, and this has a large effect on the measurements of smaller flares.  

\begin{figure*}
\centering
\includegraphics[width=0.49\linewidth]{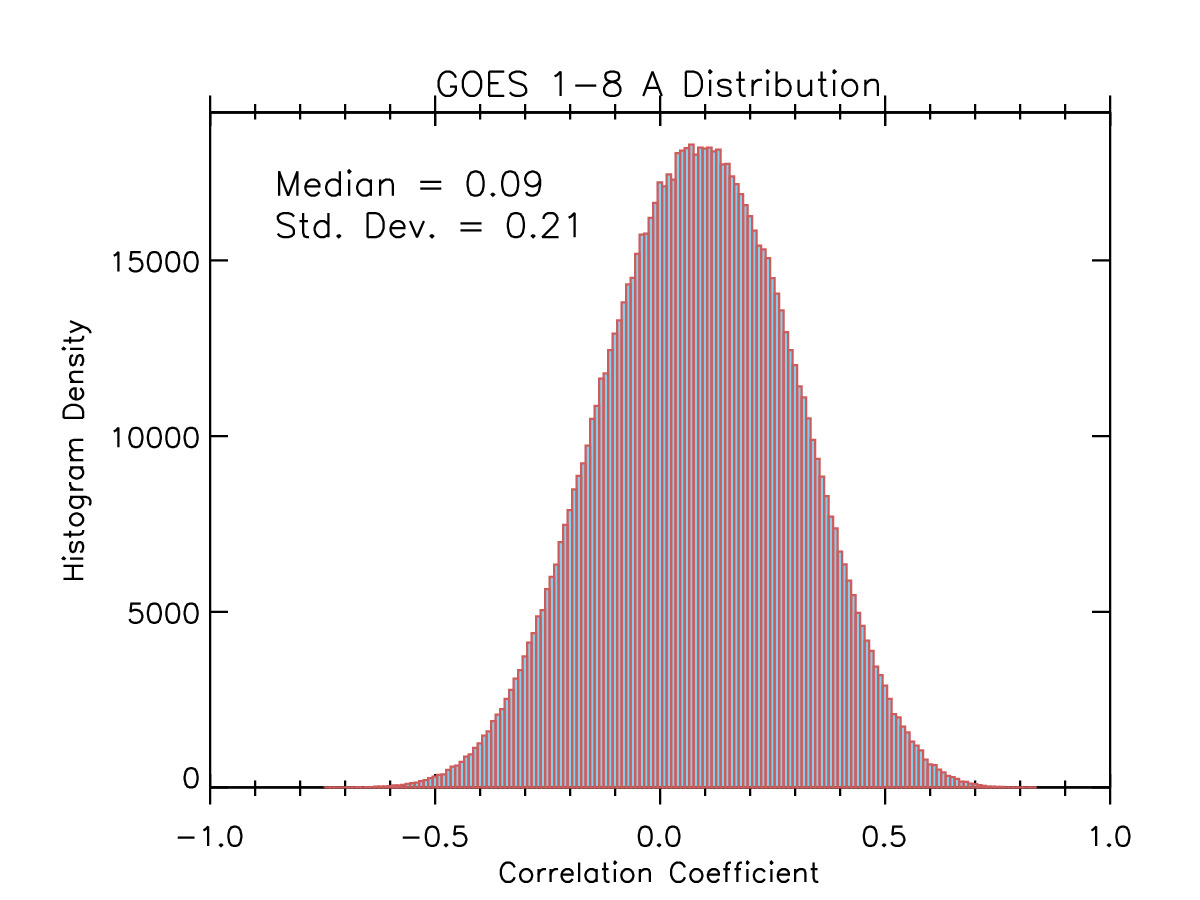}
\includegraphics[width=0.49\linewidth]{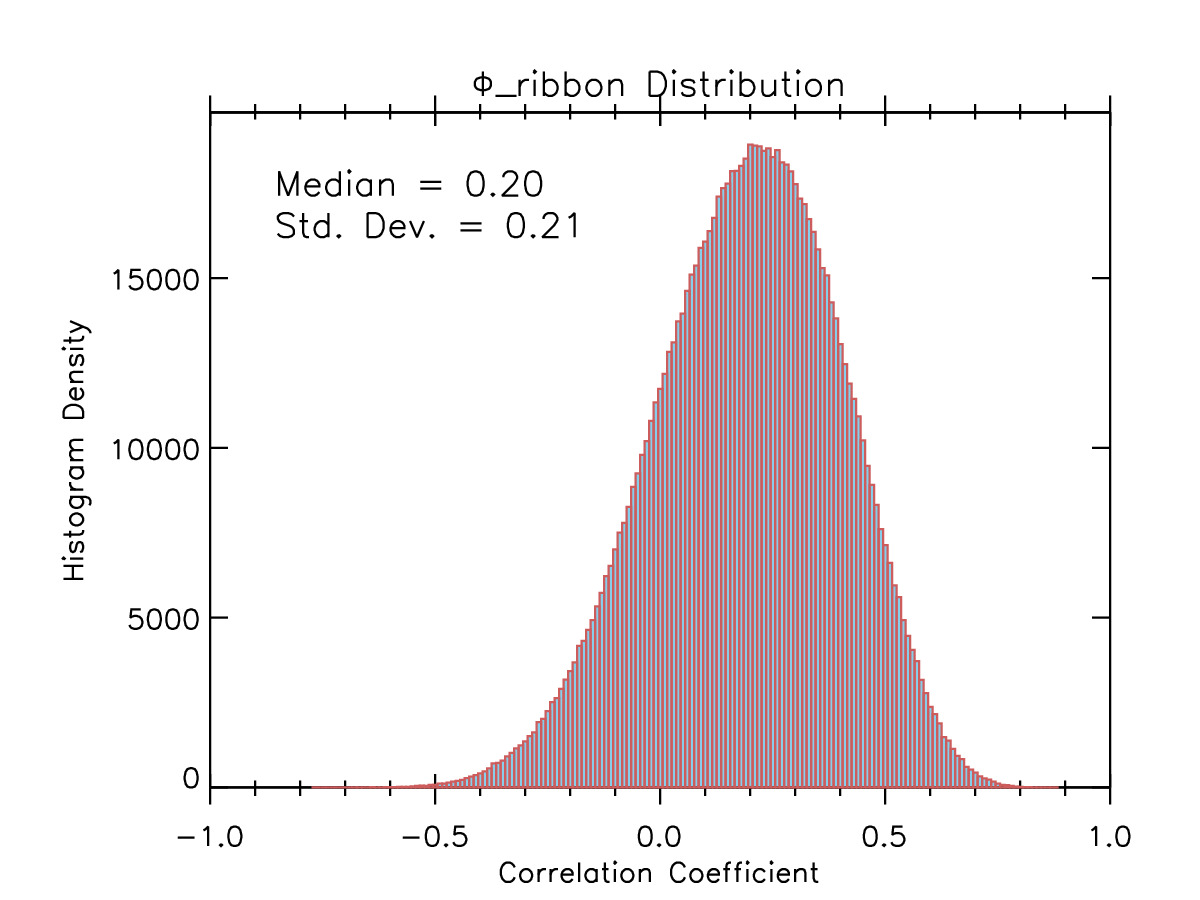}
\includegraphics[width=0.49\linewidth]{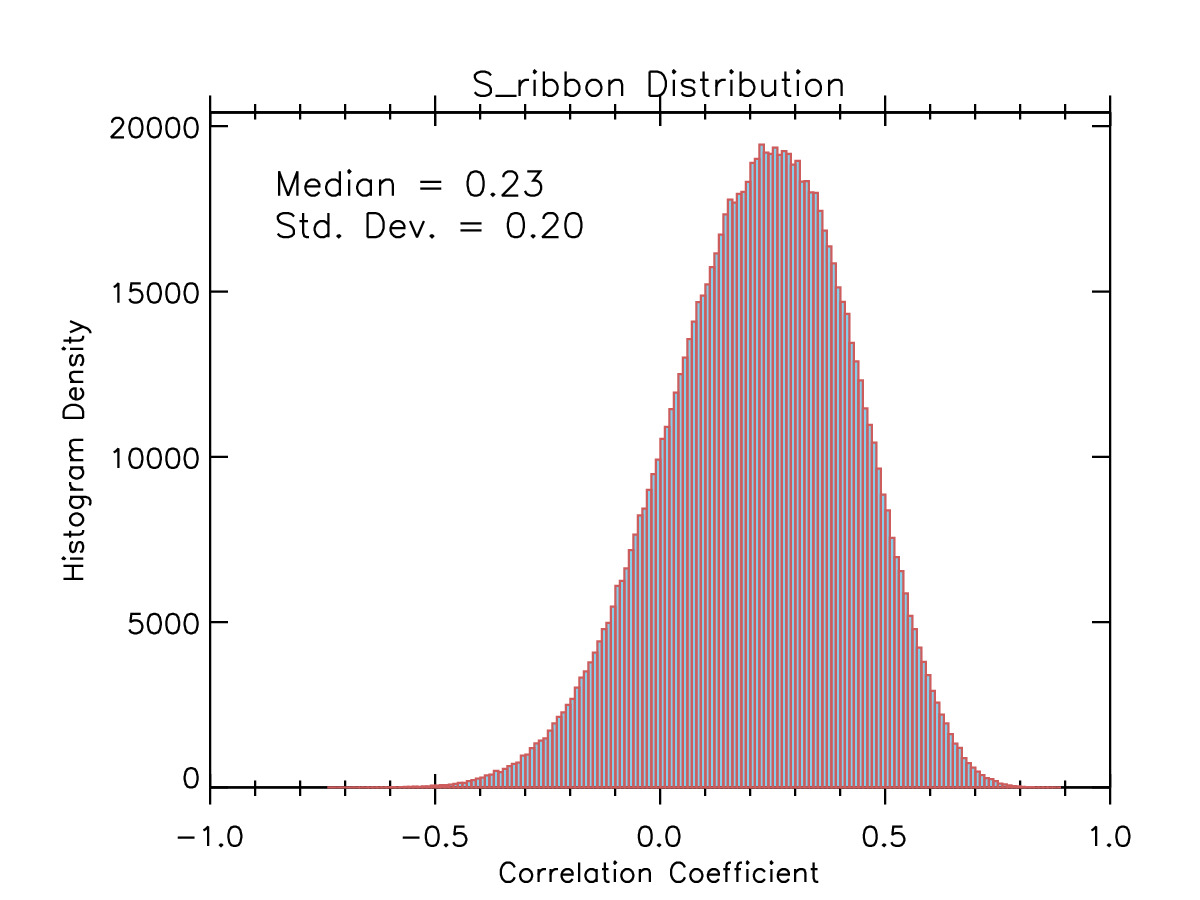}
\includegraphics[width=0.49\linewidth]{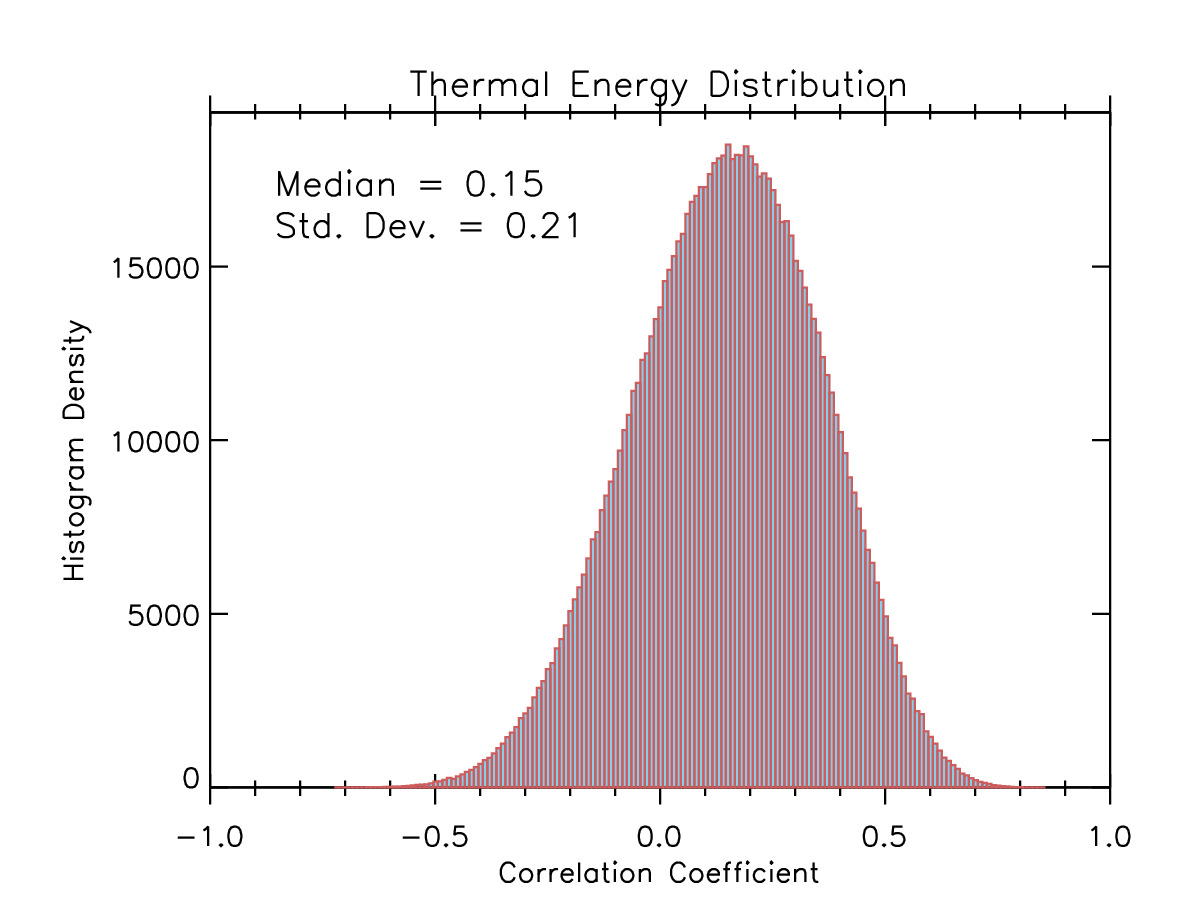}
\caption{A Monte Carlo calculation of the distribution of Pearson correlation coefficients for 25 randomly selected C-class flares, repeated $10^{6}$ times, correlating the FWHM in the 1--8\,\AA\ GOES channel with the flux in the 1--8\,\AA\ channel (top left), the magnetic reconnection flux $\Phi_{\text{ribbon}}$ (top right), the ribbon area $S_{\text{ribbon}}$ (bottom left), and the thermal energy content (bottom right).  Because the peaks are close to the correlations measured across the whole set (respectively: 0.092, 0.179, 0.209, and 0.148), we conclude that sample size is not a major issue.  \label{fig:montecarlo}}
\end{figure*}

Second, there is a selection effect.  The distribution of flares is a power law with slope $\approx -2$ (\textit{e.g.} \citealt{hudson1991}), so the occurrence of X-class flares is 100 times less common than M-class flares and 10000 times less common than C-class flares.  In our data set, there are only 15 X-class flares, which is not a large enough sample to produce statistically significant correlations.  In order to test whether the sample size is important, however, we can run a simple Monte Carlo test where we randomly select 25 C-class flares from the data set and calculate the Pearson correlation coefficients between the FWHM in the 1--8\,\AA\ channel with the 1--8\,\AA\ flux, the magnetic reconnection flux $\Phi_{\text{ribbon}}$, the ribbon area $S_{\text{ribbon}}$, and the thermal energy.  We repeat this calculation $10^{6}$ times, and create histograms of the distribution of correlation coefficients.  If the sample size is unimportant, then we should find an average correlation coefficient close to the value of the whole set.  If the sample size is important, then the two should differ significantly.  In Figure \ref{fig:montecarlo}, we show these four histograms, and note that peak of the distributions are approximately equal to the correlation coefficients measured in the whole data set.  Working with the assumption that the physical processes responsible for C, M, and X-class flares are the same, we conclude that a larger sample of X-class flares is unlikely to change the quality of the correlation (or lack thereof) between the GOES FWHM, flux, and other variables.  

Finally, there is the possibility that the energy release or reconnection event is somehow different in smaller flares than for larger flares.  We cannot rule it out based on this data alone, and it does, in fact, have some precedent.  For example, coronal mass ejections (CMEs) are more likely to occur in larger flares \citep{yashiro2009,youssef2012}, though the presence of CMEs is not related to flare duration (see \citealt{harra2016}).  There are also indications that larger flares accelerate particles more efficiently than smaller ones, such as the measurement of a higher low-energy cut-off and lower spectral index in the electron beam \citep{hannah2008}.  It may not be the case that parameters can simply be scaled in larger flares from smaller flares, and that the properties of the reconnection itself may change in larger events, which is beyond the scope of the present work.  

We summarize our basic findings here:
\begin{enumerate}
\item The GOES flux is related strongly with the temperature, EM, flare ribbon area, reconnected flux, and thermal energy content. 
\item The GOES flux scales approximately linearly with the thermal energy, a tell-tale sign of the Neupert effect \citep{lee1995}.  This contradicts previous studies that have found a super-linear correlation. 
\item The duration of GOES light curves is completely independent of the GOES class. 
\item The distribution of flare SXR durations is consistent with log-normal, with a median FWHM of around 11 minutes in the GOES 1--8\,\AA\ channel, and about 6 minutes in the GOES 0.5--4\,\AA\ channel.  It is \textit{not} a power law distribution.
\item The duration of GOES light curves is correlated with the magnetic reconnection flux and ribbon area of the flare in larger (X-class) flares, and possibly with the thermal energy.  They appear uncorrelated in smaller flares, which is likely due to the large errors in the measurements, though it remains possible that there are fundamental differences in the physical processes driving the flares. 
\item The duration of GOES light curves is independent of the maximum temperature, the active region area, and the EM in all flares.
\item The measured duration of a flare depends strongly on the wavelength in which the duration is calculated.  SXR, HXR, and white light emission are all produced by different emission mechanisms and at different heights in the atmosphere, so correlations found in one wavelength band may not hold for another. 
\end{enumerate}

\leavevmode \newline

\acknowledgments  
JWR was supported by a Jerome \& Isabella Karle Distinguished Scholar Fellowship for this work.  KJK was supported for this work by the Chief of Naval Research through the National Research Council.  The authors thank the two anonymous referees whose comments have strengthened this paper.  The authors thank Harry Warren for comments on a draft of this work.  All of the plots in this work were made using color blind-friendly color tables, developed by Paul Tol (\href{https://personal.sron.nl/~pault/}{https://personal.sron.nl/{\raise.17ex\hbox{$\scriptstyle\sim$}}pault/}).  This work made use of the JBIU IDL library, available at \href{http://www.simulated-galaxies.ua.edu/jbiu/}{http://www.simulated-galaxies.ua.edu/jbiu/}.

\bibliography{apj}
\bibliographystyle{aasjournal}

\end{document}